# Autoregressive conditional duration modelling of high frequency data

Xiufeng Yan

Phd, University of Sussex, 2021
School of Mathematical and Physical Sciences
University of Sussex

BN1 9QH

**Abstract**

This paper explores the duration dynamics modelling under the Autoregressive Conditional Durations (ACD) framework (Engle and Russell 1998). I test different distributions assumptions for the durations. The empirical results suggest unconditional durations approach the Gamma distributions. Moreover, compared with exponential distributions and Weibull distributions, the ACD model with Gamma distributed innovations provide the best fit of SPY durations.

**Table of content**

# List of Tables

## List of Figures

# Autoregressive Conditional Duration modelling of durations

1. **Introduction**

The rising availability of tick-by-tick transaction data in financial markets provides the possibility to directly investigate the characteristics associated with each tick transaction record. There are three main characteristics of each tick transaction record: execution price, trading volumes and execution time. A distinctive feature of the tick-by-tick transaction data is that the observations are no longer equally spaced in time. This feature challenges standard time series models such as the ARCH-GARCH family since such models are based on observations equally spaced in time. Early high frequency research applied the time-aggregated return series to circumvent this challenge (Hafner 1996, Guillaume, Pictet and Dacorogna 1995 and Olsen and Pictet 1997).

However, this procedure faces two main difficulties in practice. First, it inevitably causes the loss of the information contained in the order of the transaction flows. As argued by market microstructure research, timing of the transactions could be vital to the understanding of economics behind the price behavior. For instance, the duration between transactions could be viewed as a signal of asymmetric information among traders (Easley and O'Hara 1992, 1995 Diamond and Verrecchia 1987,1991 and Goodhart and O'Hara 1997). Thus, a coherent description of the tick-by-tick prices dynamics should base on the modelling of durations. Second, the intraday volatility periodicity makes the choice of the optimal time interval used to aggregate tick-by-tick transactions very difficult. Specifically, if a short interval is chosen, there could be many intervals that contain no new information. Consequently, a specific form of heteroskedasticity corresponding to this short interval will be induced into the time-aggregated return sample. In contrast, the use of a long interval will reduce the sample size significantly.

In order to give an explicit parametric description of the "waiting-time" that is the duration between consecutive transactions, Engle and Russell (1998) proposed the Autoregressive Conditional Duration (ACD) model based on the point process. A detailed introduction of the point process is presented in Daley and Vere-Jones (2006). The ACD model can be viewed as an analogy of the GARCH model for the duration modelling. Since its introduction, the ACD model has been widely developed and accepted as a tool in the modelling of financial data that are irregularly spaced in time. Many theoretical efforts have been made on the extension of the standard ACD model. See, e.g., Lunde (1996), Bauwens and Giot (2000), Engle (2000) and Zhang et al. (2001). Meanwhile, empirical research focused on the performance of the predictability of ACD models (Dufour and Engle 2000, Hautsch 2000 and Pacurar 2007).

There are two main questions related with the ACD models. The first one is: what is the theoretical relation among the various extensions of the ACD models and which one provides the most accurate modelling of the durations? The second one is: how can the ACD model be used to describe the prices dynamics?

In this paper, I address the first question by testing different specifications of the ACD model. I focus on the comparison of different distributional assumptions. I use transaction-aggregated return series at various frequencies to examine different distributions. I also provide a review of both theoretical and empirical research with respect to the ACD models. I discuss the statistical properties of several ACD specifications along with some practical issues in applications. I find out that, compared with other distributions, Gamma distribution gives the best ACD modelling results for transaction-aggregated durations. Besides, the comprehensive introduction of ACD models could contribute to the understanding of the scope of research regarding durations modelling.

This chapter is organized as follows. Section 2 presents the discussion with respect to the ACD models. I analyze the theoretical framework of the ACD models and discuss application issues. Section 3 describes the data and some preliminary results with respect to the durations. I focus on the unconditional distributional properties of the tick-by-tick durations. Section 4 presents the empirical result of the ACD modelling of duration dynamics. Section 5 is the conclusions.

## 2. ACD framework

In this section I analyze and present the framework of the ACD models. In order to present an exhaustive discussion, I first discuss the theoretical background of the ACD models in Section 2.1. Section 2.2 gives a discussion of the applications of ACD models.

### 2.1 Theoretical background of ACD

In this section I present the theoretical framework of the ACD models. Section 2.1.1 exhibits the point process that is the statistical background of the ACD models. I then discuss the ACD specification in Section 2.1.2 and analyze the relations between durations and other characteristics associated with each transaction such as volume and prices in Section 2.1.3.

### 2.1.1 Point Processes

Each tick-by-tick transaction record are associated with three basic characteristics: prices $p_{\tau,i}$, volumes $v_{\tau,i}$ and execution time $t_{\tau,i}$, where the index $\tau$ labels the trading day of the transaction and the index $i$ indicates the order of the transaction during trading day $\tau$. In order to explicitly present the point process, I now remove the day index $\tau$ and view the whole data set as a time series labeled just by the order index $i$.

As discussed previously, a salient feature of the tick-by-tick transaction records is that the durations between consecutive transactions are not equally spaced in time. However, they are ordered in time and therefore can be naturally considered as a point process. In probability theory, the point process is a collection of random variables that are randomly spaced in time. It is widely developed and applied in queueing theory and neuroscience. In 1998, Engle and Russell extended it to the high-frequency finance to model the trading process.

Specifically, consider the sequence of random variables $\{t_0, t_1, \ldots, t_n, \ldots\}$, where $t_n$ represent the arrival time and $t_0 \leq t_1 \ldots \leq t_n \ldots$ . In general, simultaneous occurrence of the events are allowed. In practice, "thinning" procedures usually are used to eliminate this possibility. Let the counting function $N(\tau)$ represents the total number of events that occur by time $\tau$. $N(\tau)$ is a step function that has right limit and is left continuous at each point. The characteristics associated with each arrival time are considered as "marks". In the context of high frequency financing, the trading volume and execution price of each transactions can be viewed as marks associated with the arrival time of transactions.

An important tool for the analysis of point process is the intensity function. With respect to the definition and detailed discussion of point process and intensity function, Daley and Vere-Jones (2006) and Cox and Isham (1980) present very comprehensive and exhaustive introductions. In this chapter, I focus on the point process in the context of high frequency financing. Specifically, the following specification of conditional intensity process is used to describe the point process,

$$\lambda\big(\tau\big|N(\tau), t_0, \ldots, t_{N(\tau)}\big) = \lim_{h \to 0} \frac{\Pr(N(\tau+h) > N(\tau) | N(\tau), t_0, \ldots, t_{N(\tau)})}{h}. \quad (1)$$

(1) implies that the current intensity is conditional the on past information of arrival time and the fact that there is no event since time $t_{N(\tau)}$. The conditional intensity process defined by (1) is also called the hazard function. The hazard function generally is defined as the ratio of the probability density function and survival function. It measures the instantaneous rate of the occurrence of events given that there are no events since the last previous event. Under (1), suppose $f_i$ is the conditional probability density function for arrival time $t_i$. The log likelihood, $L$, of the whole sample is

$$L = \sum_{i=1}^{N(T)} \log f_i(t_i | t_0, \ldots, t_{i-1}). \quad (2)$$

Observe that (2) can be rewritten in terms of the conditional intensity as

$$L = \sum_{i=1}^{N(T)} \log \lambda(t_i | i-1, t_0, \ldots, t_{i-1}) - \int_{t_0}^{T} \lambda\big(s | N(s), t_0, \ldots, t_{N(s)}\big) ds. \quad (3)$$

The equation (2) and equation (3) are used to for the maximum likelihood estimation of the model. It is very clear that the parametric specification of the conditional intensity process in (1) is critical to a successful point process modelling of high frequency trading pattern. Haustch (2000) provided a thorough review of the specifications of intensity functions used in point process for financial modelling. From the perspective of probability theory, the conditional intensity fully describes the corresponding point process. E.g., Daley and Vere-Jones (2006) and Lancaster (1990). In this Chapter, I focus on the parameterization of (1) by Engle and Russell (1998) in the ACD framework.

**2.1.2 ACD models**

The ACD model parameterizes the point process by specifying the process of durations between consecutive events. Specifically, denote the *i*th duration between two events that occur at $t_{i-1}$ and $t_i$ as $w_i = t_i - t_{i-1}$. Thus, the sequence of durations now is $\{w_0, w_1, \ldots, w_n, \ldots\}$ and is nonnegative obviously. Let $z_i$ be the marks such as volume and prices associated with the *i*th event. The durations and marks can be jointly presented as the sequence $\{(w_i, z_i)\}_{i=1,2,\ldots,n,\ldots}$. Thus, the joint conditional density of the *i*th element $(w_i, z_i)$ can be presented as

$$(w_i, z_i)|F_{i-1} \sim f(w_i, z_i|w_1, w_2, \ldots, w_{i-1}, z_1, z_2, \ldots, z_{i-1}; \theta), \qquad (4)$$

where $F_{i-1}$ denotes the information set available at time $t_{i-1}$ and $\theta \in \Theta$ is the set of parameters. Under the assumption of conditional independence, (4) can also be expressed as the product of marginal conditional density of the durations and the conditional density of the marks. Thus,

$$f(w_i, z_i|\bar{w}_{i-1}, \bar{z}_{i-1}; \theta) = f_w(w_i|\bar{w}_{i-1}, \bar{z}_{i-1}; \theta_w) g(z_i|\bar{w}_i, \bar{z}_{i-1}; \theta_z), \qquad (5)$$

where $\bar{w}_{i-1} = (w_1, w_2, \ldots, w_{i-1})$ and $\bar{z}_{i-1} = (z_1, z_2, \ldots, z_{i-1})$ are the sets of past information of the variables $W$ and $Z$ and $f_w$ is the marginal conditional density function of the duration $w_i$ with parameters $\theta_w$ and $g$ is the conditional density function of the mark $z_i$ with parameters $\theta_z$. (5) describes the relation between durations and marks associated with events. It forms the theoretical foundation of modelling tick-by-tick prices dynamics under the ACD framework. For instance, Engle (2000) built the Ultra-High-Frequency GARCH model based on the ACD modelling of durations.

From (4) and (5), one can immediately write the log likelihood as

$$L(\theta_w, \theta_z) = \sum_{i=1}^{N(T)} [\ln f_w(w_i|\bar{w}_{i-1}, \bar{z}_{i-1}; \theta_w) + g(z_i|\bar{w}_i, \bar{z}_{i-1}; \theta_z)]. \qquad (6)$$

Weak exogeneity refers to the case where the statistical estimation can be obtained by considering the conditional model (Engle, Hendry and Richard 1983 and Johansen 1991). Under the assumption that the durations are weakly exogenous to the marks, the right part and left part of (6) are usually maximized separately (Engle 2000). This property effectively simplifies the model estimation procedures. However, the weak exogeneity is usually imposed as an assumption since there is no widely accepted statistical test for it in the context of ACD.

With respect to the specification of conditional marginal density of the durations, Engle and Russell (1998) suggested the following specification. Denote the conditional expectation of the $i$th duration $w_i$ as

$$E(w_i|\bar{w}_{i-1}, \bar{z}_{i-1}) \equiv \psi_i. \qquad (7)$$

The ACD model assumes that

$$\psi_i = \psi(\bar{w}_{i-1}; \theta_\psi) \qquad (8)$$

$$w_i = \psi_i \epsilon_i, \qquad (9)$$

where $\epsilon_i$ is an iid sequence with density $p(\epsilon; \phi)$ and $\psi$ is function of $\psi_i$ with parameters $\theta_\psi$. Observe that (9) implicitly requires that $E(\epsilon_i) = 1$. This assumption is without loss of generality since one can use $w_i = \psi_i \epsilon_i / E(\epsilon_i)$ equivalently. In estimation of the model, this assumption is usually released to allow for flexibility.

The specification of (8) requires the density function $p(\epsilon; \phi)$ to have a nonnegative support since the durations are nonnegative in nature. Besides, (8) suggests that $f_w(w_i|\bar{w}_{i-1}, \bar{z}_{i-1}; \theta_w) = f_w(w_i|\bar{w}_{i-1}; \theta_w)$. In other words, the conditional expectation of durations depends upon only the past durations. Consequently, the conditional density of the durations can be expressed as

$$f_w(w_i|\overline{w}_{i-1}; \theta_w) = f_w(w_i|\overline{w}_{i-1}; \theta_\psi, \phi),$$

$$f_w(w_i|\overline{w}_{i-1}; \theta_\psi, \phi) = \frac{1}{\psi_i} p\left(\frac{w_i}{\psi_i}; \phi\right). \tag{10}$$

Thus, the log likelihood function is,

$$L(\theta_\psi, \phi) = \sum_{i=1}^{N(T)} \ln f_w(w_i|\overline{w}_{i-1}; \theta_\psi, \phi),$$

$$\sum_{i=1}^{N(T)} \ln f_w(w_i|\overline{w}_{i-1}; \theta_\psi, \phi) = \sum_{i=1}^{N(T)} [\ln p\left(\frac{w_i}{\psi_i}; \phi\right) - \ln\psi_i]. \tag{11}$$

Once the innovation term $\epsilon_i$ and the conditional expectation $\psi_i$ are specified, then maximum likelihood estimates of $\theta_\psi$ and $\phi$ can be obtained by general numerical optimization procedures. Finally, denote the survival function associated with the density $p(\epsilon; \phi)$ of $\epsilon$ as S, the conditional intensity of the ACD model can be deduced from (1),

$$\lambda(t|N(\tau), t_0, \dots, t_{N(\tau)}) = \frac{p(\tau)}{S(\tau)} \left(\frac{\tau - t_{N(\tau)}}{\psi_{N(\tau)+1}}\right) \frac{1}{\psi_{N(\tau)+1}}. \tag{12}$$

The deduction details of equation (12) could be found in Engle and Russell (1998). The setup from (7) to (9) are very general and therefore allows a flexible structure of the ACD models. Specifically, a variety of ACD models can be obtained by combining different specifications for the conditional expectation $\psi_i$ and the conditional density of $\epsilon_i$.

### 2.1.2.1 Duration specification

In this section I discuss the specification of the conditional expectation $\psi_i$. As mentioned in the previous section, the range of potential ACD specifications is so wide that a complete review of those specifications is beyond the scope of this paper. The book of Haustch (2004) provides a very exhaustive survey of existing ACD specifications. I focus on the standard ACD specification (Engle and Russell 1998) and the log-ACD of Bauwens and Giot (2000).

The conditional expectation of the duration $\psi_i$ in a standard ACD ($m,q$) model is specified as follows,

$$\psi_i = \omega + \sum_{j=1}^{m} \alpha_j w_{i-j} + \sum_{j=1}^{q} \beta_j \psi_{i-j}. \tag{13}$$

(13) is a linear parameterization of (8). Specifically, the conditional expectation depends linearly upon the past durations and the past expected durations. In order to ensure that the conditional expectations to be nonnegative, parameters $\omega$, $\alpha$ and $\beta$ are required to have, $\omega > 0, \alpha \geq 0$ and $\beta \geq 0$. Obviously, the ACD model can be viewed as an analog of the GARCH (Bollerslev 1986) of the durations. The weak (covariance) stationarity conditions for the ACD model are just similar to those of the GARCH model. Specifically, $\sum_{j=1}^{m} \alpha_j + \sum_{j=1}^{q} \beta_j < 1$. A detailed discussion regarding the stationarity of ACD model can be found in Engle and Russell (1995). Under the assumption of weak stationarity, the unconditional mean and the conditional variance can be deduced from (9) and (13) straightforwardly,

$$E(w_i) = E(\psi_i)E(\epsilon_i) = \frac{\omega}{1-\sum_{j=1}^{m}\alpha_j - \sum_{j=1}^{q}\beta_j} \quad (14)$$

$$Var(w_i|\overline{w}_{i-1}) = \psi_i^2 Var(\epsilon_i). \quad (15)$$

The unconditional variance of durations $Var(w_i^2)$ can also be deduced from (9) and (13). However, the general expression needs a direct computation of $E(\psi_i^2)$ that is rather cumbersome for a general ACD $(m,q)$. For the ACD (1,1) model, the unconditional variance of durations can be derived as follows,

$$E(w_i^2) = E(\psi_i^2)E(\epsilon_i^2) = \frac{\omega^2 E(\epsilon_i^2)}{1-\beta^2-2\alpha\beta-\alpha^2 E(\epsilon_i^2)} + \frac{(\alpha+\beta)2\omega^2 E(\epsilon_i^2)}{(1-\beta^2-2\alpha\beta-\alpha^2 E(\epsilon_i^2))(1-\alpha-\beta)}, \quad (16)$$

$$E(w_i^2) - E^2(w_i) = E^2(w_i)[\frac{E(\epsilon_i^2)(1-\beta^2-2\alpha\beta-\alpha^2)-(1-\beta^2-2\alpha\beta-\alpha^2 E(\epsilon_i^2))}{1-\beta^2-2\alpha\beta-\alpha^2 E(\epsilon_i^2)}]. \quad (17)$$

If $\epsilon_i$ is assumed to follow the exponential distribution, then $E(\epsilon_i^2) = 2$. Consequently, the unconditional variance in (17) can be simplified as

$$E(w_i^2) - E^2(w_i) = Var(w_i) = E^2(w_i)\left(\frac{1-\beta^2-2\alpha\beta}{1-\beta^2-2\alpha\beta-2\alpha^2}\right). \quad (18)$$

(18) has an important implication that if $\alpha > 0$ the unconditional standard deviation of durations will be larger than the unconditional mean of durations. This phenomenon is usually referred to as "excessive dispersion". For discussion and analysis regarding the existence of higher moments of the ACD $(m,q)$ model, see e.g., Carrasco and Chen (2002).

A useful feature of the ACD model is that it can also be formulated as the familiar form of an ARMA (*max(m,q),q*) model. Specifically, denote the martingale difference $\eta_i$ as $\eta_i = w_i - \psi_i$, then the ACD (*m,q*) model can be written as

$$w_i = \omega + \sum_{j=1}^{\max(m,q)} (\alpha_j + \beta_j) w_{i-j} - \sum_{j=1}^{q} \beta_j \eta_j + \eta_i. \tag{20}$$

The ARMA-presentation in (20) can be used to derive the conditions for the stationarity and invertibility of the ACD model. The derivations and results are almost the same as work for GARCH models (Nelson 1992). The stationarity and invertibility requires that all roots of $1 - \alpha(L) - \beta(L)$ and $1 - \beta(L)$ lie outside the unit circle respectively, where $\alpha(L)$ and $\beta(L)$ are the polynomials in terms of the lag operator $L$.

The ARMA-presentation of the ACD model in (20) also can be used to derive the autocorrelation function of ACD models. For instance, the first order autocorrelation is now

$$\rho_1 = Cov(w_i, w_{i-1}) = \frac{\alpha_1(1 - \beta_1^2 - \alpha_1 \beta_1)}{1 - \beta_1^2 - 2\alpha_1 \beta_1}. \tag{21}$$

For more general description of the autocorrelation structure of the ACD model such as the Yule-Walker equations, Bauwens and Giot (2000) provided a detailed discussion.

As discussed previously, there are many other possible specifications for the conditional expectation of the duration $\psi_i$. Another important specification of the durations is the logarithmic-ACD model of Bauwens and Giot (2000). They argued that the imposed nonnegativity of parameters in (13) are too restrictive. For instance, if one wants to incorporate other economic explanatory variables into (13) with coefficients expected to be negative, the resulting conditional expectation of the duration might be negative. In a log-ACD (*m,q*) model, the autoregressive equation is specified using the logarithm of the conditional expectations of durations. Specifically,

$$w_i = e^{\psi_i} \epsilon_i,$$

$$\ln \psi_i = \omega + \sum_{j=1}^{m} \alpha_j \ln w_{i-j} + \sum_{j=1}^{q} \beta_j \ln \psi_{i-j}. \tag{22}$$

Another logarithmic form of the conditional expectation of durations $\psi_i$ is

$$\ln \psi_i = \omega + \sum_{j=1}^{m} \alpha_j \frac{w_{i-j}}{\psi_{i-j}} + \sum_{j=1}^{q} \beta_j \ln \psi_{i-j}. \tag{23}$$

The logarithmic form of the conditional expectations of durations $\psi_i$ in (22) and (23) releases the nonnegative constraints of parameters $\alpha$ and $\beta$ in the standard ACD model. In order to ensure the weak stationarity, (22) requires that $|\alpha + \beta| < 1$ while (23) requires that $|\beta| < 1$. Empirical evidence in general favors the specification of (23). See, e.g., Bauwens and Giot (2000) and Dufour and Engle (2000).

It is possible to derive the unconditional moments and the autocorrelation function of the log-ACD model from (22) and (23). However, the expression is rather cumbersome and therefore will not be presented here. Bauwens, Galli, and Giot (2003) presented a detailed derivation. Compared with the standard ACD model that has an autocorrelation function decreasing geometrically at the rate $\alpha + \beta$ (see (20)), the log-ACD model has an autocorrelation function that decreases at a rate less than $\beta$ for small lags (Bauwens, Galli and Giot, 2003). As for the specification of $\epsilon_i$, Bauwens and Giot (2000) assumed that it follows the Weibull distribution. In general, many other candidates of the density $p(\epsilon; \phi)$ can be chosen for $\epsilon_i$ as in the standard ACD model. The maximum likelihood estimation of the log-ACD is an analog of the maximum likelihood estimation of the standard ACD models.

I will elaborate the possible candidates of the distributional assumption of $\epsilon_i$ and present the maximum likelihood estimation of the ACD model in the next section.

### 2.1.2.2 Distributional assumption and maximum likelihood estimation

As discussed in the previous section, an important feature of the ACD model is that the specification in (9) allows for flexibility in the selection of the standardized duration $\epsilon_i$. In general, any distribution with nonnegative supports could be used in the modelling of durations. Among the vast possible candidates, I focus on the following three distributions that are commonly applied in practice: exponential distribution, Weibull distribution and generalized gamma distribution.

It seems natural to start with the exponential distribution since it is widely used in the modelling for the "waiting-time" in many fields. Although the assumption of exponentially distributed standardized durations seems restrictive in many circumstances, it still serves as a good starting point since it can provide quasi-maximum likelihood estimators for the ACD parameters.

Under the assumption that the standardized durations are exponentially distributed, the log-likelihood function can be derived from (11) and can be presented as

$$L(\theta_\psi, \phi) = -\sum_{i=1}^{N(T)} [\frac{w_i}{\psi_i} + ln\psi_i]. \tag{24}$$

I use the terminology "Exponential-ACD" (EACD) to refer to the standard ACD model with exponentially distributed standardized durations $\epsilon_i$. The strong analogy between the EACD and the GARCH model can be further verified by the similar quasi-maximum likelihood estimation (QMLE) properties of the two models. Specifically, the QMLE properties of the GARCH (1,1) model derived by Lee and Hansen (1994) and Lumsdaine (1996) also holds for the EACD (1,1) model. Engle and Russell (1998) and Haustch (2004) provided detailed proof for the equivalence of QMLE property of the two models. Specifically, Theorem 1 and Theorem 3 in Lee and Hansen (1994) can be rewritten in the ACD form as:

Assume the following conditions:
(i) $\Theta$ is a compact parameter space and $int\ \Theta \neq \emptyset$.
(ii) $\theta_0 = (\omega_0, \alpha_0, \beta_0) \in int\ \Theta$.
(iii) $E(w_i|F_{i-1}; \theta_0) \equiv \psi_i(\theta_0) \equiv \psi_{i,0} = \omega_0 + \alpha_0 w_{i-1} + \beta_0 \psi_{i-1}$.
(iv) $\epsilon_i = \frac{w_i}{\psi_i}$ is strictly stationary, ergodic and nondegenerate.
(v) $E(\epsilon_i^2|F_{i-1}) < \infty$ almost surely.
(vi) $sup_i E[ln\beta_0 + \alpha_0 \epsilon_i|F_{i-1}] < 0$ almost surely.
(vii) $L(\theta) = \sum_{i=1}^n l_i(\theta) = -\sum_{i=1}^N [\frac{w_i}{\psi_i} + ln\psi_i]$, where $\psi_i = \omega + \alpha w_{i-1} + \beta \psi_{i-1}$.

Then the maximizer of $L$ is consistent and asymptotically normal. The associated covariance matrix is given by the robust standard errors by Lee and Hansen (1994).

The above corollary of the Theorem 1 and Theorem 3 in Lee and Hansen (1994) has several important implications. First, it illustrates that the maximization of the log likelihood function in (24) yields a consistent estimate of $\theta$ without a clear specification of the standardized duration $\epsilon_i$. In other words, the estimates of parameters of the EACD (1,1) is consistent and asymptotically normal even under misspecification of $\epsilon_i$. Consequently, the EACD (1,1) can be directly estimated by GARCH software. The procedure is straightforward: use $\sqrt{w_i}$ as the dependent variable in the GARCH regression equation and set the conditional mean as zero.

Second, when $\alpha + \beta < 1$, the condition (vi) is satisfied, but this inequality is a sufficient but not necessary condition for the Theorem to hold. Thus, it may hold for integrated duration process in which $\alpha + \beta = 1$. Third, although it requires the strict stationarity and ergodicity of $\epsilon_i$, it does not require $\epsilon_i$ to follow an iid process. Hence, it establishes the QMLE properties for many other ACD specifications such as the semiparametric ACD model of Drost and Werker (2001).

Nevertheless, it is noticeable that the results are based on the EACD (1,1) model and cannot be extended necessarily to the general EACD (*m,q*) model. Besides, as suggested by the condition (iii), the QMLE property of the EACD (1,1) model depends largely on the correct specification of the conditional expectation $\psi_i$. Moreover, empirical research in general are against the assumption of exponentially distributed standardized duration $\epsilon_i$ (Gramming and Maurer 2000, Feng et al. 2004, Lin and Tamvakis 2004 and Dufour and Engle 2000). Thus, if the data set is relatively small, the quasi-maximum likelihood estimates of the parameters of the EACD (1,1) model might be seriously biased. Finally, the hazard function of the exponential distribution is a constant and therefore is clearly independent of the time passed since the last trade. This characteristic is undesirable in the modelling of ultra-high-frequency transactions for obvious reasons. In order to resolve these issues, it seems natural to choose a more general distribution of the standardized durations $\epsilon_i$ and to replace the quasi-maximum likelihood estimation by standard maximum likelihood estimation. This procedure is similar to the case of ARCH-GARCH model where the normal assumption of returns is often replaced by some leptokurtic distributions.

A popular alternative of the exponential distribution is the Weibull distribution. (Engle and Russell 1998 and Bauwens and Giot 2000). Consider a Weibull density $p(\epsilon; \lambda, k)$ for the standardized duration $\epsilon_i$ where $\lambda$ is the scale parameter and $k$ is the shape parameter. This assumption is equivalent to assuming that $\epsilon_i{}^k$ is exponentially distributed. I use the terminology "Weibull-ACD" (WACD) to refer to this class of ACD models. The Weibull distribution reduces to the exponential distribution when $k = 1$. The probability density of the Weibull distribution can be presented as

$$p(\epsilon; \lambda, k) = \frac{k}{\lambda}\left(\frac{\epsilon}{\lambda}\right)^{k-1} \exp\left(-\frac{\epsilon}{\lambda}\right)^k, \tag{25}$$

where $\epsilon > 0, \lambda > 0$ and $k > 0$. The hazard function of the Weibull distribution can be presented as

$$h(\epsilon) = \left(\frac{1}{\lambda}\right)^k \epsilon^{k-1} k. \tag{26}$$

Under the ACD framework, the conditional intensity of Equation (12) can be presented as

$$\lambda(\tau|N(\tau), t_0, \ldots, t_{N(\tau)}) = \left[\frac{\Gamma\left(1+\frac{1}{k}\right)}{\psi_{N(\tau)+1}}\right]^k (\tau - t_{N(\tau)})^{k-1} k, \tag{27}$$

where $\Gamma(\cdot)$ is the gamma function and $k$ is the shape parameter of the Weibull distribution. Observe that the conditional intensity function in (27) is monotonic with respect to $\tau$ as the hazard function in (26). It is increasing if $k > 1$ and is decreasing if $k < 1$. As a result, this monotonic property of the Weibull hazard function makes the occurrence of long durations, compared with exponential distributions, more or less likely dependent on whether $k < 1$ or $k > 1$. The log-likelihood function under the Weibull distribution can be derived from (11) and (13). Specifically, assume the scale parameter $\lambda$ to be unit for the simplicity of presentation, then

$$L(\theta_\psi, k) = \sum_{i=1}^{N(T)}[\ln\left(\frac{k}{w_i}\right) + k\ln\left(\frac{\Gamma\left(1+\frac{1}{k}\right)w_i}{\psi_i}\right) - \left(\frac{\Gamma\left(1+\frac{1}{k}\right)w_i}{\psi_i}\right)^k - \ln\Gamma\left(1+\frac{1}{k}\right)]. \tag{28}$$

When $k = 1$, (28) reduces to the (24).

The last family of distributions that I consider in this chapter is the generalized gamma distribution. The generalized gamma distribution nests the Weibull distribution, the Exponential distribution and the Gamma distribution and therefore provides greater flexibility in the modelling of standardized duration $\epsilon_i$. In 1996, Lunde introduced the use of generalized gamma distribution in the modelling of durations. The density of the generalized gamma distribution is

$$p(\epsilon; a, d, m) = \frac{m\epsilon^{d-1}}{a^d \Gamma(d/m)} \exp\left(-\frac{\epsilon}{a}\right)^m. \tag{29}$$

where $\Gamma(\cdot)$ is the gamma function and $a > 0, d > 0, m > 0$ and $\epsilon > 0$.

One can immediately infer from the density of generalized gamma distribution in (29) that when $d = m$ the density reduces to the density of the Weibull distribution. Further assume that $d = m = 1$, the density now is the density of the exponential distribution. Besides, if $m = 1$, the density reduces to the density of the Gamma distribution. Glaser (1980) provided a discussion of the relations between the parameters and the shape of hazard function for the generalized gamma distribution. The concrete presentation of the hazard function of the generalized gamma distribution is complicated since it involves an incomplete gamma integral and therefore has no closed form. Numerical techniques are usually applied to depict the pattern of the hazard functions. For further details regarding the hazard function of the generalized gamma distribution, see e.g., Glaser (1980). An important property of the hazard function of the generalized gamma distribution is that it is not monotonic. For instance, given that $dm < 1$, if $m > 1$ the hazard function has a U-shape and elseif $m \leq 1$ the hazard function is decreasing. Given that $dm > 1$, if $m > 1$ the hazard function is increasing and elseif $m \leq 1$ the hazard function has an inverted U-shape. This property of the hazard function of generalized gamma distributions allows for more flexibility in the modelling of the instantaneous rate of occurrence of transactions.

The derivation of the log-likelihood function of "Generalized Gamma-ACD" (GG-ACD) from (11) is straightforward, substituting the density function in (11) with (29),

$$L(\theta_\psi, a, d, m) = \sum_{i=1}^{N(T)} \left[\ln \frac{m}{aw_i} + d\ln \frac{w_i}{\psi_i}\frac{\Gamma\left(\frac{d+1}{m}\right)}{\Gamma\left(\frac{d}{m}\right)} + \left(-\frac{w_i}{\psi_i}\frac{\Gamma\left(\frac{d+1}{m}\right)}{\Gamma\left(\frac{d}{m}\right)}\right)^m - \ln\Gamma\left(\frac{d+1}{m}\right)\right]. \tag{30}$$

The detailed derivation could be found in Lunde (1996). Observe that (30) reduces to (29) if $a = 1$ and $d = m$. It further reduces to equation (24) if $a = m = d = 1$.

To conclude this section, I present the conditional density of $w_i$ dependent on past information $\bar{w}_{i-1}$ with respect to the generalized gamma distribution under the general

ACD framework of (7) - (9) for the completeness of the introduction.

$$f(w_i|\overline{w}_{i-1}; a, d, m, \theta_\psi) = \frac{m}{aw_i \Gamma\left(\frac{d+1}{m}\right)} \left[\frac{w_i}{\psi_i} \frac{\Gamma\left(\frac{d+1}{m}\right)}{\Gamma\left(\frac{d}{m}\right)}\right]^d \exp\left(-\left[\frac{w_i}{\psi_i} \frac{\Gamma\left(\frac{d+1}{m}\right)}{\Gamma\left(\frac{d}{m}\right)}\right]^m\right). \tag{31}$$

### 2.1.3 Durations and marks

It is often the case that we are not only interested in the durations between transactions but also the marks associated with the transactions. The ACD models discussed previously are proposed for describing the dynamics of durations given past information. As suggested by (5), they describe the marginal density of the duration $w_i$. Thus, it remains to present a framework that could jointly describe the durations and marks. Especially, it is often the case that given the durations and the past information, one wants to test some market microstructure hypothesis.

In the Section 2.1.3.1, I present and discuss a framework of modelling tick-by-tick transaction dynamics based on the point process. Section 2.1.3.2 focus on the volatility modelling with in the ACD framework. Section 2.1.3.3 discusses the incorporation of possible explanatory economic variables.

### 2.1.3.1 General setup

As discussed previously, I view the arrival of transactions as a point process. Specifically, denote the timing of transactions as $\{t_0, t_1, \dots, t_i, \dots t_N\}$ and denote the $i$th duration between two transactions that occur at $t_{i-1}$ and $t_i$ as $w_i = t_i - t_{i-1}$. Let $z_i$ be the marks such as volume and prices associated with the $i$th transaction. The durations and marks can be jointly presented as the sequence $\{(w_i, z_i)\}_{i=1,2,\dots,N}$. Thus, the joint conditional density of the $i$th element $(w_i, z_i)$ can be presented as

$$(w_i, z_i)|F_{i-1} \sim f(w_i, z_i|w_1, w_2, \dots, w_{i-1}, z_1, z_2, \dots, z_{i-1}; \theta_i). \tag{32}$$

(32) is a slight modification of the (4). The main difference is that the parameters $\theta_i$'s are now conditional on the past information and therefore can be potentially different for different transaction observations. The conditional density can be rewritten as the product of marginal conditional density of the durations and the conditional density of the marks as in (5).

$$f(w_i, z_i | \bar{w}_{i-1}, \bar{z}_{i-1}; \theta_i) = f_w(w_i | \bar{w}_{i-1}, \bar{z}_{i-1}; \theta_{w,i}) g(z_i | \bar{w}_i, \bar{z}_{i-1}; \theta_{z,i}), \quad (33)$$

where $\bar{w}_{i-1} = (w_1, w_2, \ldots, w_{i-1})$ and $\bar{z}_{i-1} = (z_1, z_2, \ldots, z_{i-1})$ denote the past information of $w_i$ and $z_i$. In the real world, the realized transactions are nearly zero at every point in the time. Therefore, it is natural to measure the probability of the occurrence of transactions at specific time in terms of the limit. The conditional intensity function of (1) are usually applied to describe such probabilities. Under the specification of (32) and (33), then conditional intensity of (1) can be rewritten as

$$\lambda_i(t | \bar{w}_{i-1}, \bar{z}_{i-1}) = \lim_{h \to 0^+} \frac{\Pr(N(t+h) > N(t) | \bar{w}_{i-1}, \bar{z}_{i-1})}{h}, t \in [t_{i-1}, t_i]. \quad (34)$$

$N(t)$ is the function of the number of transactions that occur by time $t$. Another presentation of (34) can be obtained by using (33). For $t \in [t_{i-1}, t_i]$, the probability of a transaction should be conditional to both past information such as $\bar{w}_{i-1}$ and $\bar{z}_{i-1}$ and the information that there has been no transaction since $t_{i-1}$. This is usually expressed as the ratio of probability density function and survival function. By (33) and (34), one can have

$$\lambda_i(t | \bar{w}_{i-1}, \bar{z}_{i-1}) = \frac{f_w(t - t_{i-1} | \bar{w}_{i-1}, \bar{z}_{i-1}; \theta_{w,i})}{\int_{\tau \geq t} f_w(\tau - t_{i-1} | \bar{w}_{i-1}, \bar{z}_{i-1}; \theta_{w,i}) ds}, t \in [t_{i-1}, t_i]. \quad (35)$$

Many questions of economic interests can be explored in the setup of (32) - (35). For instance, the conditional marginal density of the next mark $z_i$ given $\bar{z}_{i-1}$ is

$$f_z(z_i | \bar{w}_{i-1}, \bar{z}_{i-1}; \theta_i) = \int_\tau f(w_i, z_i | \bar{w}_{i-1}, \bar{z}_{i-1}; \theta_i) d\tau. \quad (36)$$

Once the $f_w$ and $g$ in (33) are specified, the log-likelihood function can be obtained immediately,

$$\begin{aligned} L(\theta) &= \sum_{i=1}^N \ln f(w_i, z_i | \bar{w}_{i-1}, \bar{z}_{i-1}; \theta_i) \\ &= \sum_{i=1}^N [\ln f_w(w_i | \bar{w}_{i-1}, \bar{z}_{i-1}; \theta_{w,i}) + \ln g(z_i | \bar{w}_i, \bar{z}_{i-1}; \theta_{z,i})]. \end{aligned} \quad (37)$$

The log-likelihood function in (37) should be maximized jointly by the sum of the first term of $\ln f_w(w_i | \bar{w}_{i-1}, \bar{z}_{i-1}; \theta_{w,i})$ and the second term of $\ln g(z_i | \bar{w}_i, \bar{z}_{i-1}; \theta_{z,i})$. Separating

the maximization of the first term and the second term comes at the cost of the efficiency of the MLE. In practice, a two-step procedure is usually adopted for a large sample. Specifically, one first takes the estimates of $\theta_w$ by maximizing the first term in (37) and further gives the $\theta_w$ into the maximization of second term in (37) to obtain the estimates of $\theta_z$. The estimates in the two-step estimation are consistent but inefficient. The main reasoning behind this procedure is that, given a large sample, the loss of efficiency might be acceptable (Engle 2000).

### 2.1.3.2 Durations and Volatility

Among the marks contained in $z_i$, the most important information is the execution price $p_i$ associated with the transaction. The price reflects the volatility of the underlying asset and the market. In this section we first discuss the relations between durations and volatility. I review the market microstructure theory that outlines the possible explanation of the relation between durations and volatility such as Diamond and Verrecchia (1987), Easley and O'Hara (1992), and Admati and Pfleiderer (1988).

Further, I analyze the GARCH volatility modelling under the ACD framework. The Ultra-High-Frequency (UHF) GARCH volatility model of Engle (2000) will be presented as an example. The joint modelling of durations and marks can also be obtained by exploiting the conditional intensity function of (35). A detailed introduction can be found in Hautsch (2004).

In 1987, Diamond and Verrecchia argued that "no trade means bad news". The theory can be briefly explained by the following scenario. Suppose there are some informed traders who currently do not own any stock and are prevented from short selling. Thus, they cannot profit from the asymmetric information and will refuse to trade at the current market prices. The specialist whose role is to facilitate trading for certain stocks for some stock exchanges such as NYSE learns from the durations and further lower their prices. Consequently, the existence of no transactions can be viewed as a signal of bad news.

Easley and O'Hara (1992) provided another economic interpretation of the durations between transactions. They focused on the arrival of new information. Suppose there are some informed traders who know whether there is new information or not. They will buy or sell depending on whether the information is good or bad respectively. Thus, long durations between transactions can be interpreted as the evidence of no news. Their theory therefore can be briefly described as "no trade means no news". Thus, under the framework of their theory the durations should be negatively correlated with volatilities.

Admati and Pfleiderer (1988) gave a very interesting alternative point of view. They argued that the short durations between transactions are the result of some liquidity traders. The trading behavior of these liquidity traders is not based on asymmetric information. As a result, the volatility should be low when durations are short. In contrast, relatively long durations could be interpreted as a sign of the leaving of these liquidity traders. Consequently, there will be a high proportion of informed traders. The volatility should be high since these transactions reflect the arrival of new information. Their theory suggests that the durations should be positively correlated with volatilities.

It is worthy to mention that the literatures of Easley and O'Hara (1992) and Admati and Pfleiderer (1988) are more relevant to the liquidity rather than volatility. The discussion above is an attempt to connect their microstructure theories with the ACD modelling of durations and volatility.

Given the theoretical background provided by above market microstructure literatures, it remains to present a parametric framework to jointly describe prices volatility and durations. For a tick-by-tick transaction data set, let the sequence of execution prices associated with each transaction be $\{p_0, p_1, \ldots, p_i, \ldots p_N\}$. The corresponding transaction time is $\{t_0, t_1, \ldots, t_i, \ldots t_N\}$. The $i$th return $r_i$ and the $i$th duration $w_i$ are defined as $r_i = ln\frac{p_i}{p_{i-1}}$ and $w_i = t_i - t_{i-1}$. Thus, the tick-by-tick return series and its corresponding duration series are $\{r_1, \ldots, r_i, \ldots, r_N\}$ and $\{w_1, \ldots, w_i, \ldots, w_N\}$ respectively.

With respect to the volatility modelling, since the return series are defined as $\{r_1, \ldots, r_i, \ldots, r_N\}$, one can therefore define the conditional variance for return $r_i$ as

$$v_{i-1}(r_i|\overline{w}_i, \overline{r}_{i-1}) = q_i. \qquad (38)$$

The definition of the conditional variance $q_i$ in (38) suggests that the conditional variance for the $i$th return depends on not only the past information of returns $\bar{r}_{i-1}$ but also on the current and past information of durations $\bar{w}_i$. Besides, (38) defines the conditional variance associated with return $r_i$ from $t_{i-1}$ to $t_i$. This definition is associated with each transaction rather than fixed time interval. Specifically, note that $q_i$ and $q_j$ now represent the conditional variance measured from durations with different length if $i \neq j$. In practice, it is often the case that the tick-by-tick durations are multiples of the minimum unit of time measurement. Consequently, the durations present a salient feature of discreteness. It seems convenient to transform the conditional variance for each return into the conditional volatility per unit of time. By (38), the transformation can be presented as

$$v_{i-1}\left(\frac{r_i}{\sqrt{w_i}}\bigg|\bar{w}_i, \bar{r}_{i-1}\right) = \sigma_i^2. \tag{39}$$

Under the definition of (38) and (39), the relation between $q_i$ and $\sigma_i$ can be presented as

$$q_i = w_i \sigma_i^2. \tag{40}$$

Thus, given past information of $\bar{w}_{i-1}$ and $\bar{r}_{i-1}$, the predicted expectation of $q_i$ is

$$E_{i-1}(q_i|\bar{w}_{i-1}, \bar{r}_{i-1}) = E_{i-1}(w_i \sigma_i^2|\bar{w}_{i-1}, \bar{r}_{i-1}). \tag{41}$$

(40) and (41) suggest a simple methodology of incorporating durations into standard techniques of volatility modelling. Since $\sigma_i^2$ now is the conditional variance per unit time, standard time series techniques based on fixed time interval such as GARCH volatility model can be applied to describe its dynamics. Specifically,

$$\frac{r_i}{\sqrt{w_i}} = \sigma_i z_i,$$

$$\sigma_i^2 = \omega + \alpha \frac{r_{i-1}^2}{w_{i-1}} + \beta \sigma_{i-1}^2, \tag{42}$$

where $z_i$ is an iid sequence with zero mean and unit variance. (42) is the simplest form of the UHF-GARCH of Engle (2000). It can be viewed as a modification of the standard GARCH (1,1) model that is adjusted to account for the irregularly spaced transactions. Recall the ACD specification of $w_i$ in (9),

$$w_i = \psi_i \epsilon_i,$$

where $\psi_i$ is the conditional expectation of $w_i$ dependent on past information of durations $\bar{w}_{i-1}$. A substitution of (9) into the (42) gives

$$r_i = \sigma_i z_i \sqrt{\psi_i \epsilon_i}. \tag{43}$$

Observe that under (42), the current duration is not informative since $\sigma_i^2$ depends on only past information of durations and returns. The implicit assumption behind this specification is that the news carried by the last return $r_{i-1}$ is fully captured by the square of the last return innovations $\sigma_{i-1}$ per unit of time and is independent of current durations. However, as suggested by the market microstructure papers, the variation of durations and variation of return volatilities might be related. In order to incorporate the information of current duration, the volatility specification of (42) can be modified to account for current durations. For instance, Engle (2000) proposed the following specification of $\sigma_i^2$.

$$\sigma_i^2 = \omega + \alpha \frac{r_{i-1}^2}{w_{i-1}} + \beta \sigma_{i-1}^2 + \gamma \frac{1}{w_i}. \tag{44}$$

(44) corresponds to the theory of Easley and O'Hara (1992). The reciprocal of the durations in (44) suggests that long durations contribute less to the variance. This implication is in line with the argument of "no trade means no news". Observe that under (44), the current return $r_i$ now depends on current durations and past durations and prices as in (33). The joint conditional density $f(w_i, z_i | \bar{w}_{i-1}, \bar{z}_{i-1}; \theta_i)$ in (33) therefore can be determined.

Another model that describes the price dynamics based on ACD modelling of durations is the Autoregressive Conditional Multinomial (ACM) model of Russell and Engle (2005). Building on (5), they applied an ACD specification for durations and a dynamic multinomial model for distribution of price changes conditional on past information and the current duration. Nevertheless, the original ACM model (Engle and Russell 2005) depends on a relatively large number of states that characterize different regimes of durations dynamics for different price changes. Consequently, many parameters are needed, which inevitably complicates the estimation. For a two-state ACM model, see Prigent et al. (2001).

Finally, other economic explanatory variables can be introduced into the conditional volatility specification in (42) such as the bid-ask spread and volumes to provide a more realistic modelling of the conditional volatility. For instance, Engle (2000) included the trading intensity and volume in (42) and found that compared with the trading intensity, volume has very little explanatory power over the conditional variance $\sigma_i^2$.

## 2.1.3.3 Durations and other explanatory variables

In this section I discuss the relation between durations and other microstructure variables. I first present the incorporation of explanatory variables into the ACD model. Then, we discuss the market microstructure variables of interests by reviewing the market microstructure literatures with respect to the timing of transactions.

Due to the flexibility of the ACD specification, economic explanatory variables also can be introduced into the modelling of durations. In general, there are two different ways to incorporate the possible explanatory variables.

First, one can add the random variable of interests directly into the specification of conditional expectation $\psi_i$. Specifically, recall (13) that gives the conditional expectation of the duration $\psi_i$ for a standard ACD $(m,q)$ model,

$$\psi_i = \omega + \sum_{j=1}^{m} \alpha_j w_{i-j} + \sum_{j=1}^{q} \beta_j \psi_{i-j}.$$

Suppose the explanatory variables that we are interested in is $\xi_i$. Include this variable in (13), then

$$\psi_i = \omega + \sum_{j=1}^{m} \alpha_j w_{i-j} + \sum_{j=1}^{q} \beta_j \psi_{i-j} + \gamma \xi_{i-1}. \tag{45}$$

An alternative form of incorporating $\xi_i$ into (3.13) is,

$$\psi_i - \gamma \xi_{i-1} = \omega + \sum_{j=1}^{m} \alpha_j w_{i-j} + \sum_{j=1}^{q} \beta_j (\psi_{i-j} - \gamma \xi_{i-1} - j). \tag{46}$$

As discussed previously, (45) and (46) cannot necessarily guarantee the nonnegativity of the conditional expectation of durations $\psi_i$ if $\xi_i$ can take negative values. There are two common solutions to this issue. First, one can consider presenting $\xi_i$ in some function forms that are nonnegative in nature such as the exponential function. Second, one can use the Log-ACD of Bauwens and Giot (2000). The Log-ACD $(m,q)$ specification is given by (22),

$$w_i = e^{\psi_i} \epsilon_i,$$

$$ln\psi_i = \omega + \sum_{j=1}^{m} \alpha_j ln\, w_{i-j} + \sum_{j=1}^{q} \beta_j \, ln\psi_{i-j}.$$

or

$$ln\psi_i = \omega + \sum_{j=1}^{m} \alpha_j \frac{w_{i-j}}{\psi_{i-j}} + \sum_{j=1}^{q} \beta_j \, ln\psi_{i-j}.$$

Observe that the use of logarithmic form of the conditional expectations of durations $\psi_i$ releases the nonnegative constraints of parameters $\alpha$ and $\beta$ in the standard ACD model. Similarly, incorporating explanatory variables into the Log-ACD model now will not suffer from the possibly negative conditional expectations of durations.

The second way of including explanatory variables is to consider the explanatory variable as a scaling function of the durations $w_i$. For instance, denote the explanatory variable of interest as $\xi_i$. Let $h$ be some nonnegative function. Then

$$w_i = h(\xi_{i-1})\psi_i\epsilon_i. \tag{47}$$

Observe that the variable $\xi$ in (47) is indexed by $i$-1. It is due to the definition that $w_i = t_i - t_{i-1}$. Thus, $w_i$ are subject to information until $t_{i-1}$. Thus, the index is labeled with one lag in order to corresponds to sequence of transaction timing $\{t_1, t_2, \dots, t_i, \dots\}$. One can of course use $\xi_i$ if the indexes are given corresponding to the sequence of durations $\{w_1, w_2, \dots, w_i, \dots\}$.

A simple and illustrative example is the inclusion of intraday periodicity. Denote the intraday periodicity of the durations as $s_i$. Then it can be incorporated into the ACD modelling of durations by

$$w_i = s_{i-1}\psi_i\epsilon_i. \tag{48}$$

I now discuss the possible economic variables that can be introduced into the ACD modelling. A natural candidate is the volume associated with each transaction, the relationship between volume and price dynamics is widely identified by both theoretical and empirical research such as Easley and O'Hara (1992), Easley et al. (1997) and Blume et al. (1994). These papers suggest that, in general, volumes convey information which are not fully captured by price dynamics. For instance, according to Easley and O'Hara (1992), the abnormally large volume might be viewed as the signal of the arrival of informed traders. In general, empirical evidence (Engle 2000, Engle and Russell 1998 and Dufour and Engle 2000) found that the volume is negatively correlated with the conditional expectation of duration $\psi_i$.

Another market microstructure variable of particular interests is the bid-ask spread. Easley and O'Hara (1992) argued that a high spread might be an indication of informed trading. Consequently, it should be related with the short durations. The reasoning is that under the framework of Easley and O'Hara (1992), the volatility of prices will increase if the proportion of informed traders increases. Consequently, the bid-ask spread will be widened. Thus, the wide bid-ask spread could be informative of information arrivals. For a theory of "no trade means no news", the wide spread therefore indicates the new information and should be related with short durations. In order to analyze this effect, the spread for each transaction can be considered as a random variable and can be incorporated into the duration modelling. The empirical evidence of Bauwens and Giot (2000) and Engle (2000) suggest that the bid-ask spread is negatively correlated to the durations.

Compared with volumes, the number of transactions that occur during a specific time interval has significantly higher explanatory powerful over the price volatility during the time interval (Ane and Geman 2000, Easley et al. 1995 and Jones et al. 1994). Jone et al. (1994) provided a market microstructure theory for this phenomenon. They argued that the number of transactions is closely related to the change of prices since transactions are more likely to happen when there is new information. In contrast, the framework of Admati and Pfleiderer (1988) suggests that the intensity of transaction should be negatively correlated with volatility.

The duration itself, of course, is a measure of trading intensity since it measures how frequently an equity is traded. For a tick-by-tick transaction record, the corresponding duration series directly reflect the transaction intensity. To analyze the effect of the trading intensity, it is necessary to conduct a deformation of the original durations. Besides, it has been reported that a large percentage of the tick-by-tick transactions has no price changes (Tsay 2002, Bertram 2005 and Dionne et al., 2005).

Intraday transaction prices often can take only finite values due to some institutional features regarding the price restrictions. For instance, financial markets like NYSE usually specifies a minimum unit of price measurement, called a tick. In other words, transaction prices must fall on a grid. Given such a grid, the durations (Engle and Russell 1998 and Giot 1999) can be defined by filtering the tick-by-tick transactions and retaining those leading to a significant change of prices. The durations defined in this sense are called the price durations. Observe that the price durations are of different length and may contain different number of tick-by-tick transactions. Thus, for example, a large number of tick-by-tick transactions over a short duration represents a high trading intensity. Clearly, volume durations can be defined in a similar way. For a price duration, trading intensity can be defined as ratio of the number of transactions contained in the duration to the length of the durations. (45) and (46) now can be applied to analyze the effect of the trading intensity. Giot and Bauwens (2000) argued that there is a highly significant negative correlation between trading intensity and the expectation of conditional durations.

Given the vast existing market microstructure literatures with respect to the relation between durations and associated market microstructure variables such as volume and prices, it is widely accepted that time is not exogenous to the price process. Nevertheless, the interdependence of those variables is still open to question. Under the framework of marked point process, such interdependence can be studied through the decomposition of (33). For instance, one can add volumes and bid-ask spread to the conditional volatility specification of the UHF-GARCH model to examine their explanatory power over price dynamics. A wide range of closely related model following this framework can be found in Gramming and Wellner (2002), Ghysels and Jasiak (1998), Bauwens and Giot (2003) and Russell and Engle (2005).

An alternative approach for studying the interdependence between durations and other microstructure variables is provided by Hasbrouck (1991). He applied the vector autoregressive (VAR) system to analyzes the impact of current execution prices on the future prices and presented a bivariate model for the relation between price changes and trade dynamics such as the sign of price movement. The timing of transactions is not considered as informative. Thus, the conditional density of the marks, $g(z_i|\overline{w}_i, \bar{z}_{i-1}; \theta_{z,i})$ in (33) depends on past information of $z_i$, which is $g(z_i|\overline{w}_i, \bar{z}_{i-1}; \theta_{z,i}) = g(z_i| \bar{z}_{i-1}; \theta_{z,i})$.

Dufour and Engle (2000) extended the model of Hasbrouck (1991) to account for the influence of durations on the price dynamics. They first applied an ACD model to describe the duration dynamics. Further, the duration is considered as a predetermined variable and therefore the coefficients for price changes and trade dynamics are allowed to be time varying. The model can be extended to incorporate marks of interest such as volume and spread easily. See e.g., Spierdijk (2004) and Manganelli (2005).

In order to capture the complicated interdependence between durations and marks, most of the joint models of durations and marks discussed above require a relatively large number of parameters. Thus, the maximization of the log-likelihood function of (37) faces the computation difficulties. In order to simplify the estimation procedures, it is usually assumed that durations have some form of exogeneity (Engle et al. 1983).

## 2.2 ACD in applications

In this section I present a discussion of the ACD models in application. As discussed previously, the ACD models are based on the point process modelling of transaction arrivals and can be applied to test many market microstructure hypotheses. It could be also used for the modelling of arrivals of a variety of financial events. In Section 2.2.1 I present the discussion of intraday periodicity of the intraday durations. Like volatility and volume, the durations also present a seasonal pattern during the trading day. Section 2.2.2 include the test procedures of the ACD modellings. Section 2.2.3 reviews the literatures of the ACD models. I categorize the literatures according to the different definitions of durations.

### 2.2.1 Intraday Periodicity

The intraday transactions during the trading day are characterized by a strong periodicity. Early research using intraday data sampled from fixed time interval (the time-aggregated returns) focused on the intraday periodicity pattern in the price dynamics (Bollerslev and Andersen 1997,1998 and Beltratti and Morana 1998 and Engle 2000). I present an estimation of the intraday volatility periodicity that can be extended to returns that are irregularly spaced in time (the transaction-aggregated returns). The intraday volatility periodicity is found to have strong impact on the GARCH modelling of price dynamics.

With respect to the durations, Engle and Russell (1998) found that in general the transactions occurring at the beginning and closing hours of the trading day are associated with short durations. In contrast, the transactions that occur in the middle of the day are with relatively long durations. This finding might not be very surprising since it is well-known that compared with the middle day, the opening and closing trading hours are with much higher trading intensity. Consequently, the durations of transactions occurring in the opening and closing hours are expected to have shorter durations. These intraday periodic patterns of durations are often explained by the institutional features and trading habits. For instance, traders are highly active during the opening hours because they want to adjust their position according to the overnight news. Similarly, at the closing, traders want to close their position based on the information during the trading day. Lunchtime are related with less trading intensity due to its nature.

The findings of Engle and Russell (1998) provide direct evidence of the existence of the intraday duration periodicity. Thus, when modelling the durations dynamics under the ACD framework, the effect of such intraday duration periodicity should be included and analyzed since ignoring these intraday patterns might distort the estimation seriously. The intraday volatility periodicity induces a very strong U-shape pattern into the autocorrelation of intraday volatilities. Standard GARCH volatility models cannot capture this effect and therefore often give estimations contradictory to the temporal aggregation predictions. Given the resemblance between the ACD model and the GARCH model, it seems natural to exclude the intraday pattern of durations before applying the ACD model as in the case of GARCH intraday volatility modelling.

The fundamental technique is proposed by Andersen and Bollerslev (1997,1998) for incorporating the intraday pattern into the intraday volatility modelling. Engle and Russell (1998) extended it to the duration modelling. Specifically, one can decompose the intraday durations as the product of a deterministic component that accounts for the intraday periodicity pattern and a stochastic component that models the duration dynamics. Let $s(t_i)$ be the intraday periodicity component at $t_i$ and let $w_i = t_i - t_{i-1}$ be the $i$th duration. Then,

$$w_i = \widetilde{w}_i \, s(t_{i-1}). \tag{49}$$

The $\widetilde{w}_i$ in (3.49) can be interpreted as the seasonal normalized durations. Now the conditional expectation $\psi_i$ of $w_i$ in (7) can be written as

$$\psi_i = E(w_i | \overline{w}_{i-1}, \bar{z}_{i-1}) s(t_{i-1}). \tag{50}$$

There are two approaches that are commonly used to specify the deterministic intraday periodicity component $s(t_i)$ (Andersen and Bollerslev 1997, Engle and Russell 1998, Bauwens and Giot 2000 and Engle 2000). The first one is to use a piecewise linear or cubic spline function to describe the intraday periodicity of durations. Further, the seasonal normalized durations $\widetilde{w}_i$ can be obtained by taking the ratio between the raw durations and the corresponding function values. For instance, one can first average the durations over ten minutes intervals for each trading day. Further, the mean of durations for each ten minutes interval can be estimated with the whole sample. Cubic splines are then applied on these intervals to smooth the intraday periodicity function.

The second approach involves the use of flexible Fourier series approximation. See, e.g., Andersen and Bollerslev (1998). This approach is based on the work of Gallant (1981). Specifically, the intraday periodic trend $s(t)$ is specified to follow the form of

$$s(t) = s(\delta^s, \bar{t}, q) = \bar{t}\delta^s + \sum_{j=1}^{q} [\delta^s_{c,j} \cos(\bar{t}2\pi j) + \delta^s_{s,j} \sin(\bar{t}2\pi j)], \tag{51}$$

where $\delta^s, \delta^s_{c,j}$ and $\delta^s_{s,j}$ are the intraday seasonal coefficients that need to be estimated. $s(t)$ is the intraday periodic trend at time $t$ of the trading day and $\bar{t}$ is the normalized intraday time trend which is defined as the ratio of the number of seconds from opening until $t$ to the length of the trading day in seconds. By this definition, $\bar{t} \in [0,1]$. The Appendix B of Andersen and Bollerslev (1997) presents a detailed discussion with respect to the estimation of (51).

In principle, the conditional expectation $\psi_i$ and intraday periodicity component $s(t_i)$ in (49) and (50) are estimated jointly by maximum likelihood estimation. Nevertheless, when studying intraday data with ultra-high frequency (e.g., tick-by-tick transaction data), numerical difficulty rises inevitably. In order to achieve convergence for a joint estimation of $\psi_i$ and $s(t_i)$, it is often the case that the algorithm costs a relatively long time. Due to this reason, a two-step procedure is commonly applied to simplify the estimation. That is, one first estimates the intraday periodicity component $s(t_i)$ and use the seasonal normalized $\widetilde{w}_i$ durations to estimate the ACD models. As discussed previously, the two-step estimates are consistent but not efficient. The asymptotic property of the two-step estimator can be found in Engle (2000) and Engle (2002). The derivation of the asymptotic property is based on the results of Newey and McFadden (1994) for the GMM (generalized methods of moments) estimation. For a large sample, the joint estimation and two-step estimation usually give very similar results (Engle and Russell 1998 and Bauwens and Giot 2000) as expected.

Veredas et al (2001) applied a different strategy to estimate the intraday pattern. They proposed a semiparametric estimator where the intraday periodicity is jointly estimated nonparametrically with ACD specifications for the duration dynamics. Specifically, they introduced a joint estimation of $\widetilde{w}_i$ and $s(t_i)$ in Equation (49) where $\widetilde{w}_i$ is modelled by the ACD specification and meanwhile $s(t_i)$ is unspecified and therefore is estimated nonparametrically.

There are also many alternative techniques used to capture the intraday pattern. Tsay (2002) applied quadratic functions with indicator variables that identify the timing of the durations. Drost and Werker (2004) chose only one indicator variable that identifies the lunchtime for the model of Tsay (2002). The main reason is that, for their data, trading intensity are almost constant except for the durations near the lunchtime.

Dufour and Engle (2000) introduced dummy variables for the intraday pattern into their vector autoregressive (VAR) system. Interestingly, they found little evidence for the intraday pattern except that the first thirty minutes of the trading day have significantly different dynamics from the rest trading hours.

It is noticeable that, although the practice of filtering raw durations by intraday pattern is widely accepted, the effectiveness of these techniques can only be roughly examined by some stylized facts such as the U-shape trading pattern. Moreover, it also complicates the diagnostic of the ACD specifications. Concerns regarding this issue can be found in Bauwens et al. (2004) and Meitz and Terasvirta (2006). Indeed, as argued by Bollerslev (1997), given the lack of economic theory that could guide a plausible parametric form of the intraday periodicity patterns, it seems natural to estimate them by nonparametric methods.

### 2.2.2 Testing the ACD

In this section, I briefly discuss the procedures to test the ACD models. Given the vast different tests that have been developed since the introduction of ACD models by Engle and Russell (1998), an exhaustive review is beyond the scope of this paper. Therefore, I focus on three different categories of these tests: residual diagnostics, density forecast evaluations and Lagrange multiplier tests.

The simplest and most direct way to evaluate the goodness of fit for the ACD models is to analyze the distributional and dynamic property of the residuals. Recall the standard specification of the ACD model from (7) to (9),

$$E(w_i|\bar{w}_{i-1}, \bar{z}_{i-1}) \equiv \psi_i,$$

$$\psi_i = \psi(\bar{w}_{i-1}; \theta_\psi),$$

$$w_i = \psi_i \epsilon_i.$$

Thus, the residuals are given by

$$\hat{\epsilon}_i = \frac{w_i}{\hat{\psi}_i}, i = 1,2,3, \dots, N, \tag{52}$$

where $\hat{\psi}_i$'s are the estimates for the conditional expectations $\psi_i$'s under the ACD specification. Suppose the specification is correct, the series $\{\hat{\epsilon}_i\}$ should be iid clearly. Thus, one can use Ljung-Box statistics based on the centered residuals to examine whether the specification can fully capture the intertemporal dependence in the duration process.

Moreover, if the distributional assumption for $\epsilon_i$ under the ACD specification is correct,

the residuals $\{\hat{\epsilon}_i\}$ should follow the distribution that corresponds to the assumption with unit mean. Thus, graphic checks such as the Quantile-Quantile plot and general goodness of fit statistics such as the EDF statistics can be applied to examine the residuals.

Alternatively, one can also use moment conditions corresponding to the specified distribution to evaluate the goodness of fit. For instance, if $\epsilon_i$ is assumed to follow the standard exponential distribution, then the variance and mean of $\{\hat{\epsilon}_i\}$ should be nearly equivalent to each other. Engle and Russell (1998) proposed the statistics $\sqrt{n}(\hat{\sigma}_{\hat{\epsilon}}^2 - 1)/\sigma_\epsilon$, where $\hat{\sigma}_{\hat{\epsilon}}^2$ is the sample variance of the residuals and $\sigma_\epsilon$ is the standard deviation of the random variable $(\epsilon_i^2 - 1)$. For an exponentially distributed $\epsilon_i$, $\hat{\sigma}_{\hat{\epsilon}}^2$ should be very close to 1 and the value of $\sigma_\epsilon$ is $2\sqrt{2}$. They also provided the asymptotic property of this test statistics.

Although the residuals check provides very illustrative evidence of the goodness of fit, it is also possible to examine the ACD specification by evaluating the in-sample density forecasts. Diebold et al. (1998) provided a framework of evaluating the goodness of fit based on the probability integral transform

$$q_i = \int_{-\infty}^{x_i} f_i(u) du, \tag{53}$$

where $f_i(x_i)$ is the sequence of one-step ahead probability density forecasts and $x_i$ is the corresponding random process.

They showed that, if the model specification is correct, then $q_i$ should be uniformly distributed and should be iid. Therefore, one can test the series of $q_i$ against the uniform distribution to evaluate the in-sample density forecasts. Specifically, a goodness of fit test can be conducted by categorizing the probability integral transform $q_i$ and calculating a chi-squared statistic that is based on the frequencies of the different categories. The chi-squared statistic can be presented as following,

$$\chi^2 = \sum_{i=1}^{T} \frac{(n_i - n\hat{p}_i)^2}{n\hat{p}_i}, \tag{54}$$

where $T$ is the total number of categories, $n_i$ is the number of observations in category $i$ and $\hat{p}_i$ is the estimated probability to observe a realization of $q_i$ in the category $i$. Further details can be found in Bauwens et al. (2000) and Dufour and Engle (2000).

It is often the case that one not only is interested in testing whether the model specification is correct or not but also is interested in, when rejected, identifying the source of misspecification. In the case of testing ACD specifications, there are two main sources of possible misspecifications. The first one is the distributional assumption and the second one is the conditional expectation specification. Moreover, as discussed previously, the quasi-maximum likelihood estimation of the ACD model relies heavily on correct specification of the conditional expectation. For research that focus on the comparison of different specifications of the conditional expectation $\psi_i$, the validity of the conditional expectation specification might be more important than the correctness of the complete density.

A common procedure in econometric literatures for detecting the model misspecification is the Lagrange multiplier (LM) tests. A detailed discussion regarding the LM tests can be found in Engle (1984). I now present a brief discussion of the LM test for the ACD specifications. In order to conduct the LM test, one need to specify a more general model that can nest the specification of the null hypothesis. Assume that the ACD specification of the null hypothesis $H_0$ is a special case of the following general specification,

$$w_i = \psi_i \epsilon_i,$$
$$\psi_i = \psi_{0,i} + \theta_a' z_{ai}, \qquad (55)$$

where $\psi_{0,i}$ denotes the conditional expectation function under the null hypothesis $H_0$ depending on the parameter vector $\theta_0$, $z_{ai}$ is the vector of missing variables and $\theta_a$ is the vector of additional parameters. The prime indicates inner product. Now, the null hypothesis $H_0$ that the specification is correct can be presented as $H_0: \theta_a = 0$. Consider the quasi-maximum likelihood estimation of the ACD model given by (24),

$$L(\theta) = \sum_{i=1}^n l_i(\theta) = -\sum_{i=1}^n \left[\frac{w_i}{\psi_i} + ln\psi_i\right].$$

Let $\hat{\theta}_0$ be the quasi-maximum likelihood estimate under $H_0$. Then, the LM test can be obtained (Engle 1984) by

$$\gamma_{LM} = l's(\hat{\theta}_0) I(\hat{\theta}_0)^{-1} s(\hat{\theta}_0)' l \sim \chi^2(M_a), \qquad (56)$$

where $l$ is a vector of unit with length $n$, $s(\hat{\theta}_0)$ is the matrix of size $n \times (M_0 + M_a)$, and $I(\hat{\theta}_0)$ is the information matrix evaluated at $\hat{\theta}_0$. With respect to the matrix $s(\hat{\theta}_0)$, $M_0$ and $M_a$ denote the number of parameters in $\theta_0$ and $\theta_a$ respectively. Moreover, $s_{i,j}(\hat{\theta}_0) = \frac{\partial l_i(\hat{\theta}_0)}{\partial \theta_j}$ where $\theta_{j>0}$ is the $j$th element in $\theta = (\theta_0, \theta_a)$. In other words, the element $s_{i,j}(\hat{\theta}_0)$ in the matrix $s(\hat{\theta}_0)$ is the contribution of $\frac{\partial l_i(\hat{\theta}_0)}{\partial \theta_j}$ to the score function evaluated under $H_0$. The information matrix $I(\hat{\theta}_0)$ can be estimated according to outer product of gradients. Hautsch (2004) and Pacurar (2006) provided detailed discussion regarding the LM tests.

Along with the introduction of ACD models, Engle and Russell (1998) also proposed a method to investigate the nonlinear dependencies between residuals and the past information set. The residuals are divided into bins ranging from zero to infinity. Then, they regress the residuals on indicators showing whether the previous duration has been in one of those bins. If the residuals are iid as expected, then the regression should present no predictability and therefore the coefficients of each indicator should be zero. When rejected, the bins corresponding to indicators with significant coefficients now provide information regarding the source of misspecification.

As discussed previously, there are many literatures regarding the test of ACD models. Many of the ACD tests are developed to test specific market microstructure hypotheses. The selection of those tests depends largely on the specific question that one wants to study. For instance, the residual diagnostics can provide valuable information for choosing the appropriate distribution assumption of $\epsilon_i$. Meanwhile, if one wants to test the correctness of the specification of the conditional expectation $\psi_i$, then LM tests might be needed to give a complete picture of the goodness of fit.

### 2.2.3 ACD Papers

As discussed previously, in order to test different market microstructure hypotheses, the definition of durations is often adjusted. Thus, I categorize the empirical results of ACD literatures with respect to the definition of durations. I discuss some stylized facts that are widely identified for the durations and review the microstructure hypotheses that are often tested using ACD specifications. In general, there are two main topics that most of the researchers focus on: the model specification that can fit the dynamics of durations and the hypothesis test of different microstructure theories.

### 2.2.3.1 Trade durations

The durations can be naturally defined as the time difference between consecutive transactions as in the point process modelling of arrivals of transactions. I now call durations defined in this sense as the "trade duration". There is, of course, a wide range of trade durations literatures. See., e.g., Engle and Russell (1998), Engle (2000), Zhang et al. (2001), Bauwens and Veredas (2004), Bauwens (2006) and Manganelli (2005).

Several stylized facts of trade durations are widely identified using different trade duration data from different markets. First, the trade durations, from different markets, are found to exhibit the phenomenon of "duration clustering" which can be viewed an analog of the "volatility clustering" for the duration series. For example, the durations have significant positive autocorrelations. In other words, long (short) durations tend to be followed by long (short) durations. Besides, the autocorrelation function decays very slowly, which in turn suggests that the persistence should be considered when modelling the trade duration dynamics. In general, the effect of the duration clustering is analyzed by the correlogram of the duration series and the LjungBox statistics. A very slowly decaying autocorrelation functions might be indicative of the existence of long-memory characteristics of the duration process. Empirical evidence for the long memory characteristics of duration series can be found in Engle and Russell (1998), Jasiak (1998) and Bauwens et al. (2004). In fact, Jasiak (1998) developed the Fractional Integrated ACD (FIACD) model to capture this effect.

Second, most of the trade durations identify the phenomenon of "overdispersion". That is, the standard deviation of the trade duration exceeds the mean of trade durations. The dispersion test of Engle and Russell (1998) can be applied to test this effect formally. Or simply, Dufour and Engle (2000) used a Wald test to examine the equality between the standard deviation of the trade duration and the mean of trade durations. In general, the empirical distribution of the trade durations has a hump at very short durations and long right tail (see, e.g., Bauwens et al. 2004, Giot 2001 and Engle and Russell 1998). This feature usually is interpreted as the evidence that the exponential distribution is not appropriate for the unconditional modelling of trade durations (for the standard exponential distribution, the first and second moments are equal to unit). Nevertheless, it does not necessarily mean that the conditional trade durations are not exponentially distributed.

Third, a large proportion of the trade durations sample has a value of zero or nearly indistinguishable from zero. This phenomenon suggests that there are many transactions occurring simultaneously. Moreover, the raw tick-by-tick trade durations are subject to institutional features such as the data recording mechanism. Unfortunately, those are of no common knowledge.

A common approach that is used to eliminate the zeros in the trade duration series is to aggregate these simultaneous transactions. Price are averaged according to volumes. The market microstructure theory behind the procedure is split-transaction strategy of the specialists. That is, large orders are carried out by splitting into small orders. When transactions in the data do not occur simultaneously but are spaced in time with varying short intervals, the identification of split-transactions might be very difficult. Besides, multiple transactions occurring simultaneously might be informative since it reflects a rapid pace of the market. This is line with the findings of empirical researches that the trading intensity have very strong explanatory power over the price dynamics (Ane and Geman 2000, Easley et al. 1995 and Jones et al. 1994).

Interestingly, Zhang et al. (2001) found that the exact number of the transactions occurring simultaneously has very limited explanatory power over future transactions rate, and meanwhile the existence of such multiple transactions does have significant explanatory power over future transaction dynamics. They further incorporated this effect into their TACD model by introducing a lagged indicator representing the existence of simultaneous transactions into the specification of conditional expectation of trade durations. Another explanation for zero trade durations is provided by Veredes et al. (2001). They argued that simultaneous transactions are possibly caused by the efforts of traders to trade at the round prices. Specifically, traders tend to post limit orders to be executed at the round prices. Their theory is based on the fact that the prices of simultaneous transactions are clustered around round prices.

With respect to the modelling of trade durations, the most frequently used model is the standard ACD and the log-ACD with low orders of the lagged term such as the ACD (1,1) or log-ACD (1,1) (Engle and Russell 1998, Engle 2000, Zhang et al. 2000 and Fernandes and Gramming 2006). Interestingly, most of the authors found that the residuals still exhibit significant autocorrelations. This phenomenon is viewed as a signal of the nonlinear dependence in the duration series because the standard ACD models are linear. Bauwens et al. (2004) provided an exhaustive review and reported that almost all existing ACD models (standard ACD, log-ACD, TACD, EACD among the others with various distributional assumptions) cannot provide a satisfied modelling of the conditional expectation of durations. Besides, Dufour and Engle (2000), Engle and Russell (1998) noticed that the ACD specification with the best in-sample test results generally does not provide the best out-sample forecasts. The persistence of the ACD specification are found to be very high (Engle and Russell 1998, Engle 2000 and Dufour and Engle 2000). For example, for a standard ACD model the sum of the two parameters $\alpha$ and $\beta$ is very close to one.

Regarding testing market microstructure hypothesis using trade durations, Engle (2000) used UHF-GARCH model to investigate the relationship between volatilities and trade durations. He found a significant negative correlation between conditional expectation of trade durations and volatility, which is in line with the theory of Easley and O'Hara (1992) where "no trade means no news". Besides, they found that, when adding the lagged bid-ask spread as an explanatory variable into the conditional volatility specification, the coefficient of the spread is positive. In other words, a high spread is sign of rising volatility. Similar results can also be found in Feng et al. (2004), where they also found a negative correlation between the trade durations and realized volatility.

In contrast, Gramming and Wellner (2002) studied the interdependence between trading intensity and volatility. They found that the trading intensity are negatively correlated with lagged volatility, which is consistent with the Admati and Pfleiderer (1988). According to their results, the trade durations are positively correlated with lagged volatility. Engle and Russell (2005) applied an ACM-ACD model on the tick-by-tick data of Airgas on NYSE and found that the long durations are associated with declining prices. This finding is in line with the theory of Diamond and Verrechia (1987), where "no news is the bad news".

There are several remarks that be summarized by the above discussion with respect to the trade duration modelling.

First, the empirical results suggest that, while there are a wide range of ACD specifications, their performance in the modelling of conditional expectation of trade durations are far away from satisfactory. The uncertainty of misspecification further complicates this issue. It naturally rises the question to select the most appropriate distributional assumptions.

Second, most of the researches using data from NYSE, usually blue chips with very high liquid. The transaction record is subject to institutional features that are hard to analyze and eliminate.

Finally, while in general the empirical results regarding the interdependence of volatility and durations support the theory of Easley and O'Hara (1992), partially contradictory empirical evidence can be found in a way similar to the theoretical microstructural models.

### 2.2.3.2 Price durations

As discussed in Section 2.1.3.2 with respect to the relationship between duration and volatility, sometimes selecting procedures are applied to the raw tick-by-tick duration series to test some market microstructure hypotheses.

In general, a large proportion of the tick-by-tick trade durations are associated with unchanged prices. Thus, sometimes special attentions are paid to transactions that are associated with certain change in the prices. That is, given the raw tick-by-tick transaction data, one keeps the transactions with certain price movements, the time difference between those transactions are defined correspondingly as the durations.

Specifically, consider certain amount of price movement, *C*. Suppose the sequence of tick-by-tick transaction arrival time is $\{t_0, t_1, \ldots, t_n\}$ and the associated price series is $\{p_{t_0}, p_{t_1}, \ldots, p_{t_n}\}$. Let $\tau_0 = t_0$, the series $\{\tau_0, \tau_1, \ldots, \tau_m\}$ are defined inductively by $\tau_i = t_s$ where $s = \min\{j \mid |p_{t_j} - p_{\tau_i}| \geq C, t_j > \tau_i, j = 1,2,3,\ldots,n\}$ for $i>0$. For instance, since $\tau_0 = t_0$, $\tau_1$ would be the first $t_j$ such that $|p_{t_j} - p_{t_0}| \geq C$ and $\tau_2$ would be the first $t_j$ such that $|p_{t_j} - p_{\tau_1}| \geq C, t_j > \tau_1$. I will call the duration series in which each duration is associated with certain price changes as the "price durations". One can immediately infer from the definition that the price duration is closely related with the volatility since each duration is associated with a price movement that is equal to or larger than *C*. The conditional expectation $\psi_i$ of price durations then can be roughly understood as an indicator of the time needed to expect a price movement larger than *C*.

Engle and Russell (1998) presented a clear parametric relation between the price duration and the condition instantaneous volatility. Specifically, let the information set until *i*th transaction be $F_i$. Then the conditional instantaneous intraday volatility is related with the conditional hazard of the price durations by

$$\sigma^2(t|F_{i-1}) = \left(\frac{C}{p(t)}\right)^2 h(w_i|F_{i-1}), \tag{57}$$

where $\sigma(t|F_{i-1})$ is the conditional instantaneous volatility, $p(t)$ is the price and $h(w_i|F_{i-1})$ is the conditional hazard function.

In general, empirical research finds that the price durations share many similar characteristics with the trade durations such as "duration clustering", overdispersion, right-skewed empirical distribution (Engle and Russell 1998, Bauwens and Giot 2000, Bauwens et al. 2004 and Fernandes and Gramming 2006). Interestingly, under the ACD specifications, compared with trade durations, the residuals of price durations are much more regular with respect to the serial dependence. Empirical evidence regarding this phenomenon can be found in Bauwens et al. (2004) and Fernandes and Gramming (2006).

With respect to the model specification, Bauwens et al. (2004) argued that the simple ACD model such as the standard ACD and log-ACD outperforms the complicated models such as the TACD and SVD. They also suggested that, compared with the model specification, the distributional assumption seems to have stronger influence on the forecast performance of the ACD models.

As discussed previously, the definition of price durations itself can be viewed as a measurement of the volatility. Therefore, the ACD modelling for price durations can be viewed as a volatility modelling method.

Papers with respect to this consideration can be found in Giot (2000), Gerhard and Hautsch (2002) and Giot (2002). For instance, Giot (2000) applied an ACD specification to the price durations of IBM and directly used the estimated ACD conditional expectation function to compute the intraday volatility by (57). Giot (2002) further built an Intraday value at risk model where the volatilities are computed according to the price durations modelled by log-ACD specifications. However, empirical results suggest that the price duration based model fails most of the time for all stocks under consideration. Giot (2002) argued that the unsatisfactory results might be due to the normal assumption imposed on intraday returns.

With respect to the market microstructural hypothesis tests, the empirical results from price durations are in general very similar to the results from trade durations. For instance, Engle and Russell (1998) and Bauwens and Giot (2000) found out that the price durations are negatively correlated with volatility, trading intensity and volumes.

### 2.2.3.3 Volume durations and others

It is clear that the tick-by-tick trade durations can be thinning by volumes in the same way as by prices. It was first introduced by Jasiak et al. (1999) as a measure of liquidity since it measures the time needed for given amount of volume to be traded. In this sense, it could be understood as the time cost of liquidity.

Compared with the trade durations and price durations, there are much fewer research applying ACD models to volume durations (Bauwens et al. 2004, Bauwens and Veredas 2004 and Fernandes and Gramming 2006). The empirical results suggest that the volume durations have statistical characteristics different from the price durations and trade durations. The standard deviation of volume durations tends to be smaller than the mean, suggesting the "under-dispersion". Meanwhile, as discussed previously, the trade durations and price durations exhibit overdispersion.

Regarding the ACD specification, Bauwens et al. (2004) suggested the use of standard ACD and log-ACD with Burr innovations, which means that the $\epsilon_i$ in (9) is assumed to have a Burr density. He also argued that, as in the case of price duration and trade duration, sophisticated ACD model cannot outperform the simple ACD models with respect to the predictability.

The ACD modelling of irregularly spaced financial data is not limited to the intraday trading process. Interesting applications of ACD models to the risk management can be found in Christoffersen and Pelletier (2004) and Focardi and Fabozzi (2005). For instance, Christoffersen and Pelletier (2004) used it to back test the Value at Risk models. Specifically, they applied ACD specifications to model the time difference between violations of some VaR models. If the VaR is correctly specified, then the conditional expectation of durations for the violations should be a constant that is decided by the significance level of the VaR model.

3. **Data and some empirical results**

In this section I present the statistical description of the data base. In Section 3.1 I present a discussion of the market microstructural issue of "split-transactions" associated with the tick-by-tick trade durations of SPY on NYSE. In Section 3.2 I present the statistical descriptions of the transaction-aggregated durations. Section 3.3 includes the analysis of the intraday periodicity of durations. Section 3.4 presents the unconditional distributional properties of the durations.

**3.1 Market microstructural issues and transaction-aggregated durations**

My data is the tick-by-tick transaction records of the symbol SPY on NYSE. The data ranges from 01/02/2014 to 12/31/2014. There are 252 trading days during this period. The out-of-hour transactions whose occurrence lie outside the time period between 9:30 am and 4:00 pm on each day are removed. There are 79156264 tick-by-tick observations in the sample. Three random variables are associated with each observation in the tick-by-tick transaction records: timing of transaction, execution price of transaction and trading volumes.

Before my presentation of the tick-by-tick durations, there are several market microstructural issues that I will discuss. These issues might be considered as irrelevant for studies using sample from fixed time interval. However, for a tick-by-tick transaction data of a symbol with ultra-high liquidity like SPY, the interpretation of the data with respect to transaction arrivals depends largely on the understanding of those issues.

The first issue that may arise when using tick-by-tick transaction data is the difficulty of matching trades and quotes. This problem occurs if the trades and quotes are separately recorded and stored like the Trade and Quote (TAQ) database released by NYSE. Fortunately, my data is not directly quoted from the TAQ. The tick-by-tick transaction record is obtained from the commercially available data company of Tick Data and the trades and quotes are matched automatically. Nevertheless, this convenience comes at the cost of the information of bid and ask prices.

The second issue is the so called "split-transactions" effect. This phenomenon has been mentioned in the previous discussion regarding price durations. I now elaborate it and explain my methodology of dealing with this issue. Split-transaction occurs when an order (usually of large size) on one side is matched with several smaller orders on the opposite side. An immediate result is that the time difference between these transactions is extremely small. Correspondingly, the execution prices of these transactions are equal or are monotonic (increasing or decreasing, depending on how the large order is matched). For an electronic trading system, the time difference between these transactions is usually determined by the measurement accuracy. Such recording mechanism also decides whether those transactions will be considered as simultaneously occurring. In most of the research, this issue is solved by aggregating transactions occurring at the same time. However, if these transactions are carried out with extremely small time differences as in my data, the identification of split-transactions will be very difficult. Moreover, since the bid-ask quotes are not available for our data, I cannot identify the possible split-transactions according to bid-ask information.

I introduced the concept of transaction-aggregated returns. Compared with time-aggregated returns, the transaction-aggregated returns exhibit stronger aggregational normality. More importantly, it preserves the information contained in the time of transactions. Moreover, the durations between each of observation in the transaction-aggregated returns contain equal number of transactions. Thus, those durations can be considered as a direct measure of trading intensity. Given the widely recognized relation between trading intensity and volatility dynamics, it seems valuable to investigate the durations defined in this sense. I will use the terminology "*transaction-aggregated durations*" to refer to the durations that correspond to the duration between tick-by-tick aggregated returns.

Specifically, let $p_{d,n}$ be the execution price of the *n*th transaction at day *d* and let $t_{d,n}$ be the associated time of transaction. Therefore, the tick-by-tick trade duration is $w_{d,n}$, where $w_{d,n} = t_{d,n} - t_{d,n-1}$ (denote the time of the first transaction at day *d* as $t_{d,0}$).
The *n*th tick-by-tick return at day *d* is, $r_{d,n} = \ln(\frac{p_{d,n}}{p_{d,n-1}})$. Suppose we use every consecutive *T* tick transactions to calculate our non-overlap transaction-aggregated return series. Then $(T)r_{d,i}$, the *i*th transaction-aggregated return on day *d*, is defined as
$$(T)r_{d,i} = \ln(\frac{p_{d,(T-1)i}}{p_{d,(T-1)(i-1)}}).$$

Correspondingly,

$$(T)w_{d,i} = t_{d,(T-1)i} - t_{d,(T-1)(i-1)} \ .$$

Thus, the tick-by-tick trade duration can be viewed as the special case where $T=2$. Moreover, although the duration between consecutive transactions has been widely considered as a random variable, its distributional property is insufficiently explored. This might be caused by the difficulty arising with the previously discussed market microstructural issues. The definition of transaction-aggregated duration is based on the sum of tick-by-tick durations. Therefore, study with respect to the aggregational property of tick-by-tick durations can provide valuable knowledge regarding the nature of the tick-by-tick durations. This is very similar to the case of intraday returns where the aggregational normality of intraday returns in turn provides support for the stable assumption on returns.

In fact, given a tick-by-tick transaction records, the sequence of tick-by-tick transaction time naturally forms a partition for the time interval between the first and last tick-transaction. Once a definition of duration is chosen, it actually just retains some of the tick-by-tick transactions and therefore a new partition of the time interval between the first and last transaction is formed. The tick-by-tick trade durations is a refinement of the new duration sequences. In this sense, the traditional use of fixed time interval in the analysis of a financial time series is equivalent to imposing a regular partition of the time.

## 3.2 The descriptive statistics of transaction-aggregated durations

I first present the descriptive statistics of the transaction-aggregated durations. Let $T$ represent the number of tick-by-tick transactions that are used to calculate the transaction-aggregated durations. For instance, $T=400$ indicates that the durations are now the time difference between every four hundred tick-by-tick transactions. The values of $T$ are determined by the sample size of time-aggregated transaction series based on some fixed time interval. Specifically, suppose I measure the return per one second from the tick-by-tick transaction data. There will be $390 \times 60 \times 252 = 5896800$ observations in the 1-second intraday return series. Correspondingly, it means in average each 1-second return is generated by $\frac{79156264}{5896800} \approx 13.42$ tick-by-tick transactions. In order to explore the transaction-aggregated returns at high frequency level, $T$ is chosen to be 13 and 67 to correspond to 1s and 5s time intervals.

|  | Sample size | Mean | Standard Deviation | q (0.05) | q (0.25) | q (0.5) | q (0.75) | q (0.95) | Ljung-Box statistics |
|---|---|---|---|---|---|---|---|---|---|
| Tick durations | 79156012 | 0.074495162 | 0.459798842 | 0 | 0 | 0 | 0.005 | 0.423 | 1974424 |
| T-13 durations | 6596215 | 0.893933316 | 2.601303677 | 0.001 | 0.006 | 0.196 | 1.014 | 4.035 | 342587 |
| T-67 durations | 1199206 | 4.914029642 | 9.170544074 | 0.047 | 1.17 | 3.018 | 6.432 | 16.187 | 85805 |
| T-134 durations | 595032 | 9.900267661 | 14.67155606 | 0.693 | 3.252 | 6.757 | 12.985 | 29.676 | 73310 |
| T-400 durations | 198261 | 29.7117307 | 34.47228761 | 4.579 | 12.164 | 21.995 | 39.197 | 80.22945 | 42351 |
| T-800 durations | 98942 | 59.33634262 | 47.88790054 | 11.046 | 26.195 | 45.426 | 78.793 | 153.4252 | 71763 |
| T-1200 durations | 65893 | 89.08588393 | 69.41574311 | 17.73315 | 40.68475 | 69.259 | 118.2125 | 225.13925 | 48350 |
| T-2400 durations | 32876 | 178.4973975 | 131.302421 | 39.4892 | 85.379 | 141.88 | 237.111 | 435.7287 | 20892 |

**Table 1 Descriptive statistics of transaction-aggregated durations**

The *q (p)* in Table 1 represents the quantile of the sample corresponding to the probability *p,* Ljung-Box Q statistics is calculated for the first 20 lags of the sample autocorrelation function. I keep durations with length of zero to give a complete picture of the raw tick-by-tick data.

There are several distinctive characteristics of the tick-by-tick durations of SPY that can be concluded from Table 1. First, the SPY has high liquidity. The mean of the tick-by-tick durations is around 0.0745s. Further evidence can be found in the quantiles. The 0.5-quantile of the tick-by-tick durations is zero and the 0.95-quantile is only 0.423s. The quantiles are directly calculated from the empirical distribution. For instance, the 0.95-quantile is solved from the equation $F(x) = \Pr(X \leq x) = 0.95$, where $F(x)$ is the empirical distribution function. In this case, the microstructure issues discussed previously have very strong impact on the pattern of tick-by-tick durations since the tick-by-tick durations have very small average that is very close to zero. The split-transaction effect is strong.

Second, the tick-by-tick durations exhibit strong overdispersion since the standard deviation exceeds the mean significantly. Given a mean of 0.0745 the standard deviation of the tick-by-tick standard deviation is 0.4598. This can be viewed as an evidence against the exponential distribution since the exponential distribution has a mean equal to its standard deviation.

Finally, as suggested by the Ljung-Box statistics, the null hypothesis that the first 20 lags in the autocorrelation functions have coefficients zero is clearly rejected. In contrast, the tick-by-tick durations have very strong autocorrelation, suggesting "duration clustering" effect.

In general, the transaction-aggregated durations exhibit characteristics similar to the tick-by-tick durations. For instance, the Ljung-Box statistics are of large values and the standard deviations exceed the mean in general. Nevertheless, it is noticeable that, as the number used to aggregate the tick-by-tick duration increases, the phenomenon of overdispersion gradually dies out. Specifically, the standard deviation of the *T-800* durations is 47.89, and the mean is 59.34.

Figure 1 presents the autocorrelation function of the tick-by-tick durations. As expected, I observe highly significant autocorrelations. The first lag autocorrelation is around 0.2 and the autocorrelations decay with the lags slowly for the first 50 lags. It then almost remains stable for the rest of the lags. It suggests the tick-by-tick duration process is very persistent. In order to explicitly present the decaying autocorrelation structure of the transaction-aggregated durations, I exhibit the correlogram of *T-13* and *T-67* durations for up to 500 lags. Note that since these two duration series are aggregated from the tick-by-tick duration series, now each lag represents longer time interval.

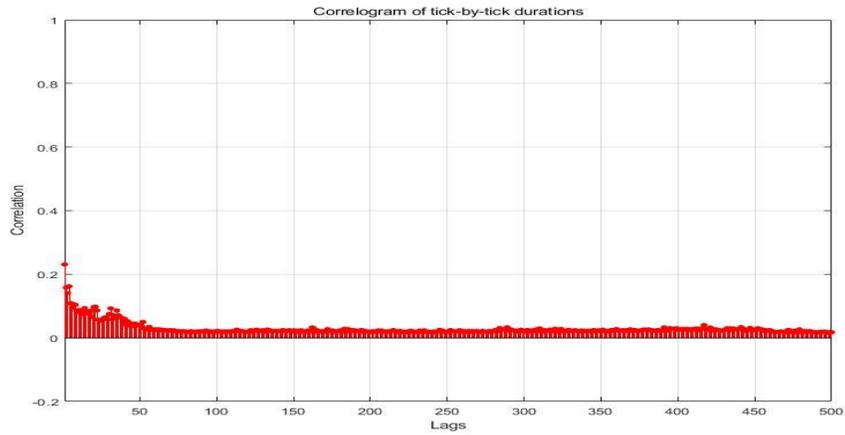

**Figure 1. Correlogram of tick-by-tick durations**

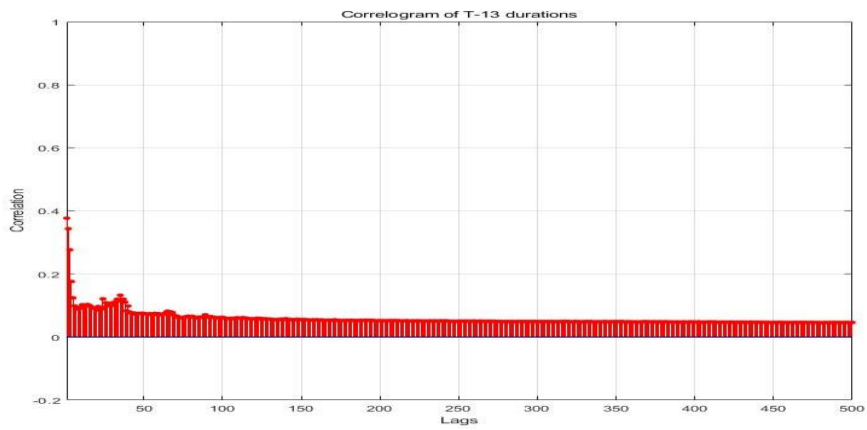

**Figure 1. Correlogram of T-13 durations**

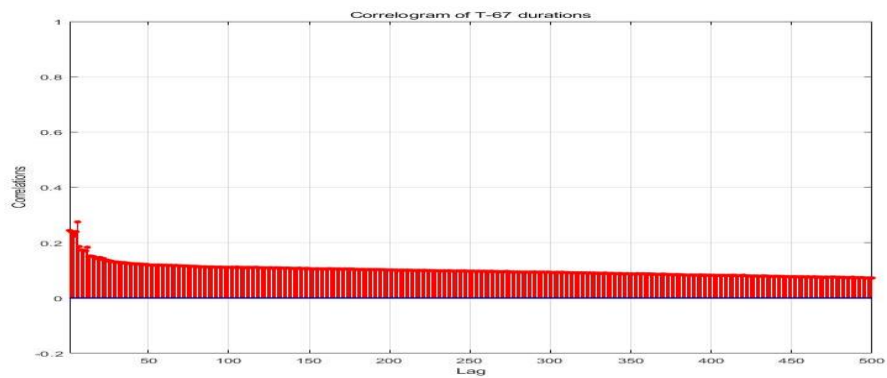

**Figure 0. Correlogram of T-67 durations**

Figures 2 and 3 present the correlogram of the *T-13* and *T-67* durations. The findings are in line with Figure 1. Observe the autocorrelations now decay with the lags at a slow, almost hyperbolic rate that is typical for long memory process. Besides, the *T-13* durations have a very strong first lag autocorrelation around 0.39.

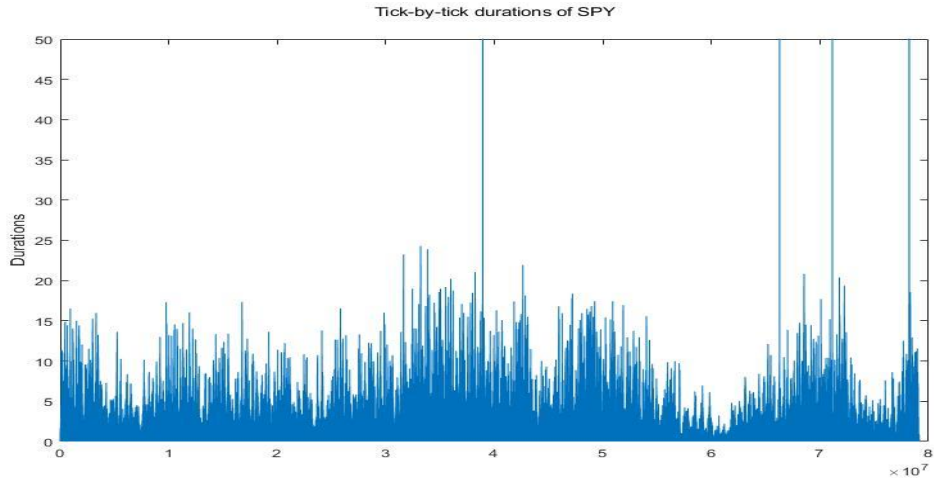

**Figure 2. The tick-by-tick duration of SPY**

Figure 4 presents the tick-by-tick duration of SPY for the whole sample. The y-axis is limited to 50s. Although the tick-by-tick durations range from zero to almost 1800 (in seconds), most of the tick durations are shorter than 1s. Besides, as suggested by figure 4, the tick durations exhibit high variability.

**3.3 Intraday seasonality of durations**

It is widely assumed that intraday transactions present very strong periodic pattern over the trading day. Consequently, before applying the ACD model to the durations of SPY, it is necessary to investigate the intraday seasonality of durations first.

In order to illustrate the intraday pattern of tick-by-tick durations, I first present the plot of average durations for 1-minute intervals over the trading day. Specifically, given a specific trading day, I count the number of tick-by-tick transactions falling in each of the 1-minute intervals and use the averages of tick durations in the 1-minute intervals to represent the tick duration dynamics over the trading day. Based on the empirical literatures, I should expect an inverted U-shape pattern of the tick durations over the trading day since the opening and closing trading hours are associated with high trading intensity. Moreover, as suggesting by Table 1 and Figure 4, the durations should be small (very close to zero) except for the "lunch time".

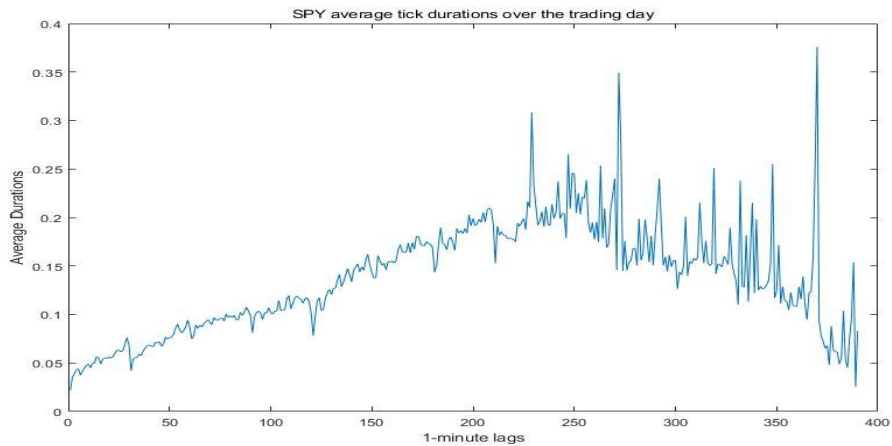

**Figure 3. The average tick-by-tick duration of SPY over the trading day**

Figure 5 presents an inverted U-shape of the tick-by-tick durations over the trading day. The average tick duration ranges from a low of 0.025s at the opening and closing of market to a high of around 0.35s at the middle day. Specifically, the tick-by-tick duration starts around 0.025s and gradually increases with the 1-minute lags. It reaches the peak around the 250$^{th}$ 1-minute lag which corresponds to the calendar time of 13:40. Then it decays slowly to 0.025 at the closing of the market. An interesting finding is that, compared with the tick durations in the morning, the tick durations in the afternoon exhibit much higher variability (observe the dramatic fluctuation of tick durations after the 200$^{th}$ 1-minute lag). As expected, the tick durations are of very small values even in the middle day.

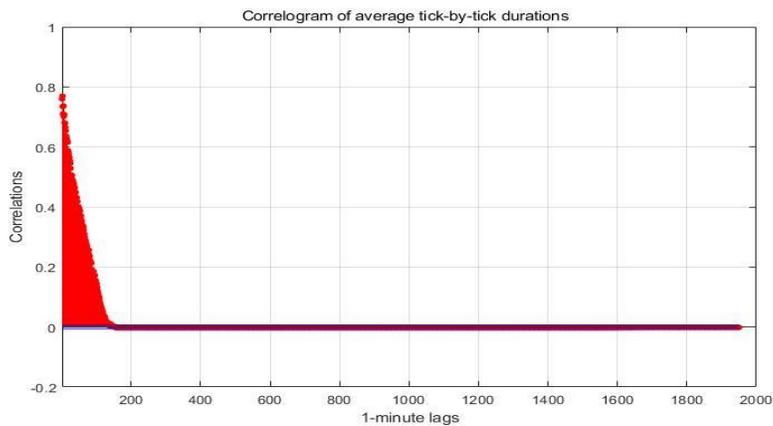

**Figure 4. The correlogram of SPY average tick-by-tick durations**

In order to investigate the dynamic feature of the average tick-by-tick durations, I further plot the correlogram of the average tick-by-tick durations for up to five days. The plot is exhibited in Figure 6. The question that I want to address is, whether the pattern depicted in Figure 5 cycles at daily basis.

Figure 6 gives a very surprising answer with respect to the intraday periodicity of durations. If the pattern depicted in Figure 5 is periodic at daily frequency, we should observe a corresponding U-shape in the correlogram as in the case of intraday periodicity of volatility. In contrast, Figure 6 suggests the autocorrelations of average tick-by-tick durations per minute decays monotonically with the 1-minute lags. The autocorrelations become insignificant for lags of order higher than 180. In other words, the average tick duration in the current minute is uncorrelated with the average tick duration three hours later.

This result is not necessarily contradictory to Figures 1, 2 and 3. Indeed, as suggested by those three figures, the autocorrelations of the transaction-aggregated durations (including the tick-by-tick durations where $T=2$) decay very slowly with lags. However, their autocorrelations do not necessarily decay slowly with time. Due to the different time interval that each duration represents, I cannot directly identify how the autocorrelation of the transaction-aggregated durations decay with time. But we can gauge this dependence by the mean of the transaction-aggregated durations. For instance, the mean of the tick-by-tick duration series is around 0.0745s. Then the time interval between the $1^{st}$ and $500^{th}$ lag in Figure 1 can be very roughly interpreted as 0.0745s×500=37.245s. Meanwhile, the average tick-by-tick durations in Figure 6 represents the average value of tick-by-tick durations in each 1-minute interval. Thus, the time difference between two consecutive lags is one minute.

The characteristics of the average tick-by-tick duration per minute can be very different from the tick-by-tick durations. In order to explicitly explore the dynamic feature of the tick-by-tick durations, one might be interested in the correlogram of the raw tick-by-tick durations for lags of higher orders. However, given the extremely small mean of tick durations (0.0745s), if I want to explore the autocorrelation structure of tick-by-tick durations for detecting some daily effect, e.g., whether the intraday pattern of Figure 5 is periodic on daily basis, an extremely large number of lags will be required. Given the tick-by-tick sample size of 79156012, the computational difficulty arises due to the high requirement of random-access memory.

Moreover, the pattern of tick-by-tick durations of SPY depends largely on the recording mechanism as discussed previously. Unfortunately, those are not common knowledge and therefore I cannot exclude their influence using my data. Thus, I exploit the autocorrelations of the transaction-aggregated durations to circumvent those difficulties.

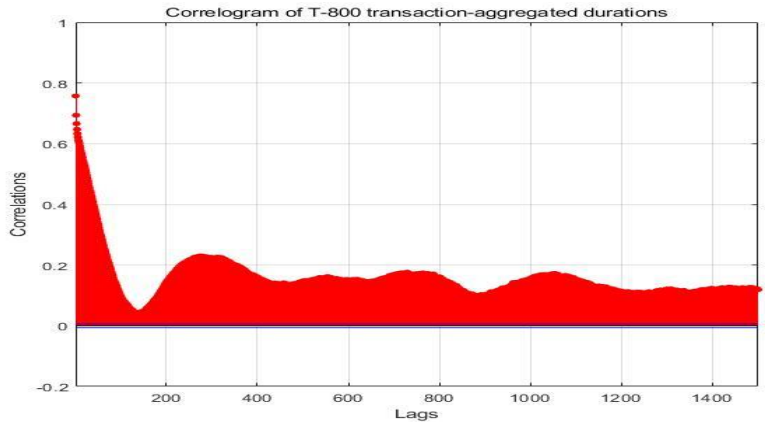

**Figure 7. The correlogram of T-800 transaction-aggregated durations**

Figure 7 presents the correlogram of the *T*-800 durations for up to 1500 lags. Given the mean of 59.34s for the *T-800* transaction-aggregated durations, Figure 7 can be roughly interpreted as the correlogram of *T-800* durations for up to four trading days. Figure 7 suggests that there is a U-shape pattern in the autocorrelations of the *T-800* durations. Although the U-shape pattern gradually dies out with the lags, it is still strikingly regular for the first 300 lags. Specifically, the distance between the first peak and the second peak of the autocorrelations is around 300 lags. Since the mean of the *T-800* transaction-aggregated durations is around one minute, 300 lags represent roughly one trading day. Figure 7 implies that, although the intraday seasonality of SPY durations is not as regular as the intraday seasonality of SPY volatilities, it still has unneglectable influence on the autocorrelation structure of the durations.

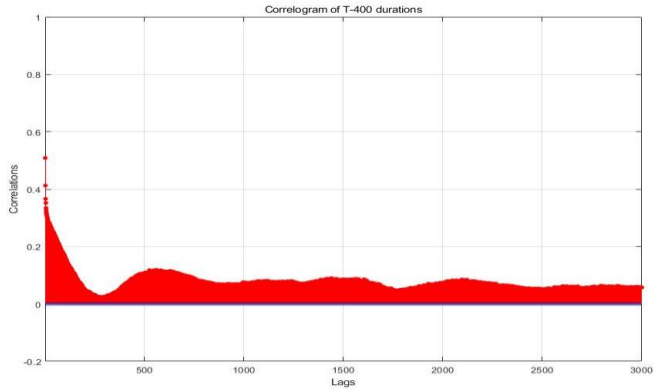

**Figure 8. The correlogram of T-400 transaction-aggregated durations**

Figure 8 gives the correlogram of the *T-400* transaction-aggregated durations for up to 3000 lags. The pattern of autocorrelations in Figure 8 is very similar to Figure 7.

Figures 7 and 8 suggest that a direct ACD modelling of the transaction-aggregated durations might be inappropriate since the ACD models impose a geometric decay on the duration autocorrelations. It cannot capture the autocorrelation pattern in Figures 7 and 8. Thus, it is necessary that I exclude the intraday seasonality of durations before applying ACD specification to model the transaction-aggregated durations.

As discussed in the Section 2.2.1, a widely accepted method that is used to exclude the intraday seasonality from the raw durations is to consider following decomposition of durations $w_i$, $w_i = \widetilde{w}_i\, s(t_i)$. $s(t_i)$ is considered as the deterministic "time-of-day" component (Engle and Russell 1998). With respect to the estimation of $s(t_i)$, the most used procedures are as follows. First, one takes averages of durations over fixed time intervals of the trading day. Second, cubic spline functions are applied to smooth the averages.

The following Figure 9 exhibits the cubic spline function with 15-minute nodes of SPY tick-by-tick durations.

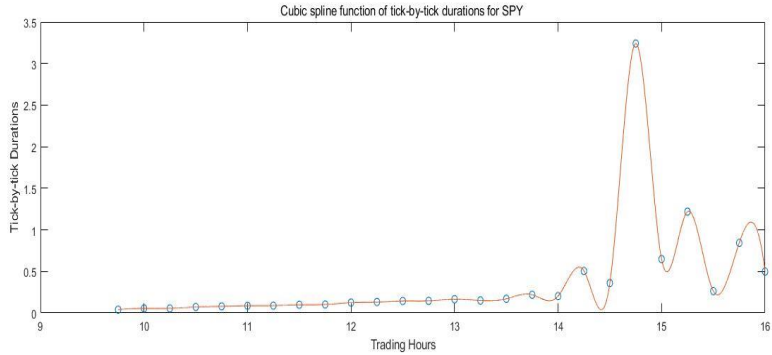

**Figure 5. Cubic spline function with 15-minute nodes of SPY tick-by-tick duration**

Once the tick-by-tick durations are adjusted according to the cubic spline function in Figure 9. I can easily calculate the seasonal adjusted transaction-aggregated durations by the sums of corresponding seasonal adjusted tick-by-tick durations.

Given the complicated intraday seasonality of tick-by-tick durations, this de-seasonal procedure might be too simple. I choose this commonly accepted procedure for the following reasons. First, as suggested by Figures 7 and 8, the autocorrelations of transaction-aggregated durations indeed present some periodic feature at daily frequencies. Second, due to the lack of economic theory that could facilitate a parametric specification of the modelling of intraday seasonality of durations, it seems plausible to choose a relatively simple nonparametric method. Third, the intraday dynamics of tick-by-tick durations are subject to strong influences of the market microstructural issues such as the "split-transactions" effect that we unfortunately are unable to identify. Therefore, it is difficult to provide an accurate modelling of the intraday seasonality.

### 3.4 Unconditional distributional properties of the durations

In this section, I provide an investigation with respect to the unconditional distributional properties of the transaction-aggregated durations. I use the raw durations instead of the de-seasonalized durations in this section since the analysis mainly serves as a guidance in the selection of distribution assumptions for ACD models. Besides, the raw durations give a complete description of the sample that I am interested in.

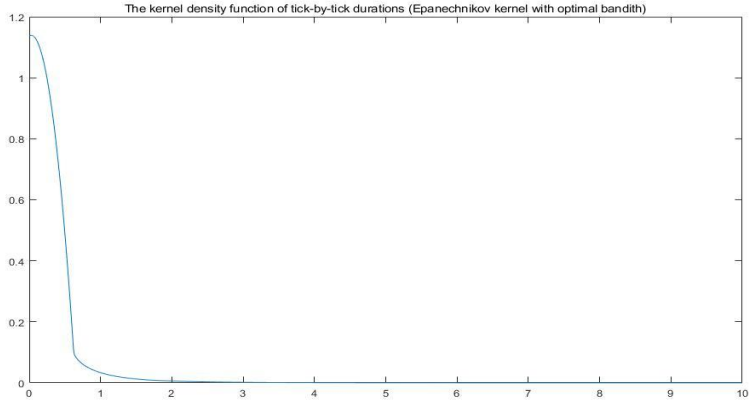

**Figure 6. The kernel density function of tick-by-tick durations**

Figure 10 exhibits the kernel density function of the tick-by-tick durations of SPY. Let $\{x_0, x_1, \dots, x_n\}$ be an iid sample following the unknown density $f$, the kernel density estimator of $f$ at $x$ is $\frac{1}{nh}\sum_{i=1}^{n} K(\frac{x-x_i}{h})$, where $K$ is the non-negative kernel function and h>0 is the smoothing parameter. I use the normal kernel $K(x) = \phi(x)$ where $\phi(x)$ is the standard normal density function. Due to the large quantity of computation required to fully exhibit the kernel density function (the maximum of tick-by-tick durations is around 1800s), I only plot the function for tick-by-tick durations less than 10s (the 0.95 quantile of the tick-by-tick sample is around 0.42s).

The kernel density function for *T*-134 durations, *T*-400 durations and *T*-800 durations are presented in the following Figures 11, 12 and 13 respectively. The range of x-axis is from 0 to 360. The kernel density functions of the transaction-aggregated durations have humps on the short durations and have relatively long right tails. Besides, the aggregational characteristics is obvious. The kernel density function of tick-by-tick durations in Figure 10 has a roughly inverted S-shape and meanwhile the kernel density functions in Figures 11, 12 and 13 are much regular and are similar to Weibull or Gamma densities. Figures 11, 12 and 13 intuitively suggest that an exponential assumption on the standardized duration $\epsilon_i$ might be wrong. Weibull distribution and generalized gamma distribution seems to be more appropriate for describing transaction-aggregated durations.

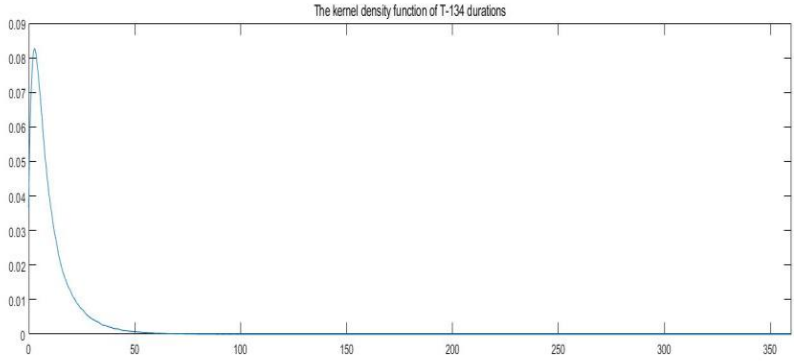

**Figure 11. The kernel density function of T-134 durations**

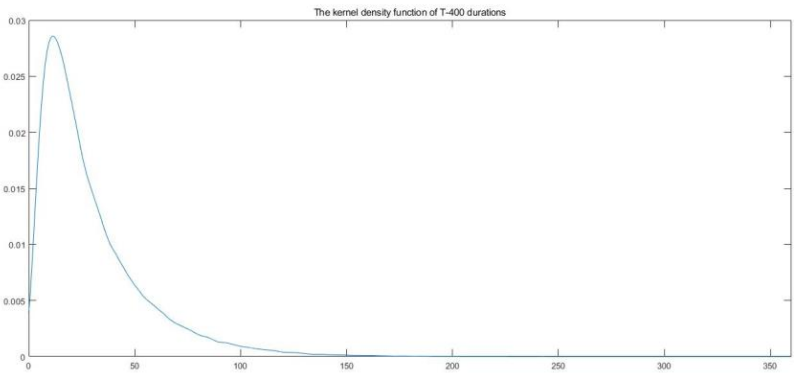

**Figure 12. The kernel density function of T-400 durations**

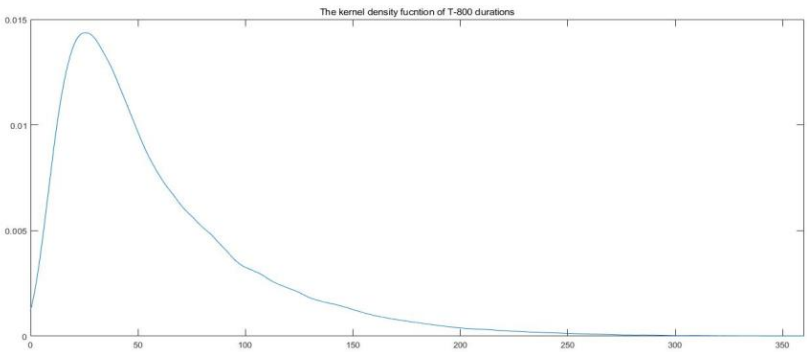

**Figure 13. The kernel density function of T-800 durations**

Given the characteristics of the tick-by-tick trade duration presented in Table 1 and Figure 10, continuous distributions that are widely used in ACD duration modelling such as Weibull distributions and exponential distributions cannot provide satisfactory description of the unconditional tick-by-tick durations.

Based on the empirical evidences from Table 1 and Figure 10 - 13, I focus on the following two issues. First, I want to analyze the effect of the market second-by-second operational details on the tick-by-tick durations. Although I cannot directly identify the effect by our data, we can gauge it by the following method. I divide the tick-by-tick durations into different categories according to their length. Then, I study the distributional properties of durations in different categories. According to market microstructure literatures, short durations are often the results of liquidity trading and meanwhile long durations are more likely to be related with informed trading (Admati and Pfleiderer 1988). Consequently, tick-by-tick durations in different categories should have very different distributional properties.

Second, I want to explore the aggregational characteristics of the tick-by-tick durations.

With respect to the first issue, I categorize the tick-by-tick durations into the followings three categories. The first category contains tick-by-tick durations with length less than 1s. Since the tick-by-tick durations have a 0.95 quantile of 0.43s, it contains most of the tick-by-tick duration data. According to Admati and Pfleiderer (1998), tick-by-tick duration in this category should be the results of liquidity transaction and therefore the market operational details play a decisive role in determining their dynamics.

It is interesting to explore the unconditional distribution of the tick-by-tick duration under such circumstances. I remove all the zero observations in this category. The second category contains tick-by-tick durations with length between 1s and 60s and the third category contains tick-by-tick durations with length larger than 60s. The following Table 2 summaries the descriptive statistics of the three subsamples.

| | Sample size | Mean | Standard Deviation | q (0.05) | q (0.25) | q (0.5) | q (0.75) | q (0.95) |
|---|---|---|---|---|---|---|---|---|
| $w \in (0, 1]$ | 31929162 | 0.099620631 | 0.191059825 | 0.001 | 0.001 | 0.009 | 0.095 | 0.559 |
| $w \in (1, 60]$ | 1418818 | 1.899863586 | 1.164725964 | 1.02 | 1.197 | 1.535 | 2.163 | 4.008 |
| $w \in (60, \infty]$ | 110 | 185.2366909 | 211.4020134 | 63.521 | 85.811 | 117.219 | 194.309 | 567.133 |

**Table 2. Descriptive statistics of tick-by-tick durations with different length**

Table 2 indicates some characteristics of the SPY tick-by-tick durations that are certainly determined by market operational details. First, the smallest unit of time that can be recorded by the market is 0.001s. This can be easily verified by the quantiles of durations less than 1s. As a result, for durations with extremely small values, the empirical distribution presents strong discreteness. For instance, the 0.5 quantile of the first subsample is 0.09 but the smallest difference between two observations is 0.001.

Second, a large proportion of the SPY tick-by-tick durations are less than 0.1s. The sample sizes of the durations in the three subsamples are 31929162, 1418818 and 110 respectively. Moreover, the 0.75 quantile of the durations that are less than 1s is 0.095.

Third, although the durations less than 1s and durations longer than 60s both exhibit "overdispersion" (sample mean less than sample standard deviation), the durations with length between 1s and 60s do not have this characteristic. Specifically, the durations in the second subsample have a sample mean of 1.899 and a standard deviation of 1.165.

Based on these empirical findings, I can conclude that the tick-by-tick duration of SPY is heavily influenced by the market microstructural effects such as the accuracy of recording systems and the "split-transactions". I now further investigate the distributional properties of the tick-by-tick trade durations. I fit the exponential distribution, Weibull distribution, the gamma distribution and the generalized pareto distribution to the first two subsamples of the tick-by-tick durations. Specifically, the four continuous distributions are estimated by MLE using the tick-by-tick durations in each subsample normalized by corresponding sample mean.

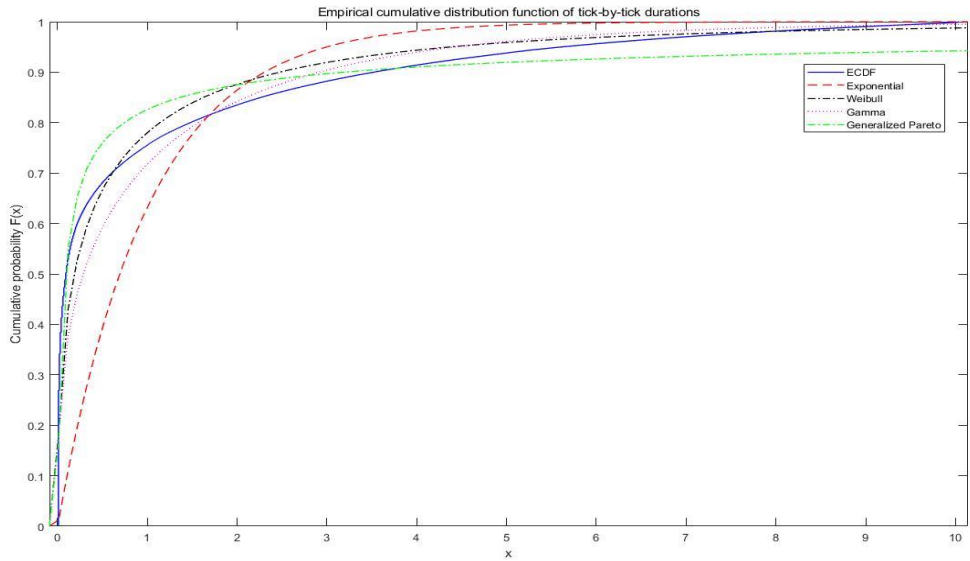

**Figure 14. The ECDF of tick-by-tick durations less than 1s**

Figure 14 presents the plot of the empirical cumulative distribution function of durations in the first subsample and the cumulative distribution functions of the four continuous distributions estimated from the subsample. The results regarding the estimation are provided in Table 3. The details regarding the maximum likelihood estimation are presented in Appendix D.

| Exponential Distribution | | |
|---|---|---|
| Parameter | Standard Error | Log likelihood |
| mu=1 | 0.000176973 | -3.19E+07 |
| | | |
| **Weibull Distribution** | | |
| Parameter | Standard Error | Log likelihood |
| (scale) a=0.4069 | 0.000165374 | -8.18E+06 |
| (shape) b=0.4623 | 6.20E-05 | |
| **Gamma Distribution** | | |
| Parameter | Standard Error | Log likelihood |
| (scale) a=0.3316 | 6.61E-05 | -1.03E+07 |
| (shape) b=3.0153 | 1.10E-03 | |
| **Generalized Pareto Distribution** | | |
| Parameter | Standard Error | Log likelihood |
| (scale) sigma=0.0571 | 2.51E-05 | -6.50E+06 |
| (shape) k=2.0664 | 5.44E-04 | |
| (location) theta=0 | 0 | |

**Table 0. The fitting of Exponential, Weibull, Gamma and Generalized Pareto distribution using durations less than 1s**

None of the four distributions can provide a satisfactory modelling of the unconditional distribution of tick-by-tick durations in the first subsample. The discreteness reported previously cannot be easily captured by the four continuous distributions that are widely used in duration modellings. Thus, ignoring such effects might be wrong for ACD duration dynamics modellings.

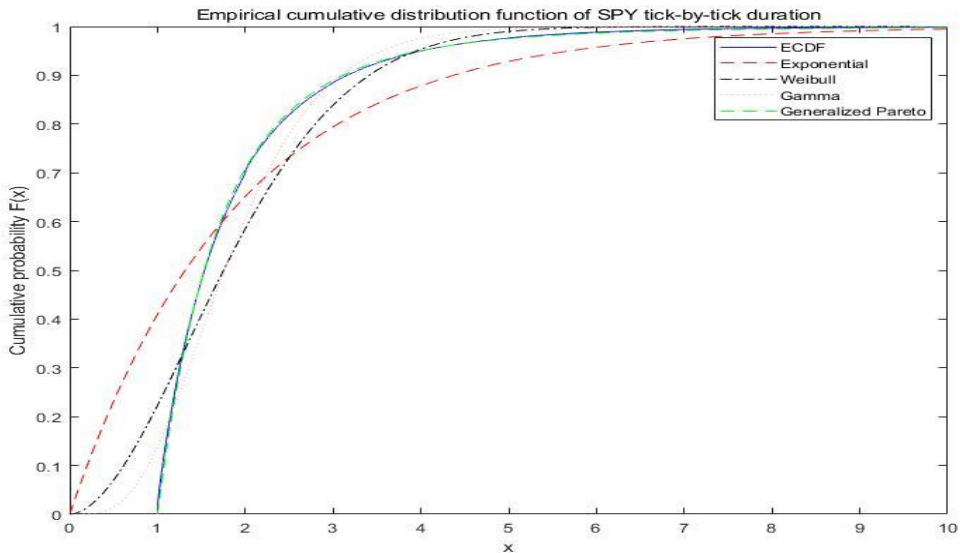

**Figure 15. The ECDF of tick-by-tick durations with length between 1s and 60s**

| Exponential Distribution | | |
|---|---|---|
| Parameter | Standard Error | Log likelihood |
| Mu=1.89986 | 0.00159499 | -2.32 E+06 |
| | | |
| Weibull Distribution | | |
| Parameter | Standard Error | Log likelihood |
| (scale) a=2.14715 | 0.00106326 | -1.94E+06 |
| (shape) b= 1.80038 | 9.36E-04 | |
| Gamma Distribution | | |
| Parameter | Standard Error | Log likelihood |
| (scale) a= 0.41307 | 0.00527462 | -1.73274E+06 |
| (shape) b=4.59937 | 0.000500552 | |
| Generalized Pareto Distribution | | |
| Parameter | Standard Error | Log likelihood |
| (scale) sigma=0.709601 | 0.00096356 | -1.23614E+06 |
| (shape) k=0.214296 | 0.00108232 | |
| (location) theta=1 | 0 | |

**Table 4. The fitting of Exponential, Weibull, Gamma and Generalized Pareto distribution using durations with length between 1s and 60s**

Figure 15 and Table 4 summarize the fitting regarding durations in the second subsample where observations vary from 1s to 60s. For clearer presentation, the x-axis is limited to 10. It can be concluded from Figure 15 and Table 4 that the four continuous distributions provide much closer approximations for the durations in the second subsample than for the durations in the first subsample. Specifically, the fit of Gamma distribution and generalized Pareto distribution seem to be promising. Although the fitting cannot be considered as statistically successful, it still implies that the characteristics of durations depends largely on the length of durations. Consequently, it intuitively suggests the use of a regime switching ACD specification for modelling tick-by-tick duration dynamics.

For the durations longer than 60s, I present the histogram to describe its distributional characteristics since there are only 110 observations. The histogram is presented in Figure 16. These durations are highly likely to be associated with some market events.

Based on the empirical results of the three subsamples, I can conclude several important remarks. First, for a symbol with high liquidity like SPY, the tick-by-tick durations are often with very small values (the limit of recording accuracy is often reached) that are results of the market microstructural effects. Second, durations with different length have very different characteristics. As suggested by microstructure literatures, the long durations might convey information very different from the information carried by short durations. Consequently, when modelling tick-by-tick durations in the ACD framework, a regime-switching specification such as the model of Hujer and Vuleti (2005) might be more realistic. Finally, given the dominating rules of market operational details in determining the pattern of tick-by-tick durations, ACD models might not be very suitable for describing the arrivals of transactions if one cannot explicitly exclude the effects of these details. For instance, the raw tick-by-tick duration sample of SPY contains 79156012 observations. Meanwhile, the number of observations equal to 0, 0.001 and 0.002 are 45807922, 8597959 and 2302916 respectively. It is highly unlikely that such discreteness can be captured by ACD specifications based on continuous distributional assumptions since there will be many jumps in the empirical distribution of durations.

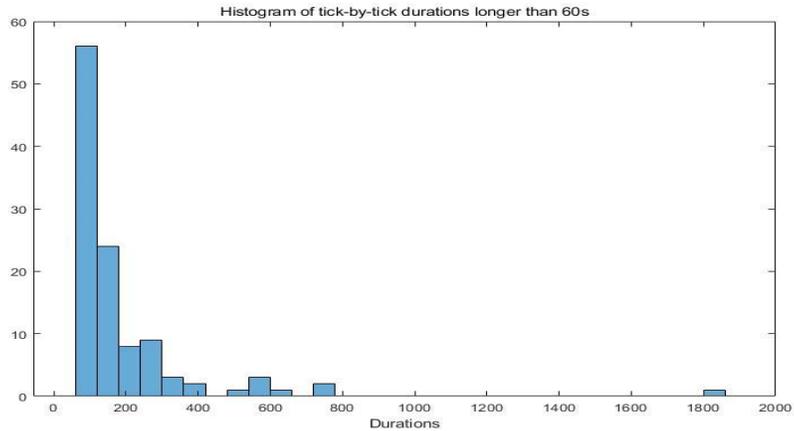

**Figure 16. The histogram of tick-by-tick durations longer than 60s**

I now use Kolmogorov statistics $D$, the Cramér–von Mises statistics $W^2$ and the Andersen-Darling statistics $A^2$, to formally test the unconditional distribution of transaction-aggregated durations. The procedures of applying EDF statistics can be in found in the Appendix A. In order to investigate the aggregational distributional property of tick-by-tick durations, I fit the exponential distribution, Weibull distribution, the Gamma distribution and the generalized Pareto distribution to transaction-aggregated durations. The first three distributions are nested in the generalized Gamma distribution and meanwhile the Weibull distribution is the limiting distribution of the Burr distribution. All zeros are removed from the sample.

Table 5 reports the EDF statistics for the goodness of fit for the transaction-aggregated durations whose descriptive statistics are given in Table 1. The distributions are estimated by MLE and Monte-Carlo simulations are applied to give critical values. Specifically, parameters of each distribution are estimated by corresponding transaction-aggregated durations and then random numbers are generated by the estimated distribution for the calculation of critical values. Details are presented in Appendix D. I do not include the critical values in Table 5 since the EDF statistics exceed the critical values significantly.

As suggested by the extremely high values of the three EDF statistics (the EDF statistics are upper tail tests), none of the four distributions can provide a satisfactory modelling of the unconditional transaction-aggregated durations. However, the generally decreasing

EDF statistics (especially for Gamma distributions) imply that there is some aggregational characteristics of the transaction-aggregated durations. Given the large sample of tick-by-tick durations, we can further investigate this aggregational characteristic by increasing $T$ that is the fixed number used to aggregate the tick-by-tick durations (the trading day with least tick-by-tick durations still have 1214888 observations).

|  | Sample size | Exponential | | | Weibull | | | Gamma | | | Generalized Pareto | | |
|---|---|---|---|---|---|---|---|---|---|---|---|---|---|
|  |  | $D$ | $W^2$ | $A^2$ | $D$ | $W^2$ | $A^2$ | $D$ | $W^2$ | $A^2$ | $D$ | $W^2$ | $A^2$ |
| T=13 | 6596215 | 0.311 | 2211.59 | Inf | 0.0812 | 12430 | Inf | 0.088 | 7669 | Inf | 0.168 | 57340 | 4E+05 |
| T=67 | 1199206 | 0.048 | 1305.789 | Inf | 0.0367 | 514 | Inf | 0.039 | 498.2 | Inf | 0.036 | 500.2 | Inf |
| T=134 | 595032 | 0.037 | 229.8506 | Inf | 0.0316 | 185.4 | Inf | 0.026 | 148.4 | Inf | 0.020 | 60.83 | Inf |
| T=400 | 198261 | 0.105 | 630.4093 | Inf | 0.0421 | 127.9 | Inf | 0.033 | 73.5 | Inf | 0.0167 | 19.64 | Inf |
| T=800 | 98942 | 0.131 | 504.7833 | 3183.103 | 0.0435 | 72.28 | Inf | 0.035 | 39.1 | Inf | 0.021 | 14.88 | 98.15 |
| T=1200 | 65893 | 0.143 | 404.4711 | 2496.856 | 0.0446 | 51.16 | Inf | 0.036 | 26.7 | Inf | 0.022 | 11.24 | 73.21 |
| T=2400 | 32876 | 0.159 | 254.6905 | 1537.22 | 0.0471 | 26.82 | Inf | 0.037 | 13.5 | Inf | 0.025 | 6.788 | 44.59 |

**Table 3. The fitting of Exponential, Weibull, Gamma and generalized Pareto distribution to transaction-aggregated durations**

|  | Sample size | Exponential | | | Weibull | | | Gamma | | | Generalized Pareto | | |
|---|---|---|---|---|---|---|---|---|---|---|---|---|---|
|  |  | $D$ | $W^2$ | $A^2$ | $D$ | $W^2$ | $A^2$ | $D$ | $W^2$ | $A^2$ | $D$ | $W^2$ | $A^2$ |
| T=4000 | 19663 | 0.171 | 175.61 | 1043.05 | 0.048 | 17.07 | 110.84 | 0.037 | 8.416 | 46.632 | 0.028 | 4.693 | 30.58 |
|  |  | [0.014] | [0.586] | [2.862] | [0.009] | [0.439] | [2.421] | [0.008] | [0.439] | [2.258] | [0.011] | [0.571] | [3.033] |
| T=8000 | 9763 | 0.187 | 103 | 600.063 | 0.04991 | 8.855 | 57.609 | 0.041 | 4.256 | 23.748 | 0.033 | 2.76 | 17.73 |
|  |  | [0.014] | [0.496] | [2.704] | [0.014] | [0.401] | [2.195] | [0.012] | [0.409] | [2.249] | [0.0148] | [0.593] | [3.032] |
| T=12000 | 6468 | 0.194 | 74.206 | 427.937 | 0.05004 | 6.007 | 38.879 | 0.039 | 2.822 | 15.851 | 0.033 | 1.927 | 12.59 |
|  |  | [0.016] | [0.421] | [2.371] | [0.015] | [0.391] | [2.091] | [0.015] | [0.438] | [2.237] | [0.015] | [0.348] | [2.061] |
| T=24000 | 3162 | 0.205 | 41.168 | 233.888 | 0.05101 | 2.634 | 17.323 | 0.038 | 1.154 | 6.5967 | 0.035 | 1.025 | 6.639 |
|  |  | [0.024] | [0.461] | [2.449] | [0.023] | [0.482] | [2.347] | [0.022] | [0.431] | [2.355] | [0.022] | [0.353] | [2.001] |
| T=40000 | 1854 | 0.217 | 26.504 | 149.038 | 0.05693 | 1.599 | 10.046 | 0.04 | 0.734 | 4.0912 | 0.04 | 0.782 | 4.904 |
|  |  | [0.032] | [0.507] | [2.535] | [0.032] | [0.539] | [2.791] | [0.033] | [0.631] | [3.423] | [0.031] | [0.401] | [2.129] |
| T=60000 | 1193 | 0.226 | 19.095 | 106.202 | 0.04383 | 0.807 | 5.3345 | 0.035 | 0.354 | 2.2714 | 0.039 | 0.497 | 3.409 |
|  |  | [0.041] | [0.625] | [3.192] | [0.035] | [0.352] | [2.003] | [0.041] | [0.555] | [2.849] | [0.037] | [0.371] | [2.105] |

**Table 4. The fitting of Exponential, Weibull, Gamma and generalized Pareto distribution to transaction-aggregated durations**

Table 6 summarizes the goodness of fit for the *T-4000, T-8000, T-12000, T-24000, T-4000* and *T-60000* transaction-aggregated durations. The critical values are reported in the brackets and the significance level is 5%. The EDF statistics are very sensitive even when the sample is relatively small. The details of the power studies with respect to EDF statistics can be found in the classic paper of Stephens (1974). There are two interesting findings that can be concluded from Table 6. First, compared with other distributions under consideration, the Gamma distributions fit the transaction-aggregated duration much better in the sense that the corresponding EDF statistics are closer to the critical values.

Second, the distributions of transaction-aggregated durations approach the Gamma distribution as the number used to aggregate to tick-by-tick durations, *T*, increases. For instance, for *T*-40000 transaction-aggregated durations, the EDF statistics are 0.04, 0.734 and 4.0912 for *D*, $W^2$ and *A* respectively that are very close to the corresponding critical values of 0.033, 0.631 and 3.423. Meanwhile, for *T*-60000 transaction-aggregated durations, the three EDF statistics are 0.035, 0.354 and 2.2714. The corresponding critical values are 0.041, 0.555 and 2.849 respectively. Thus, the null hypothesis of a Gamma distributed sample cannot be rejected at the significance level of 5%. The following Figure 17 and Figure 18 present respectively the empirical cumulative distribution functions of T-24000 and T-40000 durations together with corresponding estimated CDF of Gamma distributions.

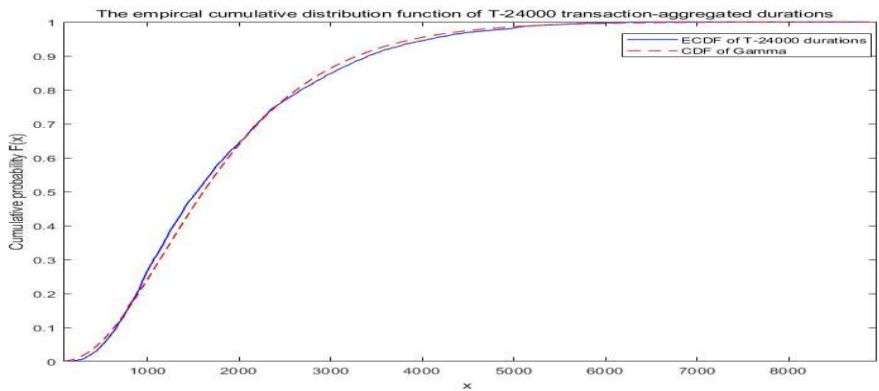

**Figure 7. The empirical cumulative distribution functions of T-24000 transaction-aggregated durations**

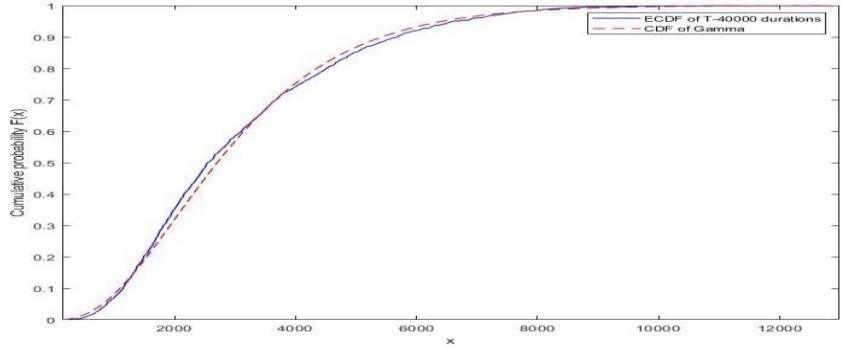

**Figure 18. The empirical cumulative distribution functions of T-40000 transaction-aggregated durations**

The above empirical results suggest the existence of the aggregational characteristics of tick-by-tick durations. It is well-known that the intraday financial transactions are not only subject to "time-of-day" effect but also "day-of-week" effect. Therefore, I further investigate the aggregational characteristics for tick-by-tick durations within each trading day. The purpose is to explore whether the intraday durations exhibit aggregational characteristics.

Specifically, I carry out the test for the null hypothesis of Gamma distribution using EDF statistics D, $W^2$ and $A^2$ on each day's transaction-aggregated durations. Transaction-aggregated durations within a trading day are considered to follow Gamma distributions if all the three statistics $D$, $W^2$ and $A^2$ cannot reject the null hypothesis at significance level of 5%. As suggested by the simple size in Table 7, the choices of T generate enough data for intraday durations.

|  | Average daily sample size | Exponential | | Weibull | | Gamma | |
| --- | --- | --- | --- | --- | --- | --- | --- |
|  |  | $N_0$ | $N_0/N$ | $N_0$ | $N_0/N$ | $N_0$ | $N_0/N$ |
| T=13 | 26175 | 0 | 0% | 0 | 0% | 0 | 0% |
| T=67 | 4758 | 0 | 0% | 0 | 0% | 0 | 0% |
| T=134 | 2361 | 2 | 0.79% | 21 | 8.33% | 16 | 6.35% |
| T=400 | 786 | 0 | 0% | 48 | 19.04% | 133 | 52.78% |
| T=800 | 392 | 0 | 0% | 81 | 32.14% | 162 | 64.29% |
| T=1200 | 261 | 0 | 0% | 112 | 44.44% | 175 | 69.44% |
| T=2400 | 130 | 0 | 0% | 160 | 63.49% | 199 | 78.97% |

**Table 5. The fitting of Exponential, Weibull and Gamma distributions to within-day transaction-aggregated duration**

The $N_0$ in Table 7 represents the number of days for which the EDF statistics cannot reject the null hypothesis that the within-day transaction-aggregated durations come from the family of corresponding distribution and *N* is the total number of trading days in our sample. Table 7 have several important implications.

First, the tick-by-tick duration presents aggregational characteristics regarding its empirical unconditional distributions. Although it cannot be described effectively by the distributions commonly used in duration modellings such as exponential, Weibull and Gamma distributions, when aggregated by tick-by-tick durations, the distribution of transaction-aggregated durations approaches a shape similar to Gamma and Weibull distributions.

Second, compared with exponential distributions, Weibull and Gamma distributions seem to be more suitable for modelling transaction-aggregated durations. Table 7 gives very clear evidence against the use of exponential distributions in modelling transaction-aggregated durations. In contrast, the Gamma distribution seems to be promising in the modelling of the unconditional transaction-aggregated durations. For instance, there are 162 trading days (64.29% of the whole sample of 252 trading days) for which the EDF statistics cannot reject the null hypothesis that the *T*-800 (each *T*-800 duration is aggregated from 800 tick-by-tick durations) durations within the trading day follow Gamma distributions. In other words, the probability of the *T*-800 durations within a trading day to follow a Gamma distribution is 62.29%. With respect to the *T*-2400 transaction-aggregated durations within a trading day, the probability of them to follow a Gamma distribution is 78.97%.

Finally, as reported previously, since tick-by-tick durations present characteristics that are directly determined by market operational details, when measured at the highest frequency (*T*-13, *T*-67 and *T*-134 durations), the transaction-aggregated durations cannot be effectively described by the continuous distributions that are commonly applied in duration modellings.

In conclusion, in this section I provide an investigation with respect to the unconditional distributional properties with respect to tick-by-tick durations. I show that the pattern of tick-by-tick durations of SPY are largely influenced by market microstructural issues discussed in section 3.3.1. Besides, I identify that the tick-by-tick durations present aggregational characteristics and Gamma distributions seem to serve as a good candidate for modelling unconditional transaction-aggregated durations.

## 4  ACD modelling of durations

In this section I present the empirical results with respect to the ACD duration modelling. I use the transaction-aggregated durations instead of the tick-by-tick durations since that, as discussed in Section 3.4, the SPY tick-by-tick durations are largely influenced by market microstructural issues. Unfortunately, I cannot explicitly exclude the effect of such issues using my data. Thus, a direct use of ACD models on the tick-by-tick duration of SPY might provide misleading results.

### 4.1 Data preparation: Intraday seasonality and re-initialization

Studies regarding intraday financial data have widely documented the strong intraday pattern over the trading day (Jain and John 1988, Andersen and Bollerslev 1997 and Bollerslev and Domowitz 1993). Before applying ACD models to describe the transaction-aggregated duration dynamics, it is necessary to exclude this intraday periodic pattern from the raw transaction-aggregated durations. Since the transaction-aggregated durations are calculated by the sum of tick-by-tick durations, excluding the intraday seasonality from the raw transaction-aggregated duration is equivalent to excluding the intraday seasonality from the raw tick-by-tick durations. In other words, I can first de-seasonalize the raw tick-by-tick durations and further build the corresponding transaction-aggregated durations by taking the sum of seasonal adjusted tick-by-tick durations.

In Section 3.3.3, I investigate the intraday pattern of tick-by-tick durations of SPY and briefly introduce the method that I use to exclude the "time-of-day" effect from the raw tick-by-tick intraday durations. I now elaborate the procedures.

I first remove all tick-by-tick durations that exceed 3s or equal to zero from the sample (the

0.99 quantile of the tick-by-tick sample without zeros is 2.226 second). Recall the assumption of (49),

$$w_i = \widetilde{w}_i \, s(t_{i-1}),$$

where $s(t_i)$ is the intraday periodicity component at $t_i$ and $w_i = t_i - t_{i-1}$ is the $i$th tick-by-tick duration. The intraday periodic component $s(t_i)$ is commonly modelled by cubic splines with nodes set on some fixed interval over the trading day. The constant on each node is given by the observed sample mean over the corresponding time interval. A clear example is presented by Figure 9 in Section 3.3.

However, as one might have noticed, the intraday tick-by-tick duration pattern in Figure 3.9 is very different from the one in Figure 3. This contradiction is mainly caused by the extremely high liquidity of SPY. Generally, the nodes of the cubic splines are setting on relatively long intervals over the trading day to make sure the spline is smooth. For instance, in Engle and Russell (1998), the nodes are setting on each hour of the trading day. Bauwens and Giot (2000) applied thirty minutes intervals. For tick-by-tick durations of SPY without zeros, the 0.99 quantile of the sample is only 2.226 second (the 0.99 quantile of the raw tick-by-tick sample with zeros is only 1.441s). Consequently, if a long interval is used, the long tick-by-tick durations that usually occur after middle day might bias the sample mean to a large extent. This is exactly the case of Figure 9 that is plotted using whole tick-by-tick sample.

I further notice that there is a dilemma when using ACD models to describe tick-by-tick duration dynamics of stocks with high liquidity like SPY. From economic point of view, the long durations are generally believed to convey important information as discussed in section 2.2.3. Nevertheless, for a symbol like SPY, most of the transactions are liquidity transactions that refer to the buy or sell for liquidity and therefore most of the tick-by-tick durations are of extremely small values. Consequently, there will be some "jumps" in the duration series. Thus, if one wants to retain as many observations as possible in the sample for ACD modellings of duration dynamics, the ACD specifications are challenged to capture these "jumps".

Besides, the characteristic of tick-by-tick durations for stocks with extremely high liquidity also suggest that the commonly used de-seasonal technique of cubic splines might need a careful review. In addition to the technique difficulty of choosing proper time intervals, a more serious issue is that the averages over time intervals of trading days might not be periodic. For instance, as discussed in Section 3.3, the averages of tick-by-tick durations over 1-minute intervals of different trading days seem to have little periodic pattern in the correlogram.

Although I believe that studies in the above two directions can yield valuable knowledge on the modelling of duration dynamics, it unfortunately is beyond the scope of this chapter. In this chapter, I simply rule out the durations that exceeds 3s as mentioned previously so that I can focus on the comparison of different distribution assumptions.

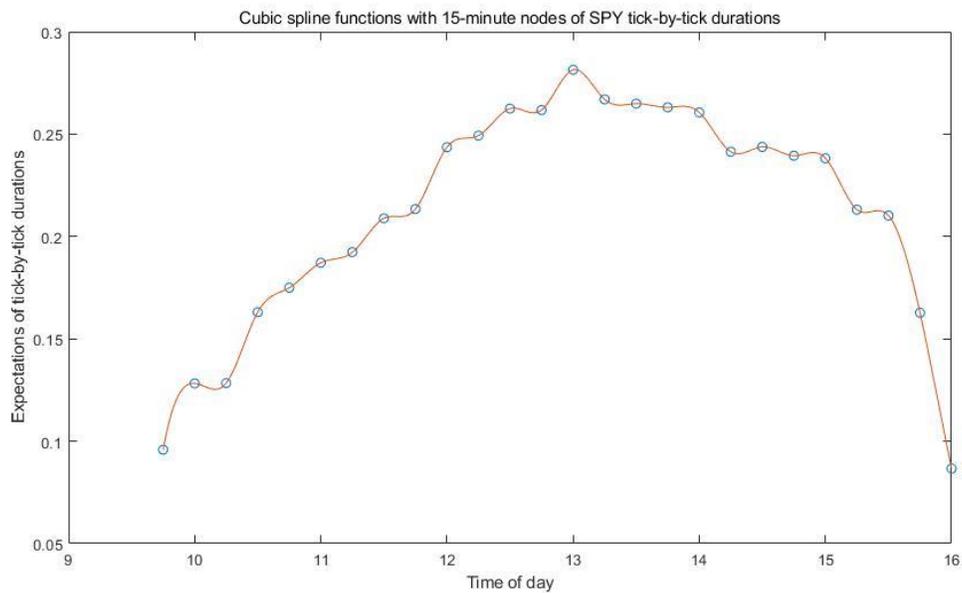

**Figure 19. Cubic spline function with 15-minute nodes of SPY tick-by-tick durations**

Figure 19 presents the cubic spline function with 15-minute nodes. The constant on each node is given by the observed sample mean of tick-by-tick durations over the corresponding 15-minute interval. Now the spline line is much more in line with the intraday tick-by-tick pattern in Figure 3. The market opening is very active as expected with transactions occurring, on average, every 0.1s. In contrast, the middle of day, around 13:00, has the longest averaged tick-by-tick durations around 0.27s. The trading intensity rises again until the closing of the market. Specifically, near 16:00, the transactions occur again almost every 0.08s.

I now exhibit the descriptive statistics of the raw transaction-aggregated durations and the seasonal filtered transaction-aggregated durations respectively.

| Raw durations | Sample size | Mean | Standard Deviation | q (0.05) | q (0.25) | q (0.5) | q (0.75) | q (0.95) | Ljung-Box statistics |
|---|---|---|---|---|---|---|---|---|---|
| Tick durations | 30542307 | 0.162 | 0.363 | 0.001 | 0.001 | 0.011 | 0.129 | 0.897 | 7891637 |
| T-13 durations | 2349290 | 2.104 | 2.162 | 0.069 | 0.533 | 1.391 | 2.977 | 6.567 | 4607113 |
| T-67 durations | 455735 | 10.847 | 7.910 | 1.675 | 4.772 | 8.898 | 15.136 | 26.474 | 2456680 |
| T-134 durations | 227801 | 21.699 | 14.405 | 4.150 | 10.443 | 18.482 | 30.140 | 49.768 | 1558918 |
| T-400 durations | 76233 | 64.831 | 38.589 | 14.885 | 34.290 | 57.538 | 89.434 | 137.979 | 629368 |
| T-800 durations | 38054 | 129.839 | 73.257 | 32.307 | 71.638 | 117.021 | 177.719 | 266.746 | 295839 |
| T-1200 durations | 25327 | 195.026 | 106.965 | 50.971 | 109.960 | 177.542 | 266.873 | 394.675 | 169913 |
| T-2400 durations | 12608 | 391.299 | 205.066 | 110.154 | 228.128 | 361.458 | 531.184 | 769.135 | 89904 |

**Table 8. Descriptive statistics of the raw transaction-aggregated durations**

| Filtered durations | Sample size | Mean | Standard Deviation | q(0.05) | q (0.25) | q (0.5) | q (0.75) | q (0.95) | Ljung-Box statistics |
|---|---|---|---|---|---|---|---|---|---|
| Tick durations | 30542307 | 0.817 | 1.790 | 0.004 | 0.008 | 0.059 | 0.697 | 4.387 | 6053335 |
| T-13 durations | 2349290 | 10.620 | 10.265 | 0.373 | 2.955 | 7.524 | 15.218 | 31.411 | 3350750 |
| T-67 durations | 455735 | 54.744 | 36.298 | 9.633 | 26.622 | 47.395 | 75.783 | 124.091 | 1808628 |
| T-134 durations | 227801 | 109.511 | 65.120 | 24.356 | 58.369 | 98.637 | 149.743 | 230.738 | 1142505 |
| T-400 durations | 76233 | 327.132 | 170.805 | 89.073 | 191.315 | 307.626 | 440.641 | 634.629 | 479633 |
| T-800 durations | 38054 | 654.958 | 320.347 | 194.702 | 400.873 | 626.931 | 873.049 | 1217.990 | 258894 |
| T-1200 durations | 25327 | 983.445 | 464.559 | 306.379 | 615.816 | 947.819 | 1303.624 | 1796.435 | 169913 |
| T-2400 durations | 12608 | 1971.052 | 881.642 | 650.639 | 1277.236 | 1927.322 | 2589.364 | 3490.888 | 55324 |

**Table 9. Descriptive statistics of the filtered transaction-aggregated durations**

As suggested by the Ljung-Box statistics in Table 9, the seasonal filtered transaction-aggregated durations still present strong autocorrelations. It suggests the periodic pattern of Figure 18 is not the only reason behind the strong autocorrelations of the raw transaction-aggregated durations. In general, compared with Table 8, the LB statistics in Table 9 decrease as expected, which suggests the cubic spline function indeed removes some trend pattern from the raw durations.

Another important issues that I address here is the re-initialization of tick-by-tick durations in each trading day. Clearly, durations are calculated consecutively from day to day. For instance, the first tick-by-tick duration of a specific trading day will always be the time difference between the first and second transactions in that trading day. Hence, I re-initialize for tick-by-tick durations within each trading day. Specifically, when applying ACD models, the conditional expectation of the first transaction-aggregated durations of each trading day will be the average of the transaction-aggregated durations over the first 15 minutes of the trading day. Consequently, the ACD process of each trading day starts at 9:45 that is in line with the cubic spline function in Figure 19.

**4.2 ACD modelling of transaction-aggregated duration dynamics**

In this section I present the empirical results of ACD modelling on transaction-aggregated durations. With respect to the conditional expectation mean specifications, I realize that there is a wide range of possibilities as discussed in Section 2. However, due to the computational difficulties arising with the complexity of those ACD models, I focus on the comparison of different distributional assumptions based on our prior findings regarding the unconditional distributional properties of tick-by-tick durations.
Thus, I choose the standard ACD ($m,q$) of (13),

$$\psi_i = \omega + \sum_{j=1}^{m} \alpha_j w_{i-j} + \sum_{j=1}^{q} \beta_j \psi_{i-j}.$$

I select the following two specifications with different orders of lags: ACD (1,1) and ACD (2,1). Regarding the distributional assumptions of the errors, the three distributions that I consider are the exponential distribution, Weibull distribution and Gamma distributions. Thus, there are six ACD specifications in total.

The diagnostics are: Log-likelihood (LL), Bayes Information Criterion (BIC) and Ljung-Box statistics (LB) with respect to 20 lags of model residuals. LL and BIC are applied to compare the different specifications and meanwhile LB is used to test the autocorrelations of residuals.

I use the *T*-13, *T*-67 *T*-134 and *T*-400 transaction-aggregated durations to fit the selected ACD specifications. Each transaction-aggregated duration series is normalized by sample means before the estimation. Given the extremely high value of LB statistics in Tables 8 and 9, it is interesting to see whether the ACD models can meet this challenge.

| *T*-13 | Exponential | | Weibull | | Gamma | |
|---|---|---|---|---|---|---|
| | EACD (1,1) | EACD (2,1) | WACD (1,1) | WACD (2,1) | GACD (1,1) | GACD (2,1) |
| $\omega$ | 0.0013 | 0.0036 | 0.0093 | 0.0029 | 0.0014 | 0.0091 |
| | (0.0002) | (0.0002) | (0.0002) | (0.0001) | (0.0002) | (0.0004) |
| $\alpha_1$ | 0.0183 | -0.0013 | 0.0557 | 0.0531 | 0.0132 | 0.0172 |
| | (0.0012) | (0.0006) | (0.0011) | (0.0009) | (0.0008) | (0.0007) |
| $\beta_1$ | 0.9445 | 0.9501 | 0.9379 | 0.9212 | 0.9762 | 0.9622 |
| | (0.0007) | (0.0005) | (0.0011) | (0.0012) | (0.0003) | (0.0003) |
| $\alpha_2$ | | 0.0087 | | 0.0427 | | 0.0105 |
| | | (0.0011) | | (0.0011) | | (0.0013) |
| | | | | | | |
| Diagnostics | | | | | | |
| Sample size | 2349290 | 2349290 | 2349290 | 2349290 | 2349290 | 2349290 |
| LL | -2011007 | -1993041 | -1966524 | -1947097 | -1773042 | -1723578 |
| BIC | 4022058 | 3986126 | 3933092 | 3894238 | 3546128 | 3447221 |
| LB | 51102 | 50502 | 48324 | 47032 | 43402 | 42456 |

**Table 10. ACD estimation of T-13 transaction-aggregated durations**

| T-67 | Exponential | | Weibull | | Gamma | |
|---|---|---|---|---|---|---|
| | EACD (1,1) | EACD (2,1) | WACD (1,1) | WACD (2,2) | GACD (1,1) | GACD (2,1) |
| $\omega$ | 0.0319 | 0.0075 | 0.0186 | 0.0056 | 0.0173 | 0.0063 |
| | (0.0008) | (0.0002) | (0.0008) | (0.0003) | (0.0004) | (0.0002) |
| $\alpha_1$ | 0.2094 | 0.1214 | 0.1497 | 0.1112 | 0.2032 | 0.1672 |
| | (0.0031) | (0.0018) | (0.0031) | (0.0022) | (0.0026) | (0.0026) |
| $\beta_1$ | 0.7593 | 0.8765 | 0.8320 | 0.8718 | 0.7762 | 0.7822 |
| | (0.0036) | (0.0024) | (0.0026) | (0.0031) | (0.0035) | (0.0034) |

|  |  | -0.0088 |  | -0.0928 |  | 0.0345 |
|---|---|---|---|---|---|---|
| $\alpha_2$ |  | (0.0030) |  | (0.0038) |  | (0.0009) |
|  |  |  |  |  |  |  |
| Diagnostics |  |  |  |  |  |  |
| Sample size | 455735 | 455735 | 455735 | 455735 | 455735 | 455735 |
| LL | -398949 | -397742 | -239314 | -233694 | -167542 | -155358 |
| BIC | 797937 | 795536 | 478667 | 467440 | 335123 | 310768 |
| LB | 10773 | 9856 | 8883 | 8241 | 7734 | 7413 |

Table 11.6 ACD estimation of T-67 transaction-aggregated durations

| T-134 | Exponential | | Weibull | | Gamma | |
|---|---|---|---|---|---|---|
|  | EACD (1,1) | EACD (2,1) | WACD (1,1) | WACD (2,1) | GACD (1,1) | GACD (2,1) |
| $\omega$ | 0.0203 | 0.0045 | 0.0303 | 0.0035 | 0.0224 | 0.009 |
|  | (0.0007) | (0.0003) | (0.0019) | (0.0005) | (0.0008) | (0.0005) |
| $\alpha_1$ | 0.3767 | 0.4655 | 0.3029 | 0.4084 | 0.2152 | 0.1856 |
|  | (0.0038) | (0.0026) | (0.0058) | (0.0034) | (0.0019) | (0.0012) |
| $\beta_1$ | 0.6078 | 0.8341 | 0.6641 | 0.9103 | 0.7362 | 0.8122 |
|  | (0.0042) | (0.0045) | (0.0069) | (0.0076) | (0.0046) | (0.0049) |
| $\alpha_2$ |  | -0.3026 |  | -0.3226 |  | -0.0305 |
|  |  | (0.0049) |  | (0.0086) |  | (0.0068) |
|  |  |  |  |  |  |  |
| Diagnostics |  |  |  |  |  |  |
| Sample size | 227801 | 227801 | 227801 | 227801 | 227801 | 227801 |
| LL | -201301 | -200758 | -90405 | -88477 | -67304 | -64311 |
| BIC | 402639 | 401565 | 180847 | 177003 | 134645 | 128671 |
| LB | 5728 | 5523 | 3628 | 3356 | 2785 | 2556 |

Table 12. ACD estimation of T-134 transaction-aggregated durations

| T-400 | Exponential | | Weibull | | Gamma | |
|---|---|---|---|---|---|---|
|  | EACD (1,1) | EACD (2,1) | WACD (1,1) | WACD (2,1) | GACD (1,1) | GACD (2,1) |
| $\omega$ | 0.0266 | 0.0030 | 0.0210 | 0.0018 | 0.0286 | 0.0173 |
|  | (0.0014) | (0.0003) | (0.0035) | (0.0003) | (0.0015) | (0.0028) |
| $\alpha_1$ | 0.4500 | 0.5393 | 0.2187 | 0.4299 | 0.2232 | 0.2172 |
|  | (0.0081) | (0.0043) | (0.0205) | (0.0064) | (0.0224) | (0.0194) |
| $\beta_1$ | 0.5275 | 0.8624 | 0.7565 | 0.9390 | 0.7282 | 0.8522 |
|  | (0.0088) | (0.0066) | (0.0242) | (0.0041) | (0.0271) | (0.0069) |
| $\alpha_2$ |  | -0.4038 |  | -0.3715 |  | -0.1105 |

|  |  | (0.0076) |  | (0.0070) |  | (0.0048) |
|---|---|---|---|---|---|---|
| **Diagnostics** |  |  |  |  |  |  |
| Sample size | 76233 | 76233 | 76233 | 76233 | 76233 | 76233 |
| LL | -68454 | -68259 | -20289 | -16527 | -12733 | -11457 |
| BIC | 136941 | 136562 | 40611 | 33098 | 25499 | 22958 |
| LB | 2382 | 2134 | 1678 | 1622 | 1328 | 1246 |

**Table 13. ACD estimation of T-400 transaction-aggregated durations**

Tables 10 – 13 summarize the ACD estimation of the four transaction-aggregated durations. The estimation is based on MLE. Standard errors are calculated by the method of White and are given in parentheses. I do not impose any nonnegative constraints on the parameters in the estimation. For this exact reason, there are negative parameters for $\alpha_2$ when the ACD (2,1) model is estimated. Nevertheless, in these cases, they are compensated by rising value of $\alpha_1$ and we do not observe any negative $\psi_i$ in the estimation. It is very intriguing since such $\alpha_2$'s is of non-trivial values. I now summaries the findings from the four tables.

With respect to the specifications, I first notice that ACD (2,1) models outperform the ACD (1,1) models dominantly. The BIC and LL of ACD (2,1) models are constantly better than ACD (1,1) models (higher LL and lower BIC). It implies that the ACD (1,1) models might be over simplistic for modelling SPY durations. Further evidences can be found in the LB statistics. Compared with LB statistics of the seasonal filtered transaction-aggregated durations in Table 9, the LB statistics of residuals of ACD models are reduced massively. Nevertheless, the LB statistics still clearly exceed the critical values of 31.41. In other words, the residuals still exhibit significant autocorrelations. This might not be very surprising since, as I mentioned previously, the SPY seasonal filtered transaction-aggregated durations exhibit very strong autocorrelations (Table 9) that imply the existence of a trend. In this case, an ARMA-GARCH specification for the logarithm of durations might provide better description of the autocorrelations. Consequently, higher orders of the lags in ACD models are preferred to give more explanatory powers over the duration dynamics. For the compactness of the Tables, I do not include the T statistics into the table (the T statistics of $w_2$ in the ACD (2,1) exceed 2 for most of the cases).

Second, the parameter $\alpha$ in general are of very small values and the $\beta$ are close to unit. For instance, for the T-13 durations, the $\alpha_1$ of EACD (1,1), WACD (1,1) and GACD (1,1) are 0.0013, 0.0093 and 0.0014 respectively and the $\beta$ of the three models are 0.9445, 0.9379 and 0.9762. It suggests that the persistence of the transaction-aggregated duration process is very strong (which is in line with the previous empirical findings in Section 3.3.2). Also, in general the $\alpha$ gradually increase and $\beta$ decreases with the frequencies of transaction-aggregated durations. Consequently, recent observations now contribute more to the conditional expectations $\psi_i$.

Finally, based on the LL and BIC statistics, the ACD models with Gamma innovations outperform the ACD models with other two innovations very obviously. For each series of transaction-aggregated durations, the GACD models always have the highest LL and the lowest BIC, suggesting that they fit the data best. It further confirms the previous empirical finding with respect to the unconditional distributional properties of transaction-aggregated durations.

The following Figures 20 and 21 exhibit the correlogram and histogram of the residuals of GACD (1,1) models for *T*-400 durations. Figure 20 clearly suggests that the residuals present significant autocorrelations at the first several lags. It should be mentioned that I select only the simplest linear ACD models. The possible nonlinear dependence cannot be captured by the linear ACD (*m,q*) model. Besides, adding explanatory variables into the conditional expectation specification might be able to improve the fit as well.

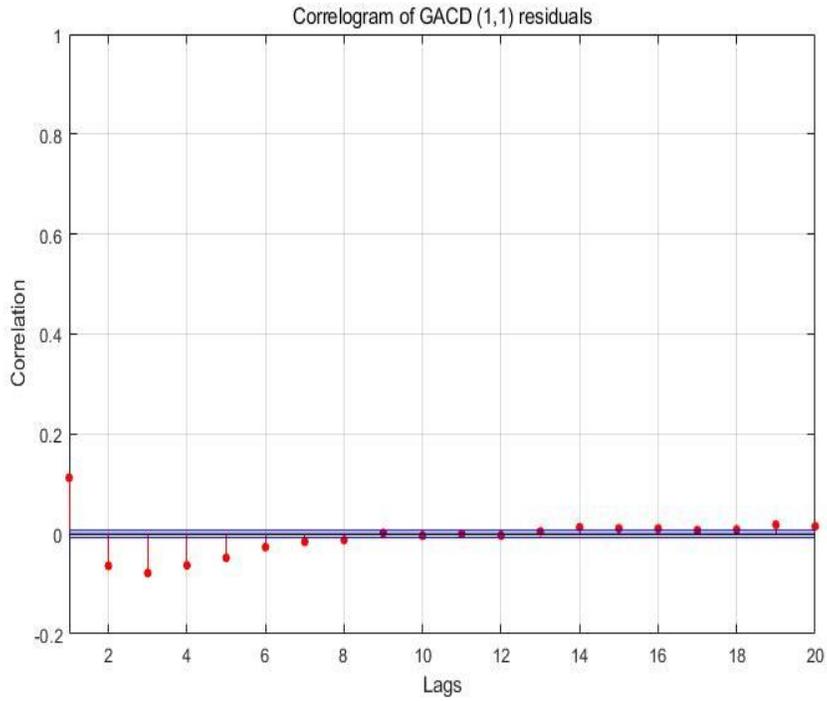

**Figure 8. The correlogram of T-400 GACD (1,1) residuals**

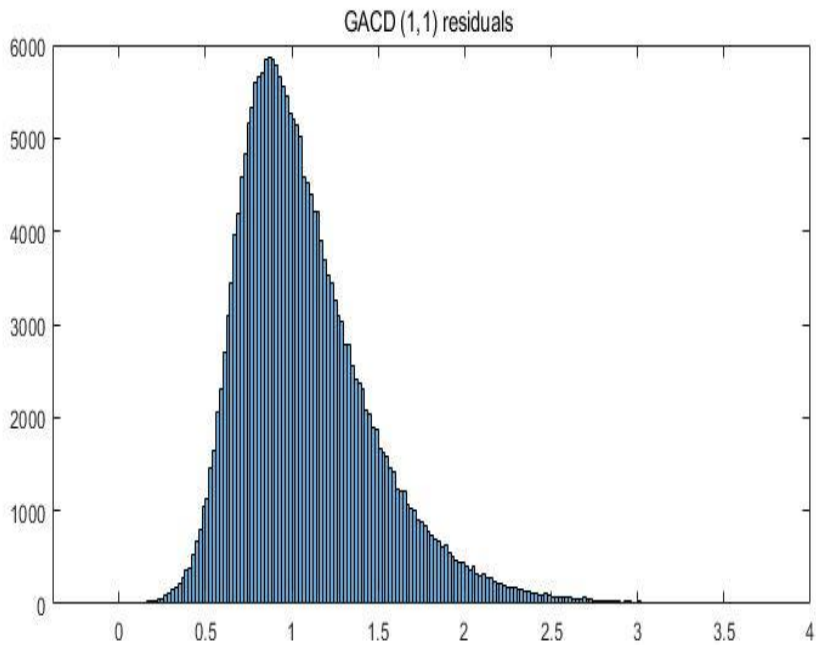

**Figure 9.The histogram of T-400 GACD (1,1) residuals**

## 5   Conclusions

In this paper I modelled of durations between transactions. I start by discussing and analyzing both theoretical and empirical works on the ACD modelling of duration dynamics. I further conduct a data analysis with respect to the tick-by-tick durations of SPY. Based on the empirical evidence, we show that the tick-by-tick transactions for stocks with ultra-high liquidity like SPY are largely affected by market microstructural issues such as the market's second-by-second operational details. Thus, studies using such data should be very careful when trying to invoke a parametric specification for modelling duration dynamics. Besides, I showed that the tick-by-tick durations have aggregational characteristics with respect to its unconditional distribution. Specifically, the empirical distributions of sums of tick-by-tick durations approximate a shape similar to the Gamma family. Standard ACD models with different distribution assumptions are applied to model the transaction-aggregated durations. The empirical results suggest that the standard ACD models cannot fully capture the interdependence structure of SPY durations. Nevertheless, the ACD specifications with Gamma distributed innovations outperforms the ACD specifications with other innovations significantly. Thus, Gamma distributions may serve as a good starting point in duration modellings. Besides, the future research direction could be the "day of week" seasonality in durations.

# Appendix A.

The classic paper of Stephens in 1974 presents a detailed discussion with respect to the empirical distribution function statistics such as the statistics $D$ (derived from $D^+$ and $D^-$), $W^2$, V, $U^2$ and $A^2$. Given some random sample $x_1, x_2, x_3,\ldots, x_n$ and the null hypothesis $H_0$ that the random sample comes from a distribution with distribution function $F(x)$, the EDF statistics compare the $F(x)$ with the empirical distribution function $F_n(x)$. In our case, we estimate the $u$ and $\sigma^2$ by the average of the sample $\hat{x}$ and $\hat{\sigma}^2 = \Sigma_i(x_i - \hat{x})/(n-1)$.

The sample are supposed to be indexed by values in ascending order, $x_1 \leq x_2 \leq x_3 \leq \ldots \leq x_n$. The test procedures are organized as follows.

(a) Estimate parameters $u$ and $\sigma^2$ as discussed previously.
(b) Calculate $z_i$'s as $z_i = F(x_i)$, where $F(x)$ is the distribution function of normal distribution with $u$ and $\sigma^2$ estimated in (a).
(c) Calculate $D$, $W^2$, V, $U^2$ and $A^2$ as follows:
  1. The Kolmogorov statistics D, D+ and D-:
  
  $D^+ = max_{1 \leq i \leq n}\left[\left(\frac{i}{n}\right) - z_i\right]$;
  
  $D^- = max_{1 \leq i \leq n}\left[-\left(\frac{i-1}{n}\right) + z_i\right]$;
  
  $D = \max(D^+, D^-)$.

  2. The Cramér–von Mises statistics $W^2$:
  
  $W^2 = \Sigma_{i=1}^n [z_i - \frac{(2i-1)}{2n}]^2 + 1/12n$.

  3. The Kuiper statistic V:
  
  V= $D^+ + D^-$.

  4. The Waston statistics $U^2$:
  
  $$U^2 = W^2 - n\left(\sum_{i=1}^n \frac{z_i}{n} - \frac{1}{2}\right)^2.$$

5. The Andersen-Darling statistics $A^2$:

$$A^2 = -\frac{\{\sum_{i=1}^{n}(2i-1)[lnz_i+\ln(1-z_{n+1-i})]\}}{n} - n.$$

I produce my table of critical values by Monte-Carol simulation. The null hypothesis $H_0$ that the sample comes from a normal distribution is rejected if the values of the statistics are larger than the critical values given corresponding significance levels. The test is the traditional upper-tail test.

| Statistics | Significance level | | |
|---|---|---|---|
| | 5% | 2.5% | 1% |
| $D$ | 0.895 | 0.955 | 1.035 |
| $V$ | 1.489 | 1.585 | 1.693 |
| $W^2$ | 0.126 | 0.148 | 0.178 |
| $U^2$ | 0.116 | 0.136 | 0.163 |
| $A^2$ | 0.787 | 0.918 | 1.092 |

**Table 10(a) Critical values of EDF statistics for the null hypothesis of normal distribution.**

# Appendix B.

The standard Central Limit Theory is the cornerstone of probability theory and statistics. It asserts that:

*If $\{X_t\}$ be a sequence of identically distributed independent random variables with $E(X_t) = u$ and $Var(X_t) = \sigma^2$ for all t. Then, $\lim\limits_{t\to\infty} P\left(\frac{\sum_{t=1}^{n} X_t - nu}{\sqrt{n}\sigma} \leq x\right) = \Phi(x)$, where $\Phi$ is the cumulative distribution function of the standard normal distribution.*

In many economic circumstances, the *i.i.d* assumption cannot be assumed to hold easily. Thus, an issue of both practical and theoretical importance is that:

Assume that $\{X_t\}_{t\in Z}$ is a process with zero mean $E(X_t) = 0$ and finite second moment $E(X_t^2) < \infty$. Define the partial sum $S_n = \sum_{t=1}^{n} X_t$ and its normalized variance $s_n^2 = \frac{Var(S_n)}{n}$. Whether or not we can apply the Central Limit Theorem to have $\lim\limits_{n\to\infty} P(S_n \leq x\sqrt{ns_n^2}) = \Phi(x)$ where $\Phi$ is the cumulative distribution function of the standard normal distribution. Moreover, if the central limit theorem holds, what is the possible convergence rate. Most of the literatures regarding the convergence rate use the Kolmogorov metric as the metric of underlying $\{X_t\}$. Define $\Delta_n(x) = |P(S_n \leq x\sqrt{ns_n^2}) - \Phi(x)|$ and further $\Delta_n = \sup\{\Delta_n(x)|x \in R\}$. Hörmann (2009) provides the bounds for normal approximation error $\Delta_n$ for dependent $\{X_t\}$. An important class of dependent $\{X_t\}$ is the ARCH/GARCH process.

In the last decades, the ARCH/GARCH model are widely applied for the modelling of time-varying volatility. In 1997, Duan introduces a general functional form of GARCH model, the so-called augmented GARCH (1,1) process, that contains many existing GARCH models as special cases. The augmented GARCH (1,1) process is given by:

$$y_t = \sigma_t \varepsilon_t,$$
$$\Lambda(\sigma_t^2) = c(\varepsilon_{t-1})\Lambda(\sigma_{t-1}^2) + g(\varepsilon_{t-1}),$$

where $\Lambda$, $c$ and $g$ are real-valued measurable functions and $\{y_t\}_{t\in Z}$ is a random variable and $\{\varepsilon_t\}_{t\in Z}$ is an *i.i.d* sequence. In order to solve $\sigma_t^2$, the definition requires the existence of $\Lambda^{-1}$. Aue et al. (2006) discuss the necessary and sufficient conditions for the augmented GARCH (1,1) model to have a strictly stationary and non-negative solution for $\sigma_t^2$. For the special case of GARCH (1,1),

$$\sigma_t^2 = \omega + \beta\sigma_{t-1}^2 + \alpha y_{t-1}^2 = \omega + [\beta + \alpha\varepsilon_{t-1}^2]\sigma_{t-1}^2 \quad .$$

It requires that $E(log|\beta + \alpha\varepsilon_0^2|) < 0$. Hörmann (2008) analyzes the dependence structure and asymptotical properties of the augmented GARCH process and shows that *m*-dependent approximations $\{Y_{tm}\}$ to the original sequence of $\{Y_t\}$ can be obtained. Specifically, $||Y_{tm} - Y_t||_2 < const \cdot \rho^m$ ($\rho < 1$), where $||\cdot||_2$ is the $L^2$ norm. He then uses the *m*-dependent approximations to construct the proof of convergence to normal distribution and corresponding convergence rate. Details of the proof can be found in Hörmann (2008).

Based on this result, consider the ARMA ($p,q$)-GARCH (1,1) model:
$$x_t = \phi_1 x_{t-1} + \cdots + \phi_p x_{t-p} + \theta_1 y_{t-1} + \cdots + \theta_q y_{t-q},$$
$$y_t = \sigma_t \varepsilon_t,$$
$$\sigma_t^2 = \omega + \beta\sigma_{t-1}^2 + \alpha y_{t-1}^2.$$

If a strictly stationary and casual solution of the equation $x_t = \phi_1 x_{t-1} + \cdots + \phi_p x_{t-p} + \theta_1 y_{t-1} + \cdots + \theta_q y_{t-q}$ exists, Brockwell and Davis (1991) show that the solution can be presented as a linear process $\sum_i \psi_i y_{t-i}$ where the coefficients $\psi_i$ exponentially decay as the lags. Then the convergence of the distribution of normalized sums of $x_t's$ to the normal distribution can be verified. According to Hörmann (2009), if $p \in (2,3]$ moments exist for $x_t$, $\Delta_n = O((logn)^{p-1} n^{1-\frac{p}{2}})$.

However, our tick-by-tick return data presents many erratic statistical features that cannot be explained by the ARMA-GARCH models. The fit of ARMA-GARCH models using tick-by-tick returns gives poor results. The required stationarity is violated since the fit gives the value of $\alpha + \beta$ that exceeds 1. More importantly, the raw tick-by-tick returns are seriously affected by the market microstructural issues such as "split-transactions" which we cannot analyze since the bid and ask prices are not included in the dataset. Thus, a theoretical modelling of the tick-by-tick price dynamics seem to be impractical.

# Appendix C

The following Matlab program "tickstnadardize" estimates the periodicity pattern for given trading interval. In practice, it is used with other programs jointly. For instance, if I want to filter the transaction-aggregated returns T-400, I will use the T-400 return series and corresponding tick-by-tick table in a for-loop to find out its periodicity pattern. The for-loop will search the whole tick-table. For time-aggregated return series, the computation is much simpler since the for-loop will be executed for fixed time interval.

```
function [r1] =tickstandardize(Ticktable,tradetimes,r)

%  tradetimes and r are one day's data for a specific day, Ticktable is

%  252*4 cell table for tick transactions

n=length(r); % number of returns in r series

for i=1:n

s2=tradetimes(i+1);

s1=tradetimes(i); % trade interval is [s1 s2] for return r(i)

v=[];

for j=1:252

 tickprices=Ticktable{1}{j};

 ticktimes=Ticktable{2}{j};

 transactions=[]; % This variable is for the transactions that occurs between s1 and s2 on day j

  for k=1:length(ticktimes)

    if ticktimes(k)>=s1 && ticktimes(k)<=s2

    transactions=[transactions;tickprices(k)];
```

```
    elseif ticktimes(k)>s2
    break
    end
  end
 standardizer=log(transactions(2:end))-log(transactions(1:end-1));
 v(j)=sum(standardizer.^2);   % use realized variance to estimate the periodicity
end
sv=(sum(v)/252)^(1/2);
r1(i)=r(i)/sv;
end
end
```

The following Matlab program "tfiteration" uses the conditional daily variance and intraday periodicity to normalize the intraday transaction-aggregated returns.

```
function [p,t,r,adjustr,ar] = tfiteration(data,tickt,tickp,T,v)
% This function uses the conditional daily variance and intraday
% periodicity to normalize the returns.
% data is the spyfiltered data,tickt,tickp are the corresppongding
% ticktable, T is the frequency index

dayindex=datasort(data{2});

for i=1:length(dayindex)-1

dayprices=data{4}(dayindex(i)+1:dayindex(i+1));
daytradetimes=data{3}(dayindex(i)+1:dayindex(i+1)); % extract the day data

[p{i},t{i},r{i},~]=tsprices(dayprices,daytradetimes,T);   % get the intraday transaction-aggregated returns.
s=[];
for j=1:length(r{i})
t1=t{i}(j);
t2=t{i}(j+1); % [t1,t2] is the timeinterval of jth intraday return
prices1=pricelock(t1,tickp,tickt);
prices2=pricelock(t2,tickp,tickt);
ticksquare=log(prices2./prices1).^2;
s(j)=mean(ticksquare./v); %s(j) is actually s^2(j)
end
```

```
adjustr{i}=(r{i}/sqrt(v(i)))./sqrt(s);
end
ar=[];
for i=1:length(dayindex)-1
ar=[ar, adjustr{i}];
end
end
```

This Matlab program "pricelock" searches the tick-by-tick transaction records for transactions occurring in specific time interval.

```matlab
function [prices] = pricelock(timing,tickp,tickt)

for i=1:length(tickp)

t=tickt{i};   %get the ith day tick time table
n=length(t);   %n is the length of transactions

right=t(t>=timing);
left=t(t<=timing);

if isempty(right)
   pright=tickp{i}(end);
else
  pright=tickp{i}(n-length(right)+1);
end

if isempty(left)
   pleft=tickp{i}(1);
else
  pleft=tickp{i}(length(left));
end

prices(i)=(pright+pleft)/2;
end
```

# Appendix D

This appendix includes the introduction of the MLE estimation for the exponential distribution, the Weibull distribution, the Gamma distribution and the generalized Pareto distribution. The optimization of the loglikelihood functions is conducted by Matlab solver based on iteration. It also includes the Matlab code for Monte-Carlo simulations that is used to produce the table of critical values for EDF statistics. The procedure is similar to the one in Appendix A.

Let $\{x_1, \ldots, x_n\}$ be the random sample from some distribution. The exponential distribution has the probability density function

$$f(x; \lambda) = \begin{cases} \lambda e^{-\lambda x} & x \geq 0, \\ 0 & x < 0. \end{cases}$$

The loglikelihood function therefore is $l(\lambda; x_0, x_1, \ldots, x_n) = n\ln(\lambda) - \lambda \sum_{i=1}^{n} x_i$.

With respect to the Weibull distribution, it has the probability density function

$$f(x; \lambda, k) = \begin{cases} \frac{k}{\lambda} \left(\frac{x}{\lambda}\right)^{k-1} \exp\left(-\frac{\epsilon}{\lambda}\right)^k & x \geq 0, \\ 0 & x < 0. \end{cases}$$

The corresponding loglikelihood function is

$$l(\lambda, k; x_0, x_1, \ldots, x_n) = \sum_{i=1}^{n} \ln\left[\frac{k}{\lambda}\left(\frac{x_i}{\lambda}\right)^{k-1} \exp\left(-\frac{x_i}{\lambda}\right)^k\right]$$

$$= n(\ln k - k\ln\lambda) + (k-1)\sum_{i=1}^{n} \ln x_i - \sum_{i=1}^{n} \left(\frac{x_i}{\lambda}\right)^k.$$

Regarding the Gamma distribution, its probability density function is

$$f(x; \alpha, \beta) = \begin{cases} \dfrac{x^{\alpha-1}\exp\left(-\frac{x}{\beta}\right)}{\beta^\alpha \Gamma(\alpha)} & x > 0, \\ 0 & x \leq 0, \end{cases}$$

where $\Gamma$ is the gamma function. Its loglikelihood function is

$$l(\alpha, \beta; x_0, x_1, \ldots, x_n) = \ln\left(\prod_{i=1}^{n} \frac{x_i^{\alpha-1} \exp\left(-\frac{x_i}{\beta}\right)}{\beta^\alpha \Gamma(\alpha)}\right)$$

$$= (\alpha - 1) \sum_{i=1}^{n} \ln x_i - \frac{1}{\beta} \sum_{i=1}^{n} x_i - n(\alpha \ln \beta + \ln \Gamma(\alpha)).$$

For the generalized Pareto distribution, it has the probability density function

$$f(x|k, \sigma, x_0) = \begin{cases} \frac{1}{\sigma}\left(1 + \frac{k(x-x_0)}{\sigma}\right)^{-1-1/k} & ; x > x_0 \text{ and } k > 0 \\ \frac{1}{\sigma} \exp\left(-\frac{x-x_0}{\sigma}\right) & ; x > x_0 \text{ and } k = 0 \\ \frac{1}{\sigma}\left(1 + \frac{k(x-x_0)}{\sigma}\right)^{-1-1/k} & ; x_0 < x < x_0 - \frac{\sigma}{k} \text{ and } k < 0. \end{cases}$$

For the simplicity, assume that $x_0 = 0$ and let $x^* = \max\{x_0, x_1, \ldots, x_n\}$. The loglikelihood function is

$$l(\sigma, k; x_1, \ldots, x_n) = \begin{cases} -n\ln\sigma + \left(\frac{1}{k} - 1\right) \sum_{i=1}^{n} \ln\left(1 - \frac{kx_i}{\sigma}\right), & k \neq 0 \\ -n\ln\sigma - \frac{1}{\sigma} \sum_{i=1}^{n} x_i, & k = 0, \end{cases}$$

where $\sigma > 0$ for $k \leq 0$ and $\sigma > k\, x^*$ for $k > 0$. The derivation of $l(\sigma, k; x_1, \ldots, x_n)$ can be found in DuMouchel (1984) and Joe (1987). The optimization algorithm of $l(\sigma, k; x_1, \ldots, x_n)$ can be found in D.Grimshaw (1993).

The following Matlab function carries out the Monte-Carlo simulation for giving critical values for EDF statistics. It is used to test the fit of duration distribution using exponential distributions and Weibull distributions. The code for Gamma distributions and generalized Pareto distribution is basically the same. The differences are the random number generating function and format of inputs. It consists of two subfunctions: stacal and critical that are presented later.

```
function [ exptable,wbltable ] = staMC(para,N,M,alpha )
% this function executes the mc
%   para is the parameters, N is the length of data(without zeros),M is the
%   size of simulation,alpha is the significance
format long
for i=1:M
exp=exprnd(para(1),para(2),1,N); % The random numbers for exp distribution
expx=sort(exp);
expz=cdf('exp',expx,para(1),para(2));
expresult(i,:)=stacal(expz);

wbl=wblrnd(para(3),para(4),1,N); % The random number matrix for wbl distribution
wblx=sort(wbl);
wblz=cdf('wbl',wblx,para(3),para(4));
wblresult(i,:)=stacal(wblz);
end

exptable=critical(expresult,alpha);
wbltable=critical(wblresult,alpha);
end
```

The Matlab function stacal is used to calculate the EDF statistics.

```
function [ result ] =stacal(z)
%   since log, values in z must be positive
n=length(z);
for i=1:n
Dp(i)=(i/n-z(i));
Dm(i)=(z(i)-(i-1)/n);
v1(i)=(z(i)-(2*i-1)/(2*n))^2;
v2(i)=(2*i-1)*(log(z(i))+log(1-z(n+1-i)));
end
D1=max(Dp);

D2=max(Dm);

D=max(D1,D2);

W=sum(v1)+1/(12*n);

V=max(Dp)+max(Dm);

U=W-n*(sum(z)/n-1/2)^2;
A=-sum(v2)/n-n;

result=[D,V,W,U,A];
end
```

The Matlab function critical is used to give critical values.

```
function [ cvalue ] = critical(stat,alpha)
%  stats is the statistics and alpha is the significance
[m,n]=size(stat);
k=length(alpha);
for i=1:n
x=sort(stat(:,i));
for j=1:k
index=round(m*(1-alpha(j)));
cvalue(i,j)=x(index);
end
end
end
```

# References

Admati, A. and Pfleiderer, P., 1988. A Theory of Intraday Patterns: Volume and Price Variability. Review of Financial Studies, 1(1), pp.3-40.

Admati, A. and Pfleiderer, P., 1988. Selling and trading on information in financial markets. American Economic Review, 78 (1988), pp. 96-103

Andersen, T. and Bollerslev, T., 1997. Intraday periodicity and volatility persistence in financial markets. Journal of Empirical Finance, 4(2-3), pp.115-158.

Andersen, T. and Bollerslev, T., 1998. Deutsche Mark-Dollar Volatility: Intraday Activity Patterns, Macroeconomic Announcements, and Longer Run Dependencies. The Journal of Finance, 53(1), pp.219-265.

Andersen, T., 1996. Return Volatility and Trading Volume: An Information Flow Interpretation of Stochastic Volatility. The Journal of Finance, 51(1), pp.169-204.

Andersen, T. G. (1994). Stochastic Autoregressive Volatility: A Framework for Volatility Modeling. Mathematical Finance, 4, 75-102.

Andersen, T., 2001. The distribution of realized stock return volatility. Journal of Financial Economics, 61(1), pp.43-76.

Andersen, T., Bollerslev, T., Diebold, F. and Labys, P., 2000. Exchange Rate Returns Standardized by Realized Volatility are (Nearly) Gaussian. Multinational Finance Journal, 4(3/4), pp.159-179.

Andersen, T., T Bollerslev, FX Diebold, P Labys 1999. Realized volatility and correlation. LN Stern School of Finance Department Working Paper.

Ané, T. and Geman, H., 2000. Order Flow, Transaction Clock, and Normality of Asset Returns. The Journal of Finance, 55(5), pp.2259-2284.

Baillie, R. and Bollerslev, T., 1991. Intra-Day and Inter-Market Volatility in Foreign Exchange Rates. The Review of Economic Studies, 58(3), p.565.

Barclay, M. and Warner, J., 1993. Stealth trading and volatility. Journal of Financial Economics, 34(3), pp.281-305.

Bauwens and Giot, 2000. The Logarithmic ACD Model: An Application to the Bid-Ask Quote Process of Three NYSE Stocks. Annales d'Économie et de Statistique, (60), p.117.

Bauwens, L. and Giot, P., 2003. Asymmetric ACD models: Introducing price information in ACD models. Empirical Economics, 28(4), pp.709-731.

Bauwens, L. and Veredas, D., 2004. The stochastic conditional duration model: a latent variable model for the analysis of financial durations. Journal of Econometrics, 119(2), pp.381-412.

Bauwens, L., Galli, F. and Giot, P., 2003. The Moments of Log-ACD Models. SSRN Electronic Journal,.

Bauwens, L., Giot, P., Grammig, J. and Veredas, D., 2004. A comparison of financial duration models via density forecasts. International Journal of Forecasting, 20(4), pp.589-609.

Bauwens, L., Preminger, A. and Rombouts, J., 2006. Regime Switching GARCH Models. SSRN Electronic Journal,.

Beltratti, A. and Morana, C., 1999. Computing value at risk with high frequency data. Journal of Empirical Finance, 6(5), pp.431-455.

Bertram, W., 2004. An empirical investigation of Australian Stock Exchange data. Physica A: Statistical Mechanics and its Applications, 341, pp.533-546.

Blume, L., Easley, D. and O'hara, M., 1994. Market Statistics and Technical Analysis: The Role of Volume. The Journal of Finance, 49(1), pp.153-181.

Blume, L., Easley, D. and O'Hara, M., 1994. Market Statistics and Technical Analysis: The Role of Volume. The Journal of Finance, 49(1), pp.153-181.

Bollerslev, T. and Domowitz, I., 1993. Trading Patterns and Prices in the Interbank Foreign Exchange Market. The Journal of Finance, 48(4), pp.1421-1443.

Bollerslev, T. and Ghysels, E., 1996. Periodic Autoregressive Conditional Heteroscedasticity. Journal of Business & Economic Statistics, 14(2), p.139.

Bollerslev, T. and Ole Mikkelsen, H., 1996. Modeling and pricing long memory in stock market volatility. Journal of Econometrics, 73(1), pp.151-184.

Bollerslev, T. and Wright, J., 2000. High Frequency Data, Frequency Domain Inference and Volatility Forecasting. SSRN Electronic Journal,.

Bollerslev, T., 1986. Generalized autoregressive conditional heteroskedasticity. Journal of Econometrics, 31(3), pp.307-327.

Bollerslev, T., Chou, R. and Kroner, K., 1992. ARCH modeling in finance. Journal of Econometrics, 52(1-2), pp.5-59.

Campbell, J., Grossman, S. and Wang, J., 1993. Trading Volume and Serial Correlation in Stock Returns. The Quarterly Journal of Economics, 108(4), pp.905-939.

Carrasco, M. and Chen, X., 2002. Mixing and Moments Properties of Various GARCH and Stochastic Volatilities Models. Econometric Theory, 18(1), pp.17-39.

Chanda, A., Engle, R. and Sokalska, M., 2005. High Frequency Multiplicative Component GARCH. SSRN Electronic Journal,.

Chernov, M. and Ghysels, E., 2000. A study towards a unified approach to the joint estimation of objective and risk neutral measures for the purpose of options valuation. Journal of Financial Economics, 56(3), pp.407-458.
92

Christoffersen, P. and Pelletier, D., 2003. Backtesting Value-at-Risk: A Duration-Based Approach. SSRN Electronic Journal,.

Clark, P., 1973. A Subordinated Stochastic Process Model with Finite Variance for Speculative Prices. Econometrica, 41(1), p.135.

Cont, R. and Bouchaud, J., 1998. Herd Behavior and Aggregate Fluctuations in Financial Markets. SSRN Electronic Journal,.

Cont, R., 2001. Empirical properties of asset returns: stylized facts and statistical issues. Quantitative Finance, 1(2), pp.223-236.

Coronel-Brizio, H. and Hernández-Montoya, A., 2005. On fitting the Pareto–Levy distribution to stock market index data: Selecting a suitable cutoff value. Physica A: Statistical Mechanics and its Applications, 354, pp.437-449.

Cox, D. and Isham, V., 2018. Point Processes. Boca Raton: Routledge.

Cumby, R., Figlewski, S. and Hasbrouck, J., 1993. Forecasting Volatilities and Correlations with EGARCH Models. The Journal of Derivatives, 1(2), pp.51-63.

Dacorogna, M., Müller, U., Nagler, R., Olsen, R. and Pictet, O., 1993. A geographical model for the daily and weekly seasonal volatility in the foreign exchange market. Journal of International Money and Finance, 12(4), pp.413-438.

Dacorogna, M., Müller, U., Pictet, O. and Olsen, R., 1997. Modelling Short-Term Volatility with GARCH and HARCH Models. SSRN Electronic Journal,.

Daley, D. and Vere-Jones, D., 2003. An Introduction to The Theory of Point Processes. New York: Springer.

Diamond, D. and V Verrecchia, R., 1991. Disclosure, Liquidity, and the Cost of Capital. The Journal of Finance, 46(4), pp.1325-1359.

Diamond, D. and Verrecchia, R., 1987. Constraints on short-selling and asset price adjustment to private information. Journal of Financial Economics, 18(2), pp.277-311.93

Diebold, F., Hahn, J. and Tay, A., 1999. Multivariate Density Forecast Evaluation and Calibration in Financial Risk Management: High-Frequency Returns on Foreign Exchange. Review of Economics and Statistics, 81(4), pp.661-673.

Drost, F. and Nijman, T., 1993. Temporal Aggregation of Garch Processes. Econometrica, 61(4), p.909.

Drost, F. and Werker, B., 1996. Closing the GARCH gap: Continuous time GARCH modeling. Journal of Econometrics, 74(1), pp.31-57.

Drost, F. and Werker, B., 2004. Semiparametric Duration Models. Journal of Business & Economic Statistics, 22(1), pp.40-50.

Dufour, A. and Engle, R. F. (2000): The ACD model: Predictability of the time between consecutive trades. Working Paper, ISMA Centre, University of Reading.

DuMouchel, W., 1983. Estimating the Stable Index $\alpha$ in Order to Measure Tail Thickness: A Critique. The Annals of Statistics, 11(4), pp.1019-1031.

Easley, D. and O'Hara, M., 1992. Time and the Process of Security Price Adjustment. The Journal of Finance, 47(2), pp.577-605.

Easley, D. and O'Hara, M., 1995, Market microstructure, Handbooks in Operations Research and Management Science, Vol.9, pp. 357-384.

Easley, D., Kiefer, N., O'hara, M. and Paperman, J., 1996. Liquidity, Information, and Infrequently Traded Stocks. The Journal of Finance, 51(4), pp.1405-1436.

Engle, R. and Gallo, G., 2006. A multiple indicators model for volatility using intra-daily data. Journal of Econometrics, 131(1-2), pp.3-27.

Engle, R. and Russell, J., 1997. Forecasting the frequency of changes in quoted foreign exchange prices with the autoregressive conditional duration model. Journal of Empirical Finance, 4(2-3), pp.187-212.

Engle, R. and Russell, J., 1998. Autoregressive Conditional Duration: A New Model for Irregularly Spaced Transaction Data. Econometrica, 66(5), p.1127.

Engle, R. and Sokalska, M., 2011. Forecasting intraday volatility in the US equity market. Multiplicative component GARCH. Journal of Financial Econometrics, 10(1), pp.54-83.

Engle, R., 1982. A general approach to lagrange multiplier model diagnostics. Journal of Econometrics, 20(1), pp.83-104.

Engle, R., 1982. Autoregressive Conditional Heteroscedasticity with Estimates of the Variance of United Kingdom Inflation. Econometrica, 50(4), p.987.

Engle, R., 2000. The Econometrics of Ultra-high-frequency Data. Econometrica, 68(1), pp.1-22.

Engle, R., 2002. Dynamic Conditional Correlation. Journal of Business & Economic Statistics, 20(3), pp.339-350.

Engle, R., Hendry, D. and Richard, J., 1983. Exogeneity. Econometrica, 51(2), p.277.

Fama, E., 1965. The Behavior of Stock-Market Prices. The Journal of Business, 38(1), p.34.

Farrell, L., 1996. Foreign Exchange Volume: Sound and fury, signifying nothing. BMJ, 313(7050), pp.174-174.

Feng, D., 2004. Stochastic Conditional Duration Models with "Leverage Effect" for Financial Transaction Data. Journal of Financial Econometrics, 2(3), pp.390-421.

Focardi, S. and Fabozzi, F., 2005. An autoregressive conditional duration model of credit‐risk contagion. The Journal of Risk Finance, 6(3), pp.208-225.

Gallant, A., Hsu, C. and Tauchen, G., 2000. Using Daily Range Data to Calibrate Volatility Diffusions and Extract the Forward Integrated Variance. SSRN Electronic J ournal,.

Gallant, A., Rossi, P.E., Tauchen G., 1993. Nonlinear dynamic structures. Econometrica, 61, pp. 871-907

Gallant, A., Rossi, P. and Tauchen, G., 1992. Stock Prices and Volume. Review of Financial Studies, 5(2), pp.199-242.

Gerhard, F. and Hautsch, N., 2001. Volatility Estimation on the Basis of Price Intensities. SSRN Electronic Journal,.

Ghashghaie, S., Breymann, W., Peinke, J., Talkner, P. and Dodge, Y., 1996. Turbulent cascades in foreign exchange markets. Nature, 381(6585), pp.767-770.

Ghose D., K.F. Kroner. 1994. Temporal aggregation of high frequency data Department of Economics, University of Arizona, Unpublished manuscript.

Ghose, D. and Kroner, K., 1995. The relationship between GARCH and symmetric stable processes: Finding the source of fat tails in financial data. Journal of Empirical Finance, 2(3), pp.225-251.

Ghysels, E., and J. Jasiak, 1998, GARCH for irregularly spaced financial data: The ACDGARCH model, Studies in Nonlinear Dynamics and Econometrics, 2, 133–149.

Ghysels, E., C. Gourieroux, and J. Jasiak, 1998, Stochastic volatility duration models, CIRANO.

Ghysels, E., Lee, H. and Siklos, P., 1993. On the (mis)specification of seasonality and its consequences: An empirical investigation with US data. Empirical Economics, 18(4), pp.747-760.

Giot, P., 2003. Implied Volatility Indices as Leading Indicators of Stock Index Returns?. SSRN Electronic Journal,.

Giot, P., 2005. Market risk models for intraday data. The European Journal of Finance, 11(4), pp.309-324.

Gnaciński, P. and Makowiec, D., 2004. Another type of log-periodic oscillations on Polish stock market. Physica A: Statistical Mechanics and its Applications, 344(1-2), pp.322-325.

Goodhart, C. and O'Hara, M., 1997. High frequency data in financial markets: Issues and applications. Journal of Empirical Finance, 4(2-3), pp.73-114.

Gopikrishnan, P., Meyer, M., Amaral, L. and Stanley, H., 1998. Inverse cubic law for the distribution of stock price variations. The European Physical Journal B, 3(2), pp.139-140.

Gramming J., Maurer K.O. (1999) Non-monotonic hazard functions and the autoregressive conditional duration model, Discussion Paper 50, SFB 373, Humboldt University Berlin.

Gramming, J., and Wellner, M. (2002). 'Modeling the interdependence of volatility and intertransaction duration processes', Journal of Econometrics, 106: 369 400.

Gu, G., Chen, W. and Zhou, W., 2007. Quantifying bid-ask spreads in the Chinese stock market using limit-order book data. The European Physical Journal B, 57(1), pp.81-87.

Guillaume, D., Dacorogna, M., Davé, R., Müller, U., Olsen, R. and Pictet, O., 1997. From the bird's eye to the microscope: A survey of new stylized facts of the intra-daily foreign exchange markets. Finance and Stochastics, 1(2), pp.95-129.

H. McInish, T. and Wood, R., 1990. A transactions data analysis of the variability of common stock returns during 1980–1984. Journal of Banking & Finance, 14(1), pp.99-112.

Harris, L. and Gurel, E., 1986. Price and Volume Effects Associated with Changes in the S&P 500 List: New Evidence for the Existence of Price Pressures. The Journal of Finance, 41(4), p.815.

Harris, L., 1987. Transaction Data Tests of the Mixture of Distributions Hypothesis. The Journal of Financial and Quantitative Analysis, 22(2), p.127.

Hasbrouck, J. and Sofianos, G., 1993. The Trades of Market Makers: An Empirical Analysis of NYSE Specialists. The Journal of Finance, 48(5), pp.1565-1593.

Hasbrouck, J., 1991. The Summary Informativeness of Stock Trades: An Econometric Analysis. Review of Financial Studies, 4(3), pp.571-595.

Hautsch, N., 2004. Modelling Irregularly Spaced Financial Data. Berlin: Springer.

Heston, S., 1993. Invisible Parameters in Option Prices. The Journal of Finance, 48(3), pp.933-947.

Hörmann, S. (2009). Berry-Esseen bounds for econometric time series. ALEA 6 377–397.

Hujer, R. and Vuletic, S., 2005. The Regime Switching ACD Framework: The Use of the Comprehensive Family of Distributions. SSRN Electronic Journal,.

Jansen, D. and de Vries, C., 1991. On the Frequency of Large Stock Returns: Putting Booms and Busts into Perspective. The Review of Economics and Statistics, 73(1), p.18.

Jasiak, J., 1999. Persistence in Intertrade Durations. SSRN Electronic Journal,.

Jirak, M., 2016. Berry–Esseen theorems under weak dependence. The Annals of Probability, 44(3), pp.2024-2063.

Jones, C., Kaul, G. and Lipson, M., 1994. Transactions, Volume, and Volatility. Review of Financial Studies, 7(4), pp.631-651.

Karpoff, J., 1987. The Relation Between Price Changes and Trading Volume: A Survey. The Journal of Financial and Quantitative Analysis, 22(1), p.109.

Lamoureux, C. and Lastrapes, W., 1990. Persistence in Variance, Structural Change, and the GARCH Model. Journal of Business & Economic Statistics, 8(2), pp.225-234.

Lancaster, T., n.d. The Econometric Analysis of Transition Data. Florida: Chapman&Hall/CRC

Lee, S. and Hansen, B., 1994. Asymptotic Theory for the Garch (1,1) Quasi-Maximum Likelihood Estimator. Econometric Theory, 10(1), pp.29-52.

Lehmann, B. and Modest, D., 1994. Trading and Liquidity on the Tokyo Stock Exchange: A Bird's Eye View. The Journal of Finance, 49(3), pp.951-984.

Lin, S. and Tamvakis, M., 2004. Effects of NYMEX trading on IPE Brent Crude futures markets: a duration analysis. Energy Policy, 32(1), pp.77-82.

Ling, S. and Li, W., 1997. On Fractionally Integrated Autoregressive Moving-Average Time Series Models with Conditional Heteroscedasticity. Journal of the American Statistical Association, 92(439), pp.1184-1194.

Ling, S. and Li, W., 1998. Limiting distributions of maximum likelihood estimators for unstable autoregressive moving-average time series with general autoregressive heteroscedastic errors. The Annals of Statistics, 26(1), pp.84-125.

Ling, S. and McAleer, M., 2003. Asymptotic theory for a vector ARMA-GARCH model. Econometric Theory, 19(02).

Lockwood, L. and Linn, S., 1990. An Examination of Stock Market Return Volatility During Overnight and Intraday Periods, 1964-1989. The Journal of Finance, 45(2), pp.591-601.

Loretan, M. and Phillips, P., 1994. Testing the covariance stationarity of heavy-tailed time series: An overview of the theory with applications to several financial datasets. Journal of Empirical Finance, 1(2), pp.211-248.

Lumsdaine, R., 1996. Consistency and Asymptotic Normality of the Quasi-Maximum Likelihood Estimator in IGARCH(1,1) and Covariance Stationary GARCH(1,1) Models. Econometrica, 64(3), p.575.

Lunde, A., 1999. A generalized Gamma autoregressive conditional duration model. Working Paper, Department of Economics, Politics and Public Administration, Aalborg University, Denmark.

Mandelbrot, B., 1963. The Stable Paretian Income Distribution when the Apparent Exponent is Near Two. International Economic Review, 4(1), p.111.

Manganelli, S., 2005. Duration, volume and volatility impact of trades. Journal of Financial Markets, 8(4), pp.377-399.

Mantegna, R. and Stanley, H., 1995. Scaling behaviour in the dynamics of an economic index. Nature, 376(6535), pp.46-49.

McCulloch, J., 1997. Measuring Tail Thickness to Estimate the Stable Index α: A Critique. Journal of Business & Economic Statistics, 15(1), p.74.

Meitz, M. and Teräsvirta, T., 2006. Evaluating Models of Autoregressive Conditional Duration. Journal of Business & Economic Statistics, 24(1), pp.104-124.

Mittink, S. and Rachev, S., 1993. Reply to comments on modeling asset returns with alternative stable distributions and some extensions*. Econometric Reviews, 12(3), pp.347-389.

Müller U., Darorogna, M., Davé R., Pictet O, Olsen R., Ward J. 1993. Fractals and intrinsic time: A challenge to econometricians, Olsen and Associates, Research Institute for Applied Economics, Working Paper.

Müller, U., Dacorogna, Davé R., Olsen R., O., Pictet, O. 1995. Volatilities of different time resolutions: Analyzing the dynamics of market components, Olsen and Associates, Research Institute for Applied Economics, Working Paper.

Müller, U., Dacorogna, M., Davé, R., Olsen, R., Pictet, O. and von Weizsäcker, J., 1997. Volatilities of different time resolutions — Analyzing the dynamics of market components. Journal of Empirical Finance, 4(2-3), pp.213-239.

Müller, U., Dacorogna, M., Olsen, R., Pictet, O., Schwarz, M. and Morgenegg, C., 1990. Statistical study of foreign exchange rates, empirical evidence of a price change scaling law, and intraday analysis. Journal of Banking & Finance, 14(6), pp.1189-1208.

Nelson, D. and Cao, C., 1992. Inequality Constraints in the Univariate GARCH Model. Journal of Business & Economic Statistics, 10(2), p.229.

Nelson, D., 1991. Conditional Heteroskedasticity in Asset Returns: A New Approach. Econometrica, 59(2), p.347.

Newey, W., 1994. The Asymptotic Variance of Semiparametric Estimators. Econometrica, 62(6), p.1349.

Nolan, J., 2018. Stable Distributions and Green's Functions for Fractional Diffusions. SSRN Electronic Journal,.

Pacurar, M., 2006. Autoregressive Conditional Duration (ACD) Models in Finance: A Survey of the Theoretical and Empirical Literature. SSRN Electronic Journal,.

Prigent, J., 2001. Option Pricing with a General Marked Point Process. Mathematics of Operations Research, 26(1), pp.50-66.

Richardson, M. and Smith, T., 1994. A Direct Test of the Mixture of Distributions Hypothesis: Measuring the Daily Flow of Information. The Journal of Financial and Quantitative Analysis, 29(1), p.101.

Ross, S., 1989. Information and Volatility: The No-Arbitrage Martingale Approach to Timing and Resolution Irrelevancy. The Journal of Finance, 44(1), pp.1-17.

Russell, J. and Engle, R., 1998. Econometric Analysis of Discrete-Valued Irregularly-Spaced Financial Transactions Data Using a New Autoregressive Conditional Multinomial Model. SSRN Electronic Journal,.

Russell, J. and Engle, R., 2005. A Discrete-State Continuous-Time Model of Financial Transactions Prices and Times. Journal of Business & Economic Statistics, 23(2), pp.166-180.

Scalas, E., Gorenflo, R. and Mainardi, F., 2000. Fractional calculus and continuous-time finance. Physica A: Statistical Mechanics and its Applications, 284(1-4), pp.376-384.

Spierdijk, L., 2004. An empirical analysis of the role of the trading intensity in information dissemination on the NYSE. Journal of Empirical Finance, 11(2), pp.163-184.

Tauchen, G. and Pitts, M., 1983. The Price Variability-Volume Relationship on Speculative Markets. Econometrica, 51(2), p.485.

Taylor, S. and Xu, X., 1997. The incremental volatility information in one million foreign exchange quotations. Journal of Empirical Finance, 4(4), pp.317-340.

Tsay, R., 2003. Analysis of Financial Time Series. Wiley.

Veredas, D., Rodriguez-Poo, J., & Espasa, A. (2002). On the (intradaily) seasonality and dynamics of a financial point process: A semiparametric approach, CORE Discussion Paper, 23, Université Catholique de Louvain.

West, K. and Cho, D., 1995. The predictive ability of several models of exchange rate volatility. Journal of Econometrics, 69(2), pp.367-391.

Wood, R., Mcinish, T. and ORD, J., 1985. An Investigation of Transactions Data for NYSE Stocks. The Journal of Finance, 40(3), pp.723-739.

Whittle, P., 1953. The Analysis of Multiple Stationary Time Series. Journal of the Royal Statistical Society: Series B (Methodological), 15(1), pp.125-139.

Young, M. and Graff, R., 1995. Real estate is not normal: A fresh look at real estate return distributions. The Journal of Real Estate Finance and Economics, 10(3), pp.225-259.

Zhang, L., 2010. Implied and realized volatility: empirical model selection. Annals of Finance, 8(2-3), pp.259-275.

Zhang, L., Mykland, P. and Aït-Sahalia, Y., 2005. A Tale of Two Time Scales. Journal of the American Statistical Association, 100(472), pp.1394-1411.